%% file: pseudo_dynamic_phasefield_model.tex
\documentclass[fleqn,authoryear]{elsarticle}

\usepackage[paperwidth=192mm,
paperheight=262mm,
top=18mm,
left=14mm,
right=14mm,
bottom=20mm]{geometry}

\usepackage{CharisSIL}
\usepackage[libertine,scaled=1.1]{newtxmath}
\usepackage{esint}
\usepackage{bm}
\usepackage{nicefrac}
\usepackage[ruled,linesnumbered,longend]{algorithm2e}
\usepackage[dvipsnames]{xcolor}

\usepackage{titlesec}
\titleformat*{\section}{\small\bfseries}
\titleformat*{\subsection}{\small\itshape}

\usepackage{subcaption}
\usepackage{booktabs}
\usepackage{lineno}
\usepackage{cases}
\usepackage{enumerate}

\usepackage{caption}
\usepackage[labelfont={bf,footnotesize},textfont=footnotesize]{subcaption}

\usepackage[colorlinks=true]{hyperref}
\makeatletter
\AtBeginDocument{\def\@citecolor{cyan}}
\AtBeginDocument{\def\@linkcolor{cyan}}
\makeatother

\DeclareMathOperator{\tr}{tr}
\DeclareMathOperator*{\argmin}{arg\,min}
\DeclareMathOperator{\sgn}{sgn}

\newcommand{\citepos}[1]{\citeauthor{#1}'s (\citeyear{#1})}
\newcommand{\contract}{\kern-2pt : \kern-2pt}
\newcommand{\dee}{\mathrm{d}}
\newcommand{\strain}{{\varepsilon}}
\newcommand{\straintensor}{{\bm{\varepsilon}}}
\newcommand{\stresstensor}{{\bm{\sigma}}}
\newcommand{\trp}{\mathrm{T}}
\newcommand{\GcEff}{{G_c^\text{\kern1pteff}}}

\begin{document}	
\begin{frontmatter}
	\title{\LARGE A pseudo-dynamic phase-field model for brittle fracture}
	
	\author{Juan Michael Sargado\corref{cor1}}
	\ead{michael.sargado@nbi.ku.dk}
	\author{Joachim Mathiesen}
	\ead{mathies@nbi.dk}
	
	\cortext[cor1]{Corresponding author}
	\address{Niels Bohr Institute, University of Copenhagen \\ Jagtvej 155 A, 2200 Copenhagen N, Denmark}

	\date{}
	
	\begin{abstract}
		\fontsize{8.5pt}{11pt}\selectfont
		The enforcement of global energy conservation in phase-field fracture simulations has been an open problem for the last 25 years. Specifically, the occurrence of unstable fracture is accompanied by a loss in total potential energy, which suggests a violation of the energy conservation law. This phenomenon can occur even with purely quasi-static, displacement-driven loading conditions, where finite crack growth arises from an infinitesimal increase in load. While such behavior is typically seen in crack nucleation, it may also occur in other situations. Initial efforts to enforce energy conservation involved backtracking schemes based on global minimization, however in recent years it has become clearer that unstable fracture, being an inherently dynamic phenomenon, cannot be adequately resolved within a purely quasi-static framework. Despite this, it remains uncertain whether transitioning to a fully dynamic framework would sufficiently address the issue. In this work, we propose a pseudo-dynamic framework designed to enforce energy balance without relying on global minimization. This approach incorporates dynamic effects heuristically into an otherwise quasi-static model, allowing us to bypass solving the full dynamic linear momentum equation. It offers the flexibility to simulate crack evolution along a spectrum, ranging from full energy conservation at one extreme to maximal energy loss at the other. Using data from recent experiments, we demonstrate that our framework can closely replicate experimental load-displacement curves, achieving results that are unattainable with classical phase-field models.
	\end{abstract}
	
\end{frontmatter}

\fontsize{8.5pt}{11pt}\selectfont

\input{introduction}
\input{phase-field_fracture}
\input{pseudo-dynamic_model}
\input{implementation}
\input{numerical_examples}
\input{concluding_remarks}

\appendix
\input{spectral_decomposition}

\footnotesize
\bibliographystyle{elsarticle-harv}
\bibliography{Reference}
\end{document}

%% file: introduction.tex
\section{Introduction}
Since their inception in the late 90's, variational phase-field models have emerged as powerful tools for analyzing complex fracture phenomena. These models are widely used at present, not only in the context of pure fracture mechanics, but more so in multi-physics settings where the interaction between crack formation and different physical or chemical processes are of great interest. Areas of application include, among others, failure modeling in heterogeneous materials and composites \citep{Biner2009,Ma2023,Macias2023}, shear and mixed-mode fracture in geologic materials \citep{Bryant2018,Fei2020,Fei2021}, early-age fracture in concrete \citep{Nguyen2020}, degradation of energy storage systems such as lithium ion batteries \citep{Zhao2016,Mesgarnejad2019}, crack evolution in piezoelectric and flexo-electric materials \citep{Miehe2010_jmps,Wilson2013,Zhang2022,Zhang2022_efm}, cyclic fatigue \citep{Alessi2018,Seles2021}, crack initiation and growth due to hydrogen embrittlement \citep{Martinez-Paneda2018,Kristensen2020}, fracture in biological materials such as bone \citep{Shen2019,Preve2024}, and crack formation during additive manufacturing \citep{Li2023,Ruan2023}.

Their popularity notwithstanding, phase-field approaches suffer from a critical limitation when used to model brittle fracture: they generally fail to ensure the conservation of energy. The issue has been recognized since the early development of variational phase-field models, yet it remains unresolved to this day. The significance of this problem becomes evident when we consider that most of the applications mentioned earlier are modeled as quasi-static processes. Energy conservation is not a concern when crack growth is stable, because the governing equations of quasi-static phase-field fracture models are derived under the assumption that total energy is conserved for \emph{small} variations in both displacement and damage fields. However, finite crack growth over a single time step involves changes that may no longer be considered ``small'', particularly since the damage field at a given point can evolve from zero to unity within a single loading step, even one of arbitrarily small size. Initially, it was thought that the failure to conserve energy was due to standard solvers being unable to escape local minima in the solution space. Efforts to address this issue have taken the form of backtracking schemes \citep{Bourdin2008, Mielke2010, Mesgarnejad2015, Luege2021}, which heuristically adjust solutions post hoc to bring them closer to the global minimizer. However, backtracking can also lead to unphysical results, such as crack propagation that violates Griffith’s criterion \citep{Griffith1921}.

Due to the limitations of quasi-static frameworks in modeling unstable fracture, some recent studies \citep[e.g.][]{ChaoCorreas2024} have proposed to adopt instead a fully dynamic approach, as unstable cracking can occur even under quasi-static loading conditions. While this shift seems like a natural solution to the energy conservation issue, the advantages of utilizing full dynamics are not as straightforward as they may appear. In quasi-static simulations, mesh refinement is often designed based on prior knowledge of regions with steep gradients. In contrast, a dynamic simulation is governed by wave propagation, meaning that significant stress gradients are not fixed in space but instead propagate across the medium over time. This necessitates the use of uniform meshes to accurately capture the propagation of these gradients, or alternatively the implementation of careful adaptive remeshing. It is well known that abrupt changes in mesh refinement can lead to spurious wave reflections, due to the limitations on the wavelengths and frequencies that can be transmitted by elements of a given size \citep{Fried1979, Zukas2000}.

\cite{Borden2012} were among the first to extend variational phase-field models to simulate brittle fracture in fully dynamic scenarios. Here the key functional of interest is the Lagrangian, defined as the difference between kinetic and potential energies. The model operates under Hamilton's principle, which states that the true solution corresponds to a stationary point of the Lagrangian, assuming small perturbations in the displacement, velocity, and damage fields. However, it is unclear whether this formulation is able to effectively suppress sudden large changes in the damage field, which would otherwise lead to the same issues regarding energy conservation as observed in quasi-static models. Recently, \cite{Niu2023} reported that in some cases, the crack evolution modeled using the fully dynamic fracture phase-field model resulted in a situation where the free energy grew to become significant compared to the external work, suggesting a possible violation of the energy conservation law. Additionally, existing phase-field models for dynamic fracture are unable to account for energy dissipation through internal friction.

In the present contribution, we propose a novel approach for simulating unstable fracture within a largely quasi-static framework, as an alternative to fully dynamic methods. We call our model \emph{pseudo-dynamic}, as it accounts for kinetic as well as internal dissipation effects in a heuristic manner, thereby eliminating the need to include velocity-related terms in the governing equations. This allows it to be seamlessly integrated with existing quasi-static phase-field implementations. The rest of the paper is organized as follows: Section \ref{sec:theory} reviews the theoretical foundations of quasi-static variational phase-field models. In Section \ref{sec:pseudo-dynamic_model}, we present our proposed pseudo-dynamic model, followed by its implementation details in Section \ref{sec:implementation}. We then apply the model to simulate brittle crack initiation and growth in three modified compact tension specimens, using experimental data from \cite{Cavuoto2022}. Section \ref{sec:numerical_examples} provides a detailed analysis of the numerical results, followed by concluding remarks in Section \ref{sec:concluding_remarks}.

%% file: phase-field_fracture.tex
\section{Theoretical background} \label{sec:theory}
We first revisit the thermodynamic framework for fracture introduced by \cite{Griffith1921} and the classical phase-field models arising from regularization of the quasi-static variational theory of \citet{Francfort1998}, which serve as a point of departure for the new model developed in this paper. We also include a short review of backtracking schemes, which represent early attempts at resolving the conundrum of energy loss associated with brutal crack growth.

\subsection{Governing equations for quasi-static brittle fracture}
Consider a body occupying domain $\Omega \subset \mathbb{R}^d$ ($d \in \{ 1,2,3 \}$), with external boundary $\partial\Omega$ and internal discontinuities in the form of discrete cracks, collectively denoted by a crack set $\Gamma$ that may evolve over time. It is assumed that crack formation is an irreversible process, so that cracks cannot heal. Thus the evolution of $\Gamma$ must be such that for two temporal values $s,t \geq 0$,
\begin{linenomath}
\begin{equation}
	\Gamma \left( s \right) \subseteq \Gamma \left( t \right) \quad \forall s < t.
	\label{eq:discreteIrreversibilityConstraint}
\end{equation}
\end{linenomath}
Situations exist where the above constraint does not apply, for instance in the case of healing bone fractures, as well as the gradual cementation of cracks in geological materials due to mineral transport and deposition. In the current work however, we restrict ourselves to phenomena for which constraint \eqref{eq:discreteIrreversibilityConstraint} holds. The total potential energy associated with the body may be written as
\begin{linenomath}
\begin{equation}
	\Psi \left( \mathbf{u}, \Gamma \right) = \Psi^b \left( \mathbf{u}, \Gamma \right) + \Psi^s \left( \Gamma \right) = \int_{\Omega\setminus\Gamma} \psi_0 \left( \straintensor \left( \mathbf{u} \right) \right) \dee\Omega + \int_\Gamma G_c \left( \mathbf{x} \right) \dee \mathcal{H}^{d-1} \left( \mathbf{x} \right),
	\label{eq:griffithFunctional}
\end{equation}
\end{linenomath}
where $\mathbf{u}$ is the displacement field and $\psi_0 \left( \straintensor \left( \mathbf{u} \right) \right)$ the material strain energy density. The first term in the right hand side of \eqref{eq:griffithFunctional} represents the elastic strain energy of the cracked body. The second term is the surface energy associated with all the cracks comprising $\Gamma$, with $G_c$ being the critical energy release rate or fracture toughness\footnote{Here we use these terms synonymously. In some papers however, the term fracture toughness refers to the mode-I stress intensity factor that corresponds to $G_c$.} from \citepos{Griffith1921} theory, and $\mathcal{H}^{d-1}$ denoting the $\left( d-1 \right)$-dimensional Hausdorff measure so that $\int_\Gamma \dee \mathcal{H}^{d-1} \left( \mathbf{x} \right) = \| \Gamma \|$ gives the total surface area of all cracks in $\Gamma$. For simplicity, we assume that the material is linear elastic prior to fracture and that the displacements give rise to small deformations, so that the elastic strain energy density is given by
\begin{linenomath}
\begin{equation}
	\psi_0 \left( \straintensor \right) = \frac{1}{2} \lambda \left( \tr \straintensor \right)^2 + \mu \straintensor\contract\straintensor
	\label{eq:elasticStrainEnergyDensity}
\end{equation}
\end{linenomath}
in which $\lambda$ and $\mu$ are respectively the Lam\'e and Kirchhoff moduli, and $\straintensor = \frac{1}{2} \left[ \nabla \mathbf{u} + \left( \nabla \mathbf{u} \right)^\trp \right]$ is the infinitesimal strain tensor. We further assume that $G_c$ is constant within $\Omega$ so that the surface energy can be calculated as $G_c \| \Gamma \|$. \citet{Griffith1921} originally developed his theory based on a symmetric setup involving an infinite domain with a single crack under mode I loading, for which the path of propagation is known. He then derived a criterion for brittle fracture using the principle of energy conservation: during crack extension of an infinitesimal amount $\delta a$, the incremental change in total potential energy must be equal to the incremental work $\delta W^\text{ext}$ done by external forces. That is,
\begin{linenomath}
\begin{equation}
	\delta W^\text{ext} = \delta\Psi^b + \delta\Psi^s.
\end{equation}
\end{linenomath}
For an infinite domain, the resulting increment in boundary displacements during crack extension may be assumed equal to zero, thus no work is done by external forces. In this case, the above can be rewritten as
\begin{linenomath}
\begin{equation}
	-\frac{\partial\Psi^b}{\partial a} = \frac{\partial\Psi^s}{\partial a}.
\end{equation}
\end{linenomath} 
In the equation above, $-\partial\Psi^b/\partial a$ is known as the rate of bulk energy release with respect to crack growth and commonly denoted as $G$. On the other hand, $\partial\Psi^s/\partial a = G_c$ follows immediately from \eqref{eq:griffithFunctional}. Substituting these results into the above expression yields $G = G_c$, which is the condition for stable fracture. When $G < G_c$, the crack cannot grow, whereas when $G > G_c$, the crack propagation is unstable. Combining the two cases in which crack growth occurs gives $G \geq G_c$, which is Griffith's criterion for brittle fracture. \citet{Sun2012} discuss the nuances that arise when $\Omega$ is a finite domain, in which case the final criterion is influenced by whether it is assumed that external forces do no work during crack growth, or that external forces remain constant during crack extension resulting in a net strain energy increase. It is worth noting that Griffith's criterion only gives the condition at which crack propagation should occur. In the event of unstable fracture, the criterion does not give any information regarding the resulting length of crack advance.

In more general problems, the crack set $\Gamma$ may contain multiple cracks, and the paths of fracture propagation are generally loading-dependent and thus unknown a priori. \citet{Francfort1998} developed a variational theory of fracture as an extension of Griffith's earlier theory, with the idea being that at any given time $t$ for some specified loading, the evolution of $\Gamma \left( t \right)$ is one that minimizes $\Psi \left( \mathbf{u}, \Gamma \right)$ subject to the irreversibility constraint given  in \eqref{eq:discreteIrreversibilityConstraint}. That is,
\begin{linenomath}
\begin{equation}
	\left( \mathbf{u} \left( t \right), \Gamma \left( t \right) \right) = \argmin\limits_{\substack{\mathbf{u} \in \mathcal{K}_\mathbf{u} \\ \Gamma \left( t \right) \supseteq \Gamma \left( s \right), \; 0 \leq s < t}} \Psi \left( \mathbf{u}, \Gamma \right).
\end{equation}
\end{linenomath}
where $\mathcal{K}_\mathbf{u}$ is the set of all kinematically admissible displacements. Owing to the difficulties of carrying out such minimization over discrete crack sets, \citet{Bourdin2000} proposed to regularize the energy functional in \eqref{eq:griffithFunctional} by means of a scalar damage field. The regularized potential energy functional then has the general expression
\begin{linenomath}
\begin{equation}
	\Psi_\ell \left( \mathbf{u}, \phi \right) = \int_\Omega \psi \left( \straintensor \left( \mathbf{u} \right), \phi \right) \dee\Omega + G_c \int_\Omega \gamma_\ell \left( \phi \right) \dee\Omega.
	\label{eq:regularizedGriffithFunctional}
\end{equation}
\end{linenomath}
The scalar field $\phi$ is commonly referred to as the crack phase-field, with $\phi = 0$ and $\phi = 1$ representing respectively the fully intact and broken states of a material. The damage-dependent bulk energy density function is often expressed in the form
\begin{linenomath}
\begin{equation}
	\psi \left( \straintensor, \phi \right) = g \left( \phi \right) \psi_0^+ \left( \straintensor \right) + \psi_0^- \left( \straintensor \right)
	\label{eq:strainEnergyDecomposition}
\end{equation}
\end{linenomath}
which allows for modeling unilateral contact via a decomposition of the elastic strain energy, with $g \left( \phi \right)$ being an energy degradation function that acts only on the positive part of $\psi_0$ representing tensile material response. The degradation function must satisfy the properties $g \left( 0 \right) = 1$ and $g \left( 1 \right) = g^\prime \left( 1 \right) = 0$. The most commonly used form is of quadratic type as appears originally in \cite{Bourdin2000}, however it is known that this type of degradation function may result in premature growth of damage. Alternative forms have been proposed such as cubic and quartic polynomials \citep{Karma2001,BordenPhD,Kuhn2015}, parametric functions of exponential type \citep{Sargado2018} and rational polynomials \citep{Wu2017}. Various models have also been proposed in the literature for partitioning the elastic energy. In the current study, we make use of spectral decomposition \citep{Miehe2010_ijnme}, in which $\psi_0^+$ and $\psi_0^-$ are defined as
\begin{linenomath}
\begin{equation}
	\psi_0^\pm \left( \straintensor \right) = \frac{\lambda}{2} \left< \strain_1 + \strain_2 + \strain_3 \right>^2_\pm + \mu \left[ \left< \strain_1 \right>_\pm^2 + \left< \strain_2 \right>_\pm^2 + \left< \strain_3 \right>_\pm^2 \right],
\end{equation}
\end{linenomath}
with $\strain_1$, $\strain_2$ and $\strain_3$ being the principal components of the strain tensor and $\left< x \right>_\pm = \left( x \pm \left| x \right| \right)/2$.

Meanwhile, $\gamma_\ell \left( \phi \right)$ is a crack density function that is essentially a $d$-dimensional regularized Dirac-$\delta$ such that $\int_\Omega \gamma_\ell \left( \phi \right) \dee\Omega = \| \Gamma \|$. Alternative expressions for $\gamma_\ell \left( \phi \right)$ have been introduced by subsequent authors, see for example \citet{Pham2011,Borden2014,Li2015}. Here we focus on the so-called 2nd order phase-field models, for which the crack density function can be written in the general form
\begin{linenomath}
\begin{equation}
	\gamma_\ell \left( \phi \right) = \frac{1}{4 c_w} \left( \frac{w \left( \phi \right)}{\ell} + \ell \left| \nabla\phi \right|^2 \right),
\end{equation}
\end{linenomath}
where $\ell$ is a regularization parameter (also known as the phase-field length scale) that controls the amount of diffusion of the fractures. The function $w \left( \phi \right)$ (referred to as the crack geometric function by \citet{Wu2017}) is related to the density of energy dissipation during a homogeneous damage process wherein $\nabla \phi = \mathbf{0}$, and $c_w = \int_0^1 \sqrt{w \left( \phi \right)} \,\dee\phi$ is a scaling parameter. Two popular choices for $w$ are
\begin{linenomath}
\begin{subequations}
\begin{numcases}{w \left( \phi \right) =}
	\phi & (AT\textsubscript{1} model) \\[5pt]
	\phi^2 & (AT\textsubscript{2} model),
\end{numcases}
\end{subequations}
\end{linenomath}
so named after the work of \cite{Ambrosio1992} regularizing the Mumford-Shah functional in image segmentation. In the current study, we utilize the crack density function originally derived by \cite{Miehe2010_ijnme}, based on the assumption that $\phi$ has a negative exponential profile outward from the crack. This yields the following form for $\gamma_\ell \left( \phi \right)$:
\begin{linenomath}
\begin{equation}
	\gamma_\ell \left( \phi \right) = \frac{\phi^2}{2\ell} + \frac{\ell}{2} \nabla\phi \cdot \nabla\phi.
	\label{eq:crackDensityFcn_Miehe}
\end{equation}
\end{linenomath}
In contrast, the original AT\textsubscript{2} crack density function is $\gamma_\ell \left( \phi \right) = \phi^2 / \left( 4\ell \right) + \ell\nabla\phi \cdot \nabla\phi$. Generally, work can be done on $\Omega$ by a distributed body force $\mathbf{b}$ as well as surface tractions $\mathbf{t}$ acting on the boundary $\partial\Omega$. Thus, the external work term can be expressed as
\begin{linenomath}
\begin{equation}
	W^\text{ext} \left( \mathbf{u} \right) = \int_\Omega \mathbf{b} \cdot \mathbf{u} \,\dee\Omega + \int_{\partial\Omega} \mathbf{t} \cdot \mathbf{u} \,\dee\partial\Omega.
\end{equation}
\end{linenomath}

To obtain the governing equations for the brittle fracture problem, we first substitute \eqref{eq:strainEnergyDecomposition} and \eqref{eq:crackDensityFcn_Miehe} into \eqref{eq:regularizedGriffithFunctional}, after which we take the variation of both the total potential energy and the external work. This yields
\begin{linenomath}
\begin{align}
	\delta\Psi_\ell \left( \mathbf{u}, \phi \right) &= \int_\Omega \Big\{ \left[ g \left( \phi \right) \stresstensor_0^+ \left( \straintensor \right) + \stresstensor_0^- \left( \straintensor \right) \right] \contract \delta \straintensor + g^\prime \left( \phi \right) \psi_0^+ \left( \straintensor \right) \,\delta\phi \Big\} \dee\Omega + G_c \int_\Omega \left( \frac{\phi}{\ell} \delta\phi + \ell \nabla\phi \cdot \nabla\delta\phi \right) \dee\Omega \label{eq:increment_potentialEnergy} \\
	\delta W^\text{ext} \left( \mathbf{u} \right) &= \int_\Omega \mathbf{b} \cdot \delta\mathbf{u} \,\dee\Omega + \int_{\partial\Omega_N} \mathbf{t} \cdot \delta\mathbf{u} \,\dee\partial\Omega \label{eq:increment_externalWork}
\end{align}
\end{linenomath}
in terms of infinitesimal increments $\delta \mathbf{u}$ and $\delta\phi$ of the displacement and phase-field respectively, and the resulting increments $\delta \straintensor$ and $\nabla\delta\phi$ in the strain tensor and phase-field gradient. In writing the above, we have made use of the fact that $\partial\psi_0 / \partial\straintensor = \stresstensor_0$ for materials characterized by a strain energy potential, and that the second term reduces to an integration over the Neumann boundary. Let $\Pi = \Psi_\ell - W^\text{ext}$. Conservation of energy dictates that $\delta\Pi = \delta\Psi_\ell - \delta W^\text{ext} = 0$ for all possible values of $\delta\mathbf{u}$ and $\delta\phi$ satisfying $\delta\mathbf{u} = \mathbf{0}$ on $\partial\Omega_D$. This leads to the following coupled boundary value problem:
\begin{linenomath}
\begin{subequations}
	\label{eq:coupledBVP}
\begin{numcases}{}
	\nabla\cdot \left[ g \left( \phi \right) \stresstensor_0^+ \left( \straintensor \right) + \stresstensor_0^- \left( \straintensor \right) \right] + \mathbf{b} = \mathbf{0} & in $\Omega$ \label{eq:linearMomentum} \\
	\mathbf{u} = \bar{\mathbf{u}} & on $\partial\Omega_D$ \label{eq:u_constraint} \\
	\left[ g \left( \phi \right) \stresstensor_0^+ \left( \straintensor \right) + \stresstensor_0^- \left( \straintensor \right) \right] \cdot \mathbf{n} = \mathbf{t} & on $\partial\Omega_N$ \label{eq:tractions} \\
	G_c \left( \ell \nabla^2 \phi - \frac{1}{\ell} \phi \right) = g^\prime \left( \phi \right) \psi_0^+ \left( \straintensor \right) & in $\Omega$ \label{eq:phaseFieldEvolution} \\
	\nabla\phi \cdot \mathbf{n} = 0 & on $\partial\Omega$. \label{eq:phaseFieldBoundaryCondition}
\end{numcases}
\end{subequations}
\end{linenomath}
Enforcement of crack irreversibility requires a regularized counterpart of \eqref{eq:discreteIrreversibilityConstraint}. An often-used condition in the literature is that the damage be monotonically increasing over time, i.e.
\begin{linenomath}
\begin{equation}
	\phi \left( \mathbf{x}, t \right) \geq \phi \left( \mathbf{x}, \tau \right) \quad \forall t \geq \tau.
	\label{eq:damageIrreversibility}
\end{equation}
\end{linenomath}
In \citet{Miehe2010_ijnme}, a penalty term is introduced into the regularized surface energy to enforce the above condition, while \citet{Miehe2010_cmame} proposed replacing $\psi_0^+ \left( \straintensor \right)$ in \eqref{eq:phaseFieldEvolution} with a history field $\mathcal{H} \left( \mathbf{x}, t \right) = \max_{s \in \left[ 0,t \right]} \psi_0^+ \left( \straintensor \right)$. Nevertheless, it should be noted that \eqref{eq:damageIrreversibility} is not fully compatible with AT\textsubscript{2} phase-field models due to the manner in which $\phi$ evolves from being uniformly zero towards the exponential profile associated with a fully developed crack, as explored in \citet{Kuhn2015} and \citet{Miehe2017}. Instead, one can choose to enforce the irreversibility of damage only once $\phi$ reaches a specified threshold \citep{Sargado2018,Sargado2021}. This helps to prevent $\phi$ from converging to an incorrect profile as the material becomes fully damaged.

It should be noted that while the variational theory was developed by \citet{Francfort1998} with global minimization as the driving principle, the system in \eqref{eq:coupledBVP} merely describes a stationarity condition for $\Pi$, in that \eqref{eq:linearMomentum} and \eqref{eq:phaseFieldEvolution} represent the conditions $\partial\Pi / \partial\mathbf{u} = \mathbf{0}$ and $\partial\Pi / \partial\phi = 0$ respectively. \citet{Larsen2024} recently pointed out that in reality, current implementations of the phase-field method do not approximate the global minimizers of \citet{Francfort1998} but are rather more closely related to a sharp-interface model based on some local variational principle. Moreover, solving \eqref{eq:coupledBVP} by means of standard gradient-based minimization algorithms can result in simulated material response that exhibits energy loss. Such phenomenon has been documented in the prior literature, for instance in the case of crack nucleation in V-notches \citep{Tanne2018}, and also the propagation of an existing crack \citep{Sargado2021}.

\subsection{Backtracking schemes}
As energy conservation should be a fundamental property of the system described by \eqref{eq:coupledBVP}, it has been argued that any apparent violation of this property is an indication that whatever solution algorithm was utilized to solve the system may have converged to a local minimizer or a saddle point of $\Psi_\ell$ rather than the desired global minimizer. To eliminate such solutions, \cite{Bourdin2008} proposed to augment \eqref{eq:coupledBVP} with an additional optimality condition. For applied loading that is monotonically increasing over time, it is understood that $\Psi_\ell$ should likewise be monotonically increasing. Thus for two points in time $s$ and $t$ satisfying $s \leq t$, the following inequality holds:
\begin{linenomath}
\begin{equation}
	\Psi^b_\ell \left( \mathbf{u}_s, \phi_s \right) + \Psi^s_\ell \left( \mathbf{u}_s, \phi_s \right) \leq \Psi^b_\ell \left( \mathbf{u}_t, \phi_t \right) + \Psi^s_\ell \left( \mathbf{u}_t, \phi_t \right),
\end{equation}
\end{linenomath}
where $\mathbf{u}_s = \mathbf{u} \left( s \right)$, $\mathbf{u}_t = \mathbf{u} \left( t \right)$ and so on. Noting that $\psi_0 \left( \straintensor \right)$ in \eqref{eq:elasticStrainEnergyDensity} is homogeneous of degree 2, we can conclude that if $\left( \mathbf{u}_t, \phi_t \right)$ are admissible solutions at time $t$, then it follows from the monotonicity argument that the pair $\left( \dfrac{s}{t} \mathbf{u}_t, \phi_t \right)$ is also admissible at time $s \leq t$, for which the total energy is given by
\begin{linenomath}
\begin{equation}
	\Psi_\ell \left( \frac{s}{t}\mathbf{u}_t, \phi_t \right) = \frac{s^2}{t^2} \Psi^b_\ell \left( \mathbf{u}_t, \phi_t \right) + \Psi^s_\ell \left( \phi_t \right).
\end{equation}
\end{linenomath}
Moreover, if $\left( \mathbf{u}_s, \phi_s \right)$ is the correct solution at time $s$, then it must minimize $\Psi_\ell$ among all admissible pairs $\left( \mathbf{u}, \phi \right)$ so that for $s \leq t$,
\begin{linenomath}
\begin{equation}
	\Psi^b_\ell \left( \mathbf{u}_s, \phi_s \right) + \Psi^s_\ell \left( \phi_s \right) \leq \frac{s^2}{t^2} \Psi^b_\ell \left( \mathbf{u}_t, \phi_t \right) + \Psi^s_\ell \left( \phi_t \right).
	\label{eq:backtrackingOptimalityCondition}
\end{equation}
\end{linenomath}
By augmenting \eqref{eq:coupledBVP} with the above condition, solutions for $\mathbf{u}$ and $\phi$ are obtained which do not exhibit any loss of total energy over time. In practice, the condition is enforced in a temporally discrete setting as follows \citep{Bourdin2008}: after obtaining the solution $\left( \mathbf{u} \left( t_i \right), \phi \left( t_i \right) \right)$ at time step $i$, the optimality condition \eqref{eq:backtrackingOptimalityCondition} is checked against all previous time steps. If the inequality is violated at some prior time step $j$, then this means that the previously obtained solution $\left( \mathbf{u} \left( t_j \right), \phi \left( t_j \right) \right)$ was in fact not the global minimizer of $\Psi_\ell$ at $t_j$. The solution process then backtracks to time step $j$ where the minimization procedure is rerun with $\phi_i$ as the initial guess for the phase-field. Similar backtracking algorithms have also been employed in subsequent works, notably \citet{Mielke2010,Mesgarnejad2015,Luege2021} which either adopt or modify/improve the procedure described above.

An obvious flaw of the backtracking scheme is that a violation of the optimality condition must first occur in order to trigger recalculation. For instance, consider three time steps $a$, $b$ and $c$ where $t_a < t_b < t_c$. Suppose that at time step $c$, a solution is obtained for which \eqref{eq:backtrackingOptimalityCondition} is violated at time step $a$. Then the solution procedure restarts at $t_a$ with $\phi = \phi \left( t_c \right)$ as an initial guess. However if the solution is terminated at time step $b$ before reaching $t_c$, then the violation is not detected and no recalculation takes place. Consequently, the reported solution from $t_a$ to $t_b$ will then be incorrect from a global minimization standpoint. Thus the fracture evolution must be simulated until total failure, in order to detect all intermediate occurrences of energy loss. Moreover, insofar as dips in the total energy are associated with unstable/brutal cracking, the backtracking scheme moves the occurrence of such brutal fracture to an earlier point in time in order to recover energy conservation. However when the unstable fracture occurs in the context of simple crack extension (rather than initiation), modifying the solution such that it occurs earlier in time results in a violation of Griffith's criterion, i.e.\ the crack extends even when $G < G_c$. Such unphysical results arise due to the fact that global minimization allows for the evolution to jump from the current configuration to one that may otherwise be inaccessible from the former owing to the two being separated by arbitrarily large energy barriers \citep{Negri2008,Larsen2010}.

%% file: pseudo-dynamic_model.tex
\section{A novel pseudo-dynamic framework for handling unstable fracture} \label{sec:pseudo-dynamic_model}
In this work, we propose an alternative means of enforcing energy conservation which does not rely on global minimization of the total energy functional. Instead, our framework is based on the fact that even in a quasi-static setting, the occurrence of unstable crack propagation is an inherently dynamic phenomenon which implies that some of the external work is transformed into kinetic energy. This in turn can be subsequently converted back into bulk strain energy and go towards the formation of new crack surfaces, or be dissipated by some other means. Our proposed framework can be aptly described as being \emph{pseudo-dynamic}, in that while we do not consider dynamic/acceleration terms in the linear momentum equation, we nevertheless try to reasonably approximate the contribution of dynamic effects towards fracture growth. In particular, we explicitly impose global energy balance between successive time steps. Given two points in time $t_a$ and $t_b$, the first law of thermodynamics requires that
\begin{linenomath}
\begin{equation}
	\Delta \Psi_\ell^e + \Delta \Psi_\ell^s + \Delta \mathcal{E}^\text{kin} + \mathcal {D} = \Delta W^\text{ext},
	\label{eq:energyConservationOverTime}
\end{equation}
\end{linenomath} 
wherein $\Delta \mathcal{E}^\text{kin}$ denotes the change in kinetic energy between $t_a$ and $t_b$, and $\mathcal{D}$ is the energy dissipation that includes internal friction, the latter being dependent on material velocity. Note that during quasi-static loading characterized by either zero or stable crack growth, both $\Delta \mathcal{E}^\text{kin}$ and $\mathcal{D}$ are zero, and \eqref{eq:energyConservationOverTime} is trivially satisfied, being equivalent to equating \eqref{eq:increment_potentialEnergy} with \eqref{eq:increment_externalWork}.

In the case of unstable cracking however, the previous statement is no longer necessarily true. For the sake of simplicity, let us consider the case where crack evolution is such that the body remains kinematically stable (i.e.\ the crack does not split the body into completely unconnected components, nor give rise to kinematic mechanisms). Then at the conclusion of such a brutal fracture event when the crack has been arrested and any vibrations damped out, said body is again in a quasi-static state. Here we choose $t_a$ as the time immediately before the beginning of unstable fracture, and $t_b$ the time after crack arrest when the body is no longer vibrating. In both instances, $\mathcal{E}^\text{kin} = 0$. However unlike with stable fracture, here we expect that $\mathcal{D} > 0$ since the dynamic nature of brutal crack propagation implies that in at least part of the domain, $\dot{\mathbf{u}} \left( \mathbf{x}, t \right) \gg 0$ within the time interval  $\left[ t_a,  t_b \right]$. In this case, the only way to reconcile \eqref{eq:energyConservationOverTime} with \eqref{eq:increment_potentialEnergy}--\eqref{eq:increment_externalWork} is to assume that all the missing energy is dissipated. This assumption seems to be valid in the case of hyperelastic materials such as silicone, where experiments have revealed crack arrest occurring at a configuration that corresponds a stationary point of the total potential energy \citep{Rosendahl2019}. In such a case, we can write
\begin{linenomath}
\begin{equation}
	\mathcal{D}_\text{qs} =\left(  \Delta W^\text{ext} - \Delta \Psi_\ell^e - \Delta \Psi_\ell^s \right)_\text{qs},
	\label{eq:quasistaticDissipation}
\end{equation}
\end{linenomath}
in which $\left( \bullet \right)_\text{qs}$ denotes quantities that are calculated using solutions for $\mathbf{u}$ and $\phi$ that satisfy \eqref{eq:coupledBVP}.

On the other hand when $\mathcal{D}$ is not equal to \eqref{eq:quasistaticDissipation}, the combination of \eqref{eq:coupledBVP} and \eqref{eq:energyConservationOverTime} results in a system that has generally no solution in the event of unstable fracture. This can be explained by noting that since \eqref{eq:increment_potentialEnergy}--\eqref{eq:increment_externalWork} does not include a kinetic component, any part of the potential energy that gets converted to kinetic energy is simply ``lost'' as there is no way for the governing equations to account for it. Consequently, a straightforward solution of \eqref{eq:coupledBVP} will yield results exhibiting drops in the total energy that cannot be attenuated or controlled. Rather than apply some form of backtracking as mentioned earlier, we instead propose to modify \eqref{eq:phaseFieldEvolution} based on the assumption that any transitory kinetic energy that is not dissipated via internal friction must eventually transform back into bulk strain energy and thus be available to drive further crack evolution. In particular, we adopt a heuristic approach by scaling the right hand side of \eqref{eq:phaseFieldEvolution} to accommodate the aforementioned kinetic component. The modified phase-field evolution equation is thus given by
\begin{linenomath}
\begin{equation}
	G_c \left( \ell \nabla^2 \phi - \frac{1}{\ell} \phi \right) = \eta \left( \mathbf{x} \right) g^\prime \left( \phi \right) \psi_0^+ \left( \straintensor \right) \label{eq:modifiedPhaseFieldEvolution}
\end{equation}
\end{linenomath}
where the overload factor $\eta \left( \mathbf{x} \right) \geq 1$ is generally a scalar field whose role is to model the spatial distribution of the reconverted kinetic energy. In the absence of any additional information, the simplest approach is to assume that $\eta \left( \mathbf{x} \right)$ is constant, i.e.\ $\eta \left( \mathbf{x} \right) = \bar{\eta}$, where $\bar{\eta}$ is an additional unknown to be determined along with $\mathbf{u}$ and $\phi$. Furthermore, we assume that the dissipated energy can be adequately described as some factor of \eqref{eq:quasistaticDissipation}. That is,
\begin{linenomath}
\begin{equation}
	\mathcal{D} = \zeta \mathcal{D}_\text{qs},
	\label{eq:lossCoefficient}
\end{equation}
\end{linenomath}
where $\zeta \in \left[ 0, 1 \right]$ is an energy loss coefficient. Thus, $\zeta = 1$ corresponds to the case of maximum energy dissipation (the standard quasi-static model), and $\zeta = 0$ to the case where energy is strictly conserved. The full augmented system is then given by
\begin{linenomath}
\begin{subequations}
	\label{eq:augmentedCoupledBVP}
	\begin{numcases}{}
		\nabla\cdot \left[ g \left( \phi \right) \stresstensor_0^+ \left( \straintensor \right) + \stresstensor_0^- \left( \straintensor \right) \right] + \mathbf{b} = \mathbf{0} & in $\Omega$ \label{eq:aug_linearMomentum} \\
		\mathbf{u} = \bar{\mathbf{u}} & on $\partial\Omega_D$ \label{eq:aug_u_constraint} \\
		\left[ g \left( \phi \right) \stresstensor_0^+ \left( \straintensor \right) + \stresstensor_0^- \left( \straintensor \right) \right] \cdot \mathbf{n} = \mathbf{t} & on $\partial\Omega_N$ \label{eq:aug_tractions} \\
		G_c \left( \ell \nabla^2 \phi - \frac{1}{\ell} \phi \right) = g^\prime \left( \phi \right) \mathcal{H} \left( \mathbf{x}, \bar{\eta} \right) & in $\Omega$ \label{eq:aug_phaseFieldEvolution} \\
		\nabla\phi \cdot \mathbf{n} = 0 & on $\partial\Omega$. \label{eq:aug_phaseFieldBoundaryCondition} \\
		\Delta \Psi_\ell^e \left( \straintensor, \phi \right) + \Delta \Psi_\ell^s \left( \phi \right) + \mathcal {D} = \Delta W^\text{ext} \left( \mathbf{u} \right).
		\label{eq:aug_energyConservationOverTime}
	\end{numcases}
\end{subequations}
\end{linenomath}
in which the history field is now defined as
\begin{linenomath}
\begin{numcases}{\mathcal{H} \left( \mathbf{x}, \bar{\eta} \right) =}
	\max\limits_{s \in \left[ 0,t \right]} \eta \, \psi_0^+ \left( \bm{\epsilon} \left( \mathbf{x},s \right) \right) & if $\phi > \phi_c$ \label{eq:histFieldWEta} \\
	\eta \,\psi_0^+ \left( \bm{\epsilon} \left( \mathbf{x},t \right) \right) & otherwise \nonumber
\end{numcases}
\end{linenomath}

Note that the above system is flexible enough to describe the different cases of unstable cracking described previously. In the case where all the kinetic energy is dissipated, $\bar{\eta} = 1$, $\zeta = 1$ and the additional unknown consists of $\mathcal{D}$ (which is then equal to $\mathcal{D}_\text{qs}$). Meanwhile, if only part of the kinetic energy is dissipated, then $\zeta < 1$ and the additional unknown consists of $\bar{\eta}$. Finally, we point out that it is also possible to utilize alternative forms for $\eta \left( \mathbf{x} \right)$, however the adopted expression should contain only a single unknown parameter, otherwise additional conditions must be added to \eqref{eq:augmentedCoupledBVP} corresponding to the excess parameters.

%% file: implementation.tex
\section{Implementation aspects} \label{sec:implementation}
The most common method of discretizing the coupled linear momentum and phase-field evolution equation is by means of $P_1$ finite elements, which we have chosen to adopt in the current work. More efficient discretizations that combine finite element and finite volume concepts have been proposed in \cite{Sargado2020} and \cite{Sargado2021}, however the calculation of a quantity such as $\int_\Omega \nabla\phi \cdot \nabla\phi \,\dee\Omega$ (which is needed for evaluating the optimality condition) is not  straightforward within a finite volume framework and would require further approximations. In the classical FE framework, the primary unknowns $\mathbf{u}$ and $\phi$ as well as their gradients are approximated in terms of the corresponding nodal degrees of freedom as
\begin{linenomath}
\begin{align}
	&\mathbf{u} = \sum_{I=1}^m \mathbf{N}_I^u \mathbf{u}_I, \qquad \phi = \sum_{I=1}^m N_I \phi_I \nonumber \\
	&\straintensor = \sum_{I=1}^m \mathbf{B}_I^u \mathbf{u}_I, \qquad \nabla\phi = \sum_{I=1}^m \mathbf{B}^\phi_I \phi_I
\end{align}
\end{linenomath}
wherein
\begin{linenomath}
\begin{equation}
	\mathbf{N}_I^u = \left[ \kern-4pt\begin{array}{cc} N_I & 0 \\[3pt] 0 & N_I \end{array} \kern-4pt\right] \qquad
	\mathbf{B}_I^u = \left[ \kern-4pt\begin{array}{cc} N_{I,x} & 0 \\[3pt] 0 & N_{I,y} \\[3pt] N_{I,y} & N_{I,x} \end{array} \kern-4pt \right] \qquad
	\mathbf{B}_I^\phi = \left[ \kern-4pt\begin{array}{c} N_{I,x} \\[3pt] N_{I,y} \end{array} \kern-4pt \right],
\end{equation}
\end{linenomath}
and $N_I$ are the basis function associated with node $I$. The test functions and their gradients are approximated similarly, by replacing $\mathbf{u}_I$ and $\phi_I$ with $\delta\mathbf{u}_I$ and $\delta\phi_I$ in the above expressions. Numerical approximation of the weak forms associated with \eqref{eq:augmentedCoupledBVP} yields a set of residual equations defined at each node $I$, plus an additional equation in the form of \eqref{eq:aug_energyConservationOverTime}. The full discrete coupled system can thus be written as
\begin{linenomath}
\begin{subequations}\label{eq:residuals}
\begin{numcases}{}
	\mathbf{r}^u_I = \int_\Omega {\mathbf{B}^u_I}^T \left[ g \left( \phi \right) \stresstensor_0^+ \left( \straintensor \right) + \stresstensor_0^- \left( \straintensor \right) \right] \dee\Omega - \int_\Omega {\mathbf{N}_I^u}^T \mathbf{b} \dee\Omega - \int_{\partial\Omega^N} {\mathbf{N}_I^u}^T \mathbf{t} \,\dee\partial\Omega  = \mathbf{0} & \label{eq:residual_u} \\
	r_I^\phi = \int_\Omega \left[ G_c \ell {\mathbf{B}_I^\phi}^T \nabla\phi + \frac{G_c}{\ell} N_I \phi \right] \dee\Omega + \int_\Omega N_I g^\prime \left( \phi \right) \mathcal{H} \left( \mathbf{x}, \bar{\eta} \right) \dee\Omega = 0 & \label{eq:residual_phi} \\
	r^\eta \kern+1pt = \Delta W^\text{ext} - \Delta\Psi^e - \Delta\Psi^s - \mathcal{D} = 0, & \label{eq:residual_energyBalance}
\end{numcases}
\end{subequations}
\end{linenomath}
where $I = 1,2,\ldots,M$ with $M$ being the number of nodes. A naive monolithic solution of the discrete system using gradient-based algorithms is known to result in blowup of the solution due to non-convexity of the total energy. Instead, the standard practice is to employ techniques such as alternate minimization that exploit the convexity of the energy with respect to either $\mathbf{u}$ or $\phi$ when the other of the two is fixed. In general, \eqref{eq:residual_u} is nonlinear and must be solved using a Newton-Raphson (N-R) scheme. That is,
\begin{linenomath}
\begin{equation}
	\left\{ \mathbf{u}_i^{p+1} \right\} = \left\{ \mathbf{u}_i^p \right\} - \left[ \mathbf{K}^{uu} \left( \straintensor_i^p, \phi_i^p \right) \right]^{-1} \left\{ \mathbf{r}^u \left( \straintensor_i^p, \phi_i^p \right) \right\}
\end{equation}
\end{linenomath}
where the stiffness matrix components are calculated as
\begin{linenomath}
\begin{equation}
	\mathbf{K}^{uu}_{IJ} \left( \straintensor_i^p, \phi_i^p \right) = \int_\Omega {\mathbf{B}^u_I}^T \left[ g \left( \phi_i^p\right) \mathbb{C}_+ \left( \straintensor_i^p \right) + \mathbb{C}_- \left( \straintensor_i^p \right) \right] \mathbf{B}^u_J \,\dee\Omega.
\end{equation}	
\end{linenomath}
For the spectral decomposition proposed in \cite{Miehe2010_ijnme}, the modulus tensors $\mathbb{C}_\pm$ have a rather complicated form and are derived in \ref{sec:modulusTensor_derivation}.

The subsystem given by \eqref{eq:residual_phi} is also nonlinear whenever we employ non-quadratic forms of the degradation function. Its solution is thus also iteratively obtained through the N-R scheme:
\begin{linenomath}
\begin{equation}
	\left\{ \bm{\phi}_i^{p+1} \right\} = \left\{ \bm{\phi}_i^p \right\} - \left[ \mathbf{K}^{\phi\phi} \left( \phi^p_i \right) \right]^{-1}
	\left\{ \mathbf{r}^\phi \left( \phi_i^p \right) \right\},
\end{equation}
\end{linenomath}
in which the components of the Jacobian matrix are given by
\begin{linenomath}
\begin{equation}
	K_{IJ}^{\phi\phi} \left( \phi_i^p \right) = \int_\Omega \left\{ G_c \ell {\mathbf{B}_I^\phi}^T \mathbf{B}_J^\phi + \left[ \frac{G_c}{\ell} + g^{\prime\prime} \left( \phi_i^p \right) \mathcal{H} \left( \mathbf{x}, \bar{\eta} \right) \right] N_I N_J \right\} \dee\Omega.
\end{equation}
\end{linenomath}

On the other hand, the energy balance criterion can only be meaningfully evaluated once convergence has been achieved for both the linear momentum and phase-field equations. Once the latter have been achieved, we can compute the individual terms in \eqref{eq:residual_energyBalance} as detailed below.

\subsection{Calculation of external work increment}
For the discrete FE problem, the external work increment for a given time step can be computed straightforwardly by getting the dot product of the average force at each node during said time step with its corresponding displacement increment, and then summing over all nodes. Thus at time step $i$,
\begin{linenomath}
	\begin{equation}
		\Delta W^\text{ext}_i = \sum_{I=1}^{M} \left( \mathbf{u}_I^i - \mathbf{u}_I^{i-1} \right) \cdot \left( \frac{\mathbf{F}_I^i + \mathbf{F}_I^{i-1}}{2} \right)
	\end{equation}
\end{linenomath}
where in particular the summation is not restricted to nodes lying on the boundary, but also includes all internal nodes. This saves us the effort of having to create separate procedures for determining the respective external work contributions of forces acting at  Dirichlet and Neumann boundaries, as well as the work done by body forces, for example gravity. Note that in the absence of such body forces, the external work contribution from internal nodes should be negligible provided reasonable tolerance criteria are imposed when solving the linear momentum equation. Nevertheless, their inclusion in the summation avoids any potentially significant accumulation of error over several time steps. The full external work at time step $i$ can obtained by adding the calculated increment to the work done at the previous time step, i.e. $W^\text{ext}_i = W^\text{ext}_{i-1} + \Delta W^\text{ext}_i$.

\subsection{Calculation of bulk and surface energy increments}
The bulk energy at time step $i$ is calculated using the converged values $\left\{ \mathbf{u}_i \right\}$ and $\left\{ \bm{\phi}_i \right\}$ via the formula
\begin{linenomath}
	\begin{equation}
		\Psi^e_i = \sum_{I=1}^M \frac{A_I}{2} {\mathbf{u}_I^i}^\trp {\mathbf{B}^u_I}^\trp \left[ g \left( \phi_I^i \right) \mathbb{C}_+ + \mathbb{C}_- \right] \mathbf{B}^u_I \mathbf{u}_I^i,
	\end{equation}
\end{linenomath}
where $A_I$ denotes the area of element $I$.  We note that for $P_1$ elements, both the stresses and strains are piecewise constant over the elements, which implies that the bulk strain energy is likewise piecewise constant. Meanwhile, the surface energy is obtained as
\begin{linenomath}
	\begin{equation}
		\Psi^s_i = \sum_{I=1}^M A_I \frac{G_c}{2} {\bm{\phi}_I^i}^\trp \left[ \frac{1}{\ell} \mathbf{M}^\phi + \ell {\mathbf{B}^\phi_I}^\trp \mathbf{B}^\phi_I \right] \bm{\phi}_I^i,
	\end{equation}
\end{linenomath}
with $\mathbf{M}^\phi$ being the consistent mass matrix for the reference $P_1$ element. The corresponding energy increments are then calculated as $\Delta \Psi^e_i = \Psi^e_i - \Psi^e_{i-1}$ and $\Delta \Psi^s_i = \Psi^s_i - \Psi^s_{i-1}$.

\subsection{Iterative procedure for determining the overload factor} \label{sec:etaSolutionProcedure}
At each time step, we first solve \eqref{eq:residual_u} and \eqref{eq:residual_phi} via an alternate minimization (AM) approach, with the assumption that $\bar{\eta} = 1$. This is nothing but the classical (quasi-static) fracture phase-field model, and after achieving convergence of results for $\mathbf{u}$ and $\phi$, we can obtain the quasi-static dissipation $\mathcal{D}_\text{qs}$ according to \eqref{eq:quasistaticDissipation}. This quantity is then multiplied by the loss coefficient $\zeta$ to yield the actual target dissipation $\mathcal{D}$ for the current time step.

What remains is to design a procedure for determining the optimal value of $\bar{\eta}$ that will result in the satisfaction of \eqref{eq:aug_energyConservationOverTime}. The most important factor to consider here is efficiency, since the simulation of a step that involves brutal crack propagation typically requires a very large number of iterations to achieve convergence of the coupled linear momentum and phase-field evolution equations compared to time steps where the crack evolves in a stable manner \citep{Storvik2021}. In general, changing the value of $\bar{\eta}$ forces a recalculation of the displacement and phase-field evolution, using the converged results from the immediate prior time step as initial guesses. A poorly designed strategy can easily result in increased solution time by several orders of magnitude depending on how many iterations are needed to obtain the optimal value of $\bar{\eta}$.

To avoid the aforementioned scenario, we adopt a two-stage approach towards the solution of the overload factor. Initially, we apply an incremental algorithm: at the $i$-th iteration, $\bar{\eta}_i = \bar{\eta}_{i-1} + h_\eta$ where $h_\eta > 0$ is a fixed chosen increment.  As before, we solve \eqref{eq:residual_u} and \eqref{eq:residual_phi} using the AM algorithm, then evaluate \eqref{eq:residual_energyBalance}. A positive result for $r^\eta$ implies that $\bar{\eta}_i$ is a lower bound for the optimal value that satisfies energy balance. More importantly, it means that the crack tip needs to advance \emph{further} and therefore due to \eqref{eq:discreteIrreversibilityConstraint}, the current intermediate solution for the crack trajectory must be part of the final crack path. We can thus force an update the history field $\mathcal{H} \left( \mathbf{x}, \bar{\eta} \right)$ using the current solutions for $\mathbf{u}$ and $\phi$; moreover we store the latter values to serve as initial guesses for succeeding iterations. On the other hand, a result of $r^\eta < 0$ implies that $\bar{\eta}_i$ is an upper bound of the optimal value, and that cracks have grown beyond what is necessary to attain energy balance in the current time step. However, a naive continuation from the last iteration in which $\bar{\eta}$ is simply decreased will not converge to the right results, as this is tantamount to letting cracks heal which is not permitted by the existing irreversibility constraints. Instead, we must restart the solution from a prior converged state with respect to $\mathbf{u}$ and $\phi$ so that the ensuing fracture network evolution does not involve any crack healing. Rather than restarting the solution from previous time step, we use as initial guess the last intermediate solution corresponding to the lower bound of the overload factor. Moreover as we now have both lower and upper bounds for $\bar{\eta}$, we switch to a selective bisection/regula-falsi approach to determine the optimal value of the overload factor. That is, $\bar{\eta}_i$ now given by
\begin{linenomath}
\begin{equation}
	\bar{\eta}_i = \min \left( \frac{\bar{\eta}_\text{LB} + \bar{\eta}_\text{UB}}{2}, \frac{\bar{\eta}_\text{LB} \cdot r^\eta_\text{UB} - \bar{\eta}_\text{UB} \cdot r^\eta_\text{LB}}{r^\eta_\text{UB} - r^\eta_\text{LB}} \right).
\end{equation}
\end{linenomath}
Subsequently, $\bar{\eta}_i$ replaces either $\bar{\eta}_\text{LB}$ or $\bar{\eta}_\text{UB}$ depending on whether the resulting value for $r^\eta$ is positive or negative. As with the earlier stage, we store the intermediate solutions for $\mathbf{u}$ and $\phi$ whenever an improvement to the lower bound of $\bar{\eta}$ is made. Note that both bisection- and false position-type updates are calculated since we prefer iterates that lead to improvements of the lower (rather than upper) bound for $\bar{\eta}$. Thus we choose the update that is nearer to $\bar{\eta}_\text{LB}$. The reason for this is again efficiency: an improvement on the lower bound brings the intermediate (restart) solution of the crack evolution closer to the optimal one, whereas an improvement in the upper bound necessitates a restart from the latest lower bound solution, essentially forcing a repeat of AM iterations already made previously. The combination of alternate minimization with the iterative procedure for determining $\bar{\eta}$ results in a nested scheme whose main steps are listed in Algorithm \ref{alg:nested_alternate_minimization}.
\begin{algorithm}
	\SetKwInOut{Given}{Given}
	\DontPrintSemicolon
	\caption{Nested algorithm for obtaining energy-conserving solutions}
	\label{alg:nested_alternate_minimization}
	
	\BlankLine
	\Given{Converged values $\left\{ \mathbf{u}_{i-1} \right\}$ and $\left\{ \bm{\phi}_{i-1} \right\}$ from previous timestep, \\
	AM tolerance criteria $\text{tol}_u$, $\text{tol}_\phi$, $\text{tol}_{r^u}$, $\text{tol}_{r^\phi}$ \\
	Loss coefficient $\zeta$, Initial overload factor increment $\kappa$, \\
	Energy balance tolerance criteria $\text{tol}_{r^\eta}$, $\text{tol}_\Gamma$, $\text{tol}_\eta$
	} 
	\BlankLine

	Calculate $\mathbf{r}^u \left( \mathbf{u}_i^0, \bm{\phi}_i^0 \right)$ and $\mathbf{r}^\phi \left( \mathbf{u}_i^0, \bm{\phi}_i^0 \right)$ \;
	
	Set $\bar{\eta} = 1$, $\left\{ \mathbf{u}_i^0 \right\} = \left\{ \mathbf{u}_{i-1} \right\}$, $\left\{ \bm{\phi}_i^0 \right\} = \left\{ \bm{\phi}_{i-1} \right\}$,  $p = 0$, $q = 0$ \;
	
	Set $\left\{ \mathbf{u}^\star \right\} = \left\{ \mathbf{u}_i^0 \right\}$, $\left\{ \bm{\phi}^\star \right\} = \left\{ \bm{\phi}_i^0 \right\}$ \;
	
	Set Solution method for $\bar{\eta}$ to `Incrementation' \;
	
	Set $\bar{\eta}$ convergence flag to FALSE \;
	\While{$\bar{\eta}$ is not converged}{
		\Repeat{$\| \delta \mathbf{u} \| \leq \text{\normalfont tol}_u$ AND
						 $\| \delta \bm{\phi} \| \leq \text{\normalfont tol}_\phi$ AND
					 	 $\| \mathbf{r}^u \| \leq \text{\normalfont tol}_{r^u}$ AND
				         $\| \mathbf{r}^\phi \| \leq \text{\normalfont tol}_{r^\phi}$}{
			Set $p \leftarrow p + 1$ \;
			Use N-R scheme to find $\left\{ \mathbf{u}_i^p \right\}$ such that $\mathbf{r}^u \left( \mathbf{u}_i^p, \bm{\phi}_i^{p-1} \right) = \mathbf{0}$ \;
			Use N-R scheme to find $\left\{ \bm{\phi}_i^p \right\}$ such that $\mathbf{r}^\phi \left( \mathbf{u}_i^p, \bm{\phi}_i^p \,\right) = \mathbf{0}$ \;
			Recompute residuals $\mathbf{r}^u \left( \mathbf{u}_i^p, \bm{\phi}_i^p \right)$ and $\mathbf{r}^\phi \left( \mathbf{u}_i^p, \bm{\phi}_i^p \,\right)$ \;
		}
	
		Calculate energy balance residual $r^\eta \left( \mathbf{u}_i, \bm{\phi}_i, \bar{\eta} \right)$ \;
		
		\eIf{($q == 0$ AND ($r^\eta < 0$ OR $\Delta\Gamma < \text{\normalfont tol}_\Gamma$) OR ($0 \leq r^\eta < \text{\normalfont tol}_{r^\eta}$) OR (Solution method is `Regula falsi' AND $\delta \bar{\eta} < \text{\normalfont tol}_\eta$)}{
			Set $\bar{\eta}$ convergence flag to TRUE \;
		}{
			\eIf{$r^\eta > 0$}{
				$\bar{\eta}_\text{\normalfont LB} \leftarrow \bar{\eta}$ $\;$ \textcolor{RoyalBlue}{// Replace lower bound} \;
				
				$f \left( \bar{\eta}_\text{\normalfont LB} \right) \leftarrow r^\eta$ \;
				
				Set $\left\{ \mathbf{u}^\star \right\} = \left\{ \mathbf{u}_i^p \right\}$, $\left\{ \bm{\phi}^\star \right\} = \left\{ \bm{\phi}_i^p \right\}$ $\;$ \textcolor{RoyalBlue}{// Store intermediate solution} \;
				
			}{
				$\bar{\eta}_\text{\normalfont UB} \leftarrow \bar{\eta}$ $\;$ \textcolor{RoyalBlue}{// Replace upper bound} \;
				
				$f \left( \bar{\eta}_\text{\normalfont UB} \right) \leftarrow r^\eta$ \;
				
				Set Solution method to `Regula Falsi'
				
				Set $\left\{ \mathbf{u}_i^p \right\} = \left\{ \mathbf{u}^\star \right\}$, $\left\{ \bm{\phi}_i^p \right\} = \left\{ \bm{\phi}^\star \right\}$ \textcolor{RoyalBlue}{$\;$// Roll back values to last saved intermediate solution}
			}
			
			\eIf{Solution method is `Incrementation'}{
				$\bar{\eta} \leftarrow \bar{\eta} + \kappa$ \;
			}{
				$\bar{\eta} \leftarrow \min \left( \dfrac{\bar{\eta}_\text{\normalfont LB} + \bar{\eta}_\text{\normalfont UB}}{2}, \dfrac{\bar{\eta}_\text{\normalfont LB} \cdot f \left( \bar{\eta}_\text{\normalfont UB} \right) - \bar{\eta}_\text{\normalfont UB} \cdot f \left( \bar{\eta}_\text{\normalfont LB} \right)}{ f \left( \bar{\eta}_\text{\normalfont UB} \right) - f \left( \bar{\eta}_\text{\normalfont LB} \right) } \right)$ \;
			}
			$q \leftarrow q + 1$ \;
		}
	}

	Update history field variable $\mathcal{H}$ using Eq.\ \eqref{eq:histFieldWEta} \;
	
	\Return converged values $\left\{ \mathbf{u}_i \right\}$ and $\left\{ \bm{\phi}_i \right\}$
\end{algorithm}

\subsection{Effect of mesh refinement on simulated surface energy}
In order to obtain meaningful results, it is important to account for how the level of mesh refinement with respect to the phase-field length scale affects the numerical approximation of the surface energy. An often-adapted rule of thumb when using low order finite elements is to have $\ell/h^e \geq 2$. This expression can be traced back to the work of \cite{Miehe2010_ijnme}, however subsequent studies have found that at least $\ell/h^e \approx 20$ is needed to achieve convergence of numerical results with respect to mesh refinement \citep[see for example][]{Giovanardi2017,Sargado2021}. This is because the theoretical solution for $\phi$ corresponding to a fully developed crack contains cusps/kinks, which cannot be captured properly by classical finite elements based on piecewise smooth shape functions. In particular with linear elements, the numerical solution exhibits a plateau having the width of a single element. This in turn leads to an overestimation of the surface energy. Moreover, such effect is mitigated but not completely removed by switching to higher order approximations \citep{Bourdin2008}. Going back to the case of linear finite elements, integration of \eqref{eq:crackDensityFcn_Miehe} over the domain $\Omega = \Omega_\parallel \times \Omega_\perp$ yields
\begin{linenomath}
\begin{equation}
	\Gamma_\ell^{h^e} = \int_{\Omega_\parallel} \int_{\Omega_\perp \left( \xi_\parallel \right)} \left( \frac{\phi^2}{2\ell} + \frac{\ell}{2} \nabla\phi \cdot\nabla\phi \right) \dee\xi_\perp \,\dee\xi_\parallel \approx \int_{\Omega_\parallel} \left( 1 + \frac{1}{2\ell} h^e \right) \dee\xi_\parallel = \left( 1 + \frac{h^e}{2\ell} \right) \Gamma
\end{equation}
\end{linenomath}
owing to the plateau effect described earlier. Consequently, $\Psi_\ell^s = G_c \Gamma_\ell^{h^e} \approx \left[ 1 + h^e/ \left( 2\ell \right) \right] G_c \Gamma$. Alternatively, we can interpret the factor $1 + h^e/ \left( 2\ell \right)$ as acting on $G_c$. That is,
\begin{linenomath}
\begin{equation}
	\GcEff \approx \left( 1 + \frac{h^e}{2\ell} \right) G_c.
	\label{eq:effectiveToughness}
\end{equation}
\end{linenomath}
Note that the above differs from the original expression derived in \cite{Bourdin2008}, as the specific form of $\gamma_\ell \left( \phi \right)$ we use in this study differs from the aforementioned reference in the set of coefficients acting on the individual terms of the crack density function. 

\begin{figure}
	\centering
	\includegraphics[width=0.35\textwidth]{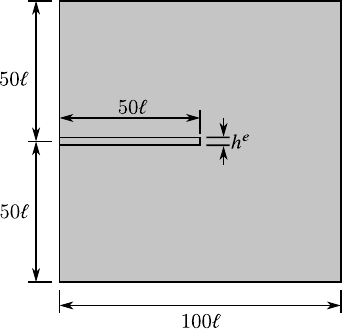}
	\caption{Single-notch specimen with initial crack modeled as row of elements where $\phi = 1$. \label{fig:singleNotchSpecimen}}
\end{figure}

\begin{figure}
	\centering
	\begin{subfigure}{0.4\textwidth}
		\includegraphics[width=\textwidth]{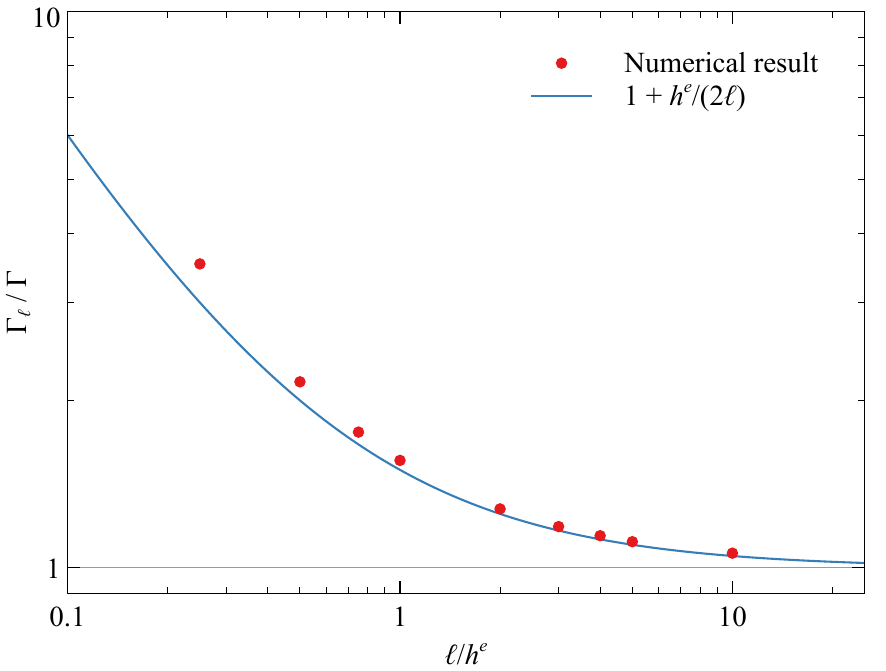}
		\caption{\label{fig:MeshEffectOnCrackLength_loglog}}
	\end{subfigure} \hspace{1cm}
	\begin{subfigure}{0.4\textwidth}
		\includegraphics[width=\textwidth]{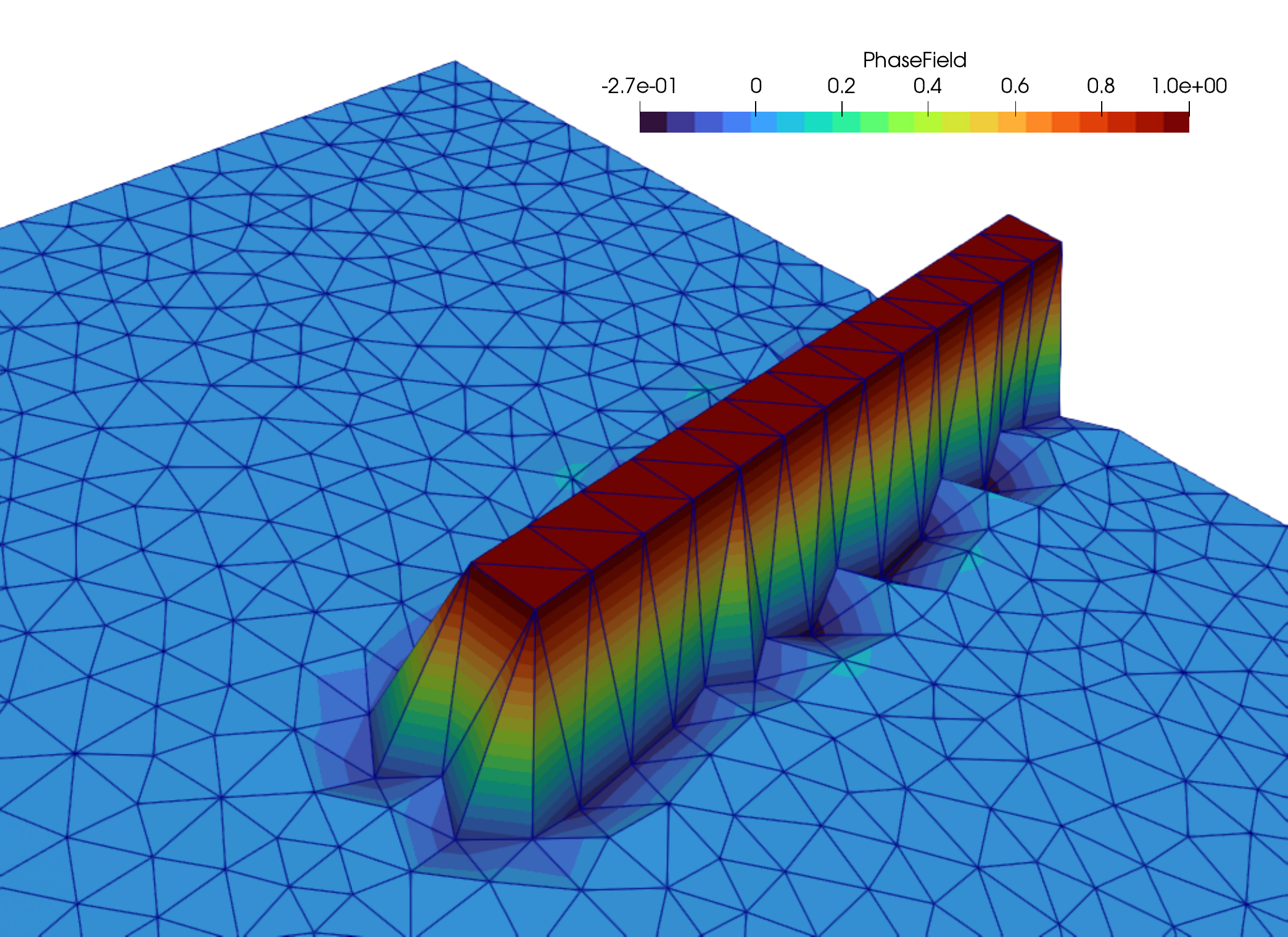}
		\caption{\label{fig:badBehavior}}
	\end{subfigure}
	\caption{Effect of mesh refinement on the length of simulated cracks, showing (a) the accurracy of crack length prediction as a function of the ratio $\ell/h^e$ and (b) the simulated phase-field profile for a fully developed crack in the case where $\ell/h^e = 0.25$. \label{fig:MeshEffectOnCrackLength}}
\end{figure}

To demonstrate the effect of mesh refinement on the calculated crack length, we revisit the single-notch problem originally investigated by \cite{Miehe2010_ijnme}. However, here we keep $\ell$ constant for the whole set of simulations, and in particular we set the specimen dimensions to $100\ell$ in order to prevent boundary conditions from adversely affecting the phase-field representation of the initial crack, which has a length of $50\ell$. The crack itself is modeled as a plateau of single-element width where $\phi = 1$, as illustrated in Figure \ref{fig:singleNotchSpecimen}.
On the boundary, we impose the condition $\nabla\phi \cdot \mathbf{n} = 0$. Furthermore, the domain is meshed uniformly with elements of characteristic size $h^e$, and the crack length is calculated by numerically integrating \eqref{eq:crackDensityFcn_Miehe} over the entire domain. 
It is easy to see from Figure \ref{fig:MeshEffectOnCrackLength_loglog} that the factor $1 + h^e/ \left( 2\ell \right)$ represents a lower bound for $\Gamma_\ell^{h^e} / \Gamma$ (or alternatively, $\GcEff / G_c$), since in reality the error is due to both the presence of a plateau at the peak and also the piecewise linear representation of a function that is otherwise smooth in the regions where $\phi < 1$.
With the data plotted in log-log scale, we can observe that there is in fact nothing remarkable about the particular ratio $\ell/h^e = 2$. Nevertheless, the discrete solution is generally no longer able to model the correct profile for $\phi$ when $h^e$ is chosen too large with respect to $\ell$, as demonstrated in Figure \ref{fig:badBehavior}. Here it can be seen that $\phi$ is no longer monotone decreasing away from the crack. Moreover, the phase-field takes on negative values which do not make sense since the theory assumes $\phi \in \left[ 0,1 \right]$. For the example above and specifically for the case of linear finite elements, we found that with $\ell/h^e = 1$ we are able to obtain a profile for $\phi$ that is still monotonically decreasing away from the crack, and at the same time lies within its assumed bounds. However the phase-field profile at the crack tip is very distorted, suggesting possible strong mesh dependence of trajectories for evolving cracks.

%% file: numerical_examples.tex
\section{Numerical simulation of fracture in V-notched plates with holes} \label{sec:numerical_examples}
\begin{figure}
	\centering
	\begin{subfigure}{0.4\textwidth}
		\includegraphics[width=\textwidth]{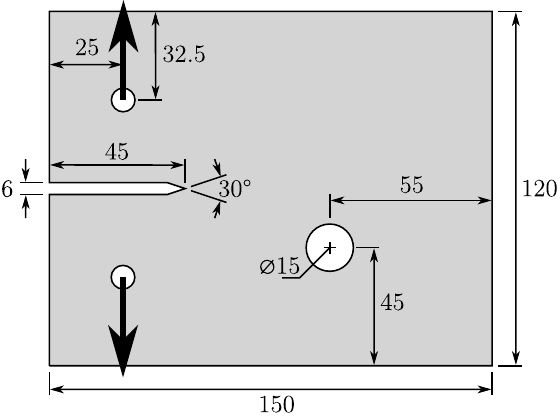}
		\caption{Single-hole specimen A\label{fig:geom_spec_1hole_A}}
	\end{subfigure}
	\begin{subfigure}{0.4\textwidth} \hspace{5mm}
		\includegraphics[width=\textwidth]{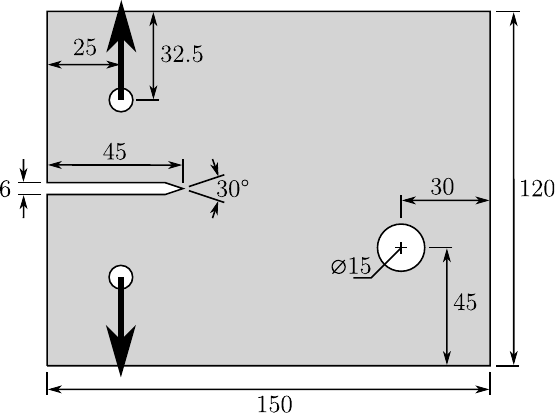}
		\caption{Single-hole specimen B\label{fig:geom_spec_1hole_B}}
	\end{subfigure} \\[5mm]
	\begin{subfigure}{0.5\textwidth}
		\includegraphics[width=\textwidth]{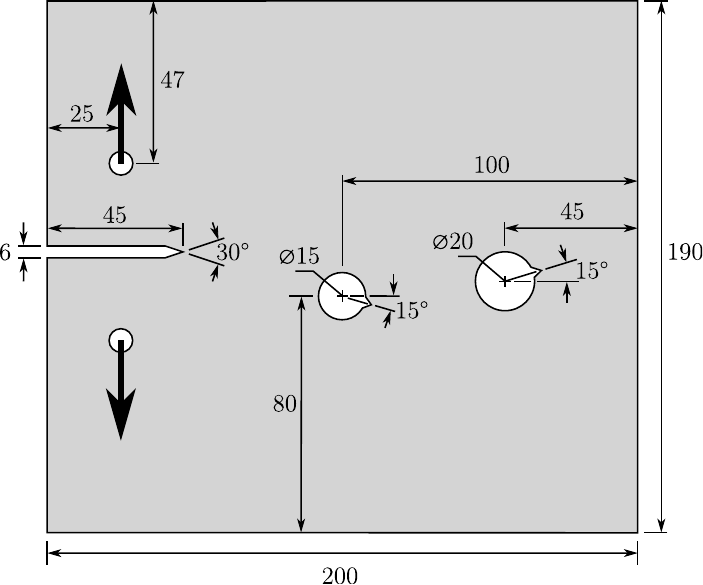}
		\caption{Double-hole specimen\label{fig:geom_spec_2hole}}
	\end{subfigure}
	\caption{Geometry of different modified compact tension specimens taken from \cite{Cavuoto2022}, with dimensions given in mm. The specimens in (a) and (b) differ in the actual location of the added hole. For the specimen shown in (c), the holes are additionally notched at their surfaces in order to control the precise location of crack nucleation. \label{fig:specimenGeometries}}
\end{figure}

Using the proposed pseudo-dynamic framework, we model fracture evolution in various compact tension (CT) specimens previously investigated by \cite{Cavuoto2022} using both physical experiments and numerical simulations. The modified CT specimens consist of V-notched plates made of photoelastic polymethyl methacrylate (PMMA), and contain additional holes that are strategically placed to induce curved crack paths as shown in Figure \ref{fig:specimenGeometries}. 
To initiate and propagate a crack in a given specimen, we apply monotonically increasing boundary conditions in the form of prescribed displacements on the specimen loading holes, at locations corresponding to the theoretical points of contact with loading pins that are connected to the testing apparatus. It is assumed that cracks nucleate at notch points\footnote{In classical compact tension tests, a fatigue crack is normally first created by subjecting the specimen to cyclic loading. The authors of \citet{Cavuoto2022} have clarified in correspondence that this step was omitted in their experiments due to integration of the loading frame with a polariscope, hence the crack can be considered to nucleate directly from the notch with zero stress intensify factor (due to the stress being weakly singular) as opposed to propagating from an existing fatigue crack under the condition $G = G_c$. \label{fn:nonstandardSetup}}
and thereafter propagate following Griffith's criterion. For the mechanical properties of PMMA (Young's modulus and Poisson's ratio), we adopt the same values utilized in \cite{Cavuoto2022}, namely $E= 3,000$ MPa and $\nu = 0.36$. However in contrast to that study, here we treat both the strength $\sigma_c$ and the critical energy release rate $G_c$ as unknown parameters, which can then be tuned in order to fit the simulations to experimental results. Finally, we assume the thickness of the specimens to be equal to 10 mm.\footnote{This value is obtained from communication with the authors of \citet{Cavuoto2022}, who explained that 10 mm was the design thickness of the plates used in the CT tests, with possible variations of $\pm 0.1$ mm in the actual manufactured specimens, as can be observed in Figure 7 of that paper.}
The experimental load-displacement (L-D) curves are obtained by digitizing the relevant plots from \citet{Cavuoto2022}, after which we divide the load values by the specimen thickness to yield normalized curves which can be directly compared with results of 2D plane-stress simulations.

The degradation function is a very important ingredient of the crack phase-field model, and particularly for models using \eqref{eq:crackDensityFcn_Miehe}, has a significant effect on solution accuracy. Here we make use of the parametric exponential family of degradation functions introduced in \citet{Sargado2018} to model the loss of material stiffness as the phase-field evolves towards unity. These take the form
\begin{linenomath}
\begin{equation}
	g \left( \phi \right) = \left( 1 - w \right) \frac{1 - e^{-k \left( 1-\phi \right)^n}}{1 - e^{-k}} + w f_c \left( \phi \right)
	\label{eq:exponentialDegradation}
\end{equation}
\end{linenomath}
in which $w$ is a weighting factor that is here set to 0.1, $n$ is a free parameter, $k = k \left( n \right)$ is a second parameter that depends on $n$, and $f_c \left( \phi \right)$ is a correction function (see the aforementioned work for further details). Regarding its use in conjunction with \eqref{eq:crackDensityFcn_Miehe}, the advantage of \eqref{eq:exponentialDegradation} over the classical quadratic degradation function is that it is able to suppress premature softening prior to crack nucleation. At the same time, it allows for greater flexibility in selecting $\ell$ within a reasonable range, which saves us from having to run prohibitively expensive simulations due to element size constraints related to $\ell$.

\begin{figure}
	\centering
	\includegraphics[width=0.45\textwidth]{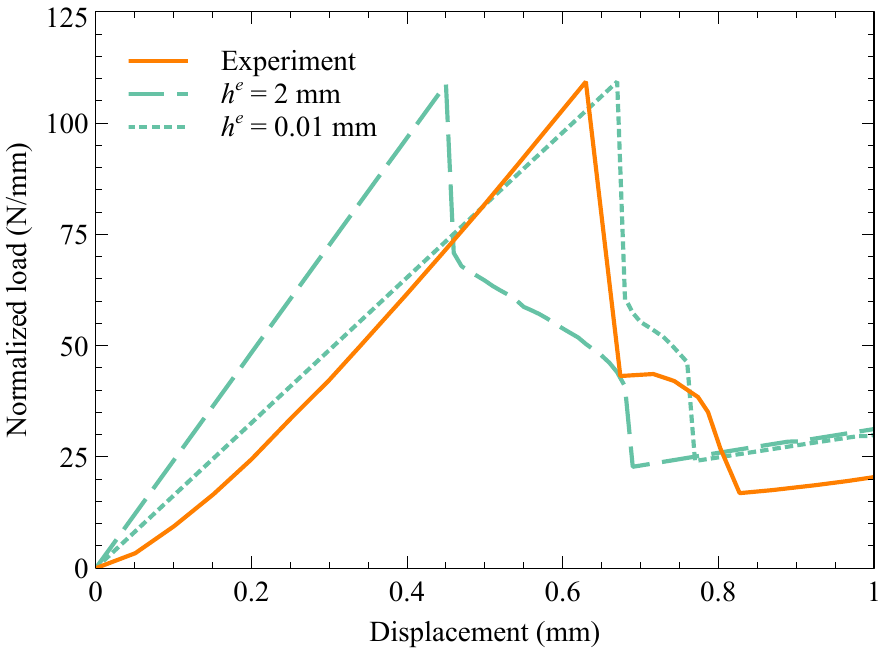}
	\caption{Influence of mesh refinement at loading point regions on the simulated load-displacement curve.\label{fig:meshEffect}}
\end{figure}

To avoid modeling nonlinear contact behavior between the loading pins and CT specimen, we apply boundary conditions directly on the specimen loading holes as mentioned earlier. Specifically, we apply a point constraint $\mathbf{u} = \mathbf{0}$ at the base of the lower hole, and a prescribed displacement $\mathbf{u} = \left\{ 0, \bar{u} \left( t \right) \right\}^\mathrm{T}$ at the apex of the upper hole, in which $\bar{u}$ is a linearly increasing function with respect to time. This would be equivalent to applying concentrated forces at these points, which results in stress singularities. Consequently, the simulated load-displacement curves are highly dependent on the level of mesh refinement in the vicinity of said concentrated loads. To avoid the occurrence of spurious damage at the loading holes, we set the critical energy release rate in the immediate region around said holes to a very high value (100 MPa-mm) so that elements surrounding the points of load application remain elastic even under very high stresses. Figure \ref{fig:meshEffect} demonstrates the variation in slope that can be obtained for the initial elastic portion of the numerical L-D curve by simply varying the characteristic size $h^e$ of elements at the regions where boundary conditions are applied (the actual problem being modeled is discussed in greater detail later in Section \ref{sec:numEx_SpecA}).
We can thus adjust the mesh refinement at the loading points so as to match the initial elastic portion of the experimental curves. Note however that when a region of concentrated loading is discretized with exceedingly small elements, the deformed mesh exhibits unphysical element interpenetration, even when the material behavior within said elements is linear elastic. Thus, care must be taken with regard to mesh refinement in order to avoid the occurrence of such effects.

Interestingly, the characteristic element size at the loading points has very little effect on the final elastic behavior of the simulated L-D curves. Additionally, we have found this region of the curve to be also insensitive to either $G_c$ and $\sigma_c$, instead being dependent only on $E$ and the overall specimen geometry.

\subsection{Adjustment procedure for experimental results \label{sec:exptAdjust}}
It can be argued that the impossibility of matching numerical results with the final elastic portion of the experimental curve shown previously (short of drastically lowering the value of $E$) motivates the need for some adjustment of the experimental data. Notably, even using the lowest values obtained from the tests results reported in \citet{Cavuoto2022} is not sufficient to get the numerical and experimental curves to coincide in this region. Moreover, in both axial and compressive tests done by \citet{Cavuoto2022}, material behavior is seen to become more compliant at higher strains, which manifests as a negative curvature in the initial elastic portion the stress-strain curve. In contrast, the same regions of the experimental L-D curves for the CT tests show hardening behavior (positive curvature).

It is more likely that these discrepancies are due to nonlinear effects arising from the non-standard setup adopted for the CT experiments, as explained in Footnote \ref{fn:nonstandardSetup}. Consequently, we have chosen to carry out an adjustment of the experimental results as an attempt to filter out such effects. In \citet{Cavuoto2022}, numerical results are fitted to the experimental data with the assumption that both the peak load and its corresponding displacement are more or less correctly represented in the experimental load-displacement curves. We adopt the same assumption here, and furthermore since we assume that the material behavior is linear elastic prior to fracture, the L-D curves should likewise exhibit initially linear behavior. That is,
\begin{linenomath}
\begin{equation}
	P_\text{ideal} \left( \Delta \right) = m \Delta, \; \forall \Delta \in \left[ 0, \Delta_\text{peak} \right]
\end{equation}
\end{linenomath}
where the slope $m$ is close to and bounded below by $P_\text{peak} / \Delta_\text{peak}$ (since probably $P_\text{peak}$ is also slightly undervalued in the experimental data). We then assume the following relationship:
\begin{linenomath}
\begin{equation}
	P_\text{adj} = f \left( P_\text{expt} \right),
\end{equation}
\end{linenomath}
wherein $P_\text{adj}$ is the adjusted value of the normalized load, and is equal to $P_\text{ideal}$ when $\Delta < \Delta_\text{peak}$. Meanwhile, $f$ is a transfer function that accounts for the nonlinear bias arising from the test setup itself. While the actual experiments cover only a limited range of force values, it is reasonable to assume that such bias tends to disappear as the magnitude of applied force becomes very large, so that $\lim_{x \rightarrow \infty} f \left( x \right) = x$. We can thus expresses the transfer function as
\begin{linenomath}
\begin{equation}
	f \left( P_\text{expt} \right) = P_\text{expt} \,f_0 \left( P_\text{expt} \right)
\end{equation}
\end{linenomath}
where now $\lim_{x \rightarrow \infty} f_0 \left( x \right) = 1$. It then remains to find a function $f_0 \left( x \right)$ such that $f_0 \left( P_\text{expt} \right) \approx P_\text{ideal} / P_\text{expt}$. Applying this to the experimental results in \cite{Cavuoto2022}, we can observe that the behavior of $P_\text{ideal} / P_\text{expt}$ is such that $f_0 \left( x \right) \rightarrow \infty$ as $x \rightarrow 0$. We found that a reasonable expression for $f_0 \left( x \right)$ is given by
\begin{linenomath}
\begin{equation}
	f_ 0 \left( x \right) = \frac{1}{\left[ \mathrm{erf} \left( \alpha x \right) \right]^\beta}
\end{equation}
\end{linenomath}
in which $\mathrm{erf} \left( x \right) = \left( 2/ \sqrt{\pi} \right) \int_0^x e^{-t^2} dt$ denotes the error function, while $\alpha$ and $\beta$ are fitting parameters with $\alpha,\beta \in \left( 0,1 \right)$. The values of $\alpha$ and $\beta$ for a particular experimental load-displacemen curve can be found by minimizing the quantity
\begin{linenomath}
\begin{equation}
	R = \sum_{i=1}^N \left[ f_0 \left( {P_\text{expt}}_i \right) - {P_\text{ideal}}_i / {P_\text{expt}}_i \right]^2,
\end{equation}
\end{linenomath}
where $\left( \Delta_i, {P_\text{expt}}_i \right)$ are the coordinates of the $i$-th digitized point from the experimental curve with $\Delta_i < \Delta_{i+1}$, and $N$ is the number of digitized points for which $\Delta_i < \Delta_\text{peak}$. Although $f_0 \left( x \right)$ is highly nonlinear, the minimization of $R$ with respect to $\alpha$ and $\beta$ can be done straightforwardly in spreadsheet software, and we have found that a differential evolution algorithm yields sufficiently good results. Having determined $\alpha$ and $\beta$, adjustment of the experimental load-displacement curve can then be done as follows:
\begin{linenomath}
\begin{equation}
	{P_\text{adj}}_i = \left\{ \begin{array}{ll}
		\dfrac{P_\text{peak}}{\Delta_\text{peak}} \Delta_i, & \Delta_i \leq \Delta_\text{peak} \\[15pt]
		\dfrac{{P_\text{expt}}_i}{\left[ \mathrm{erf} \left( \alpha {P_\text{expt}}_i \right) \right]^\beta}, & \Delta_i > \Delta_\text{peak}.
	\end{array} \right.
	\label{eq:adjustmentProcedure}
\end{equation}
\end{linenomath}
It is important to emphasize that the above adjustment is completely independent from numerical simulations, and is carried out without utilizing any specific information from the latter.

\begin{figure}
	\centering
	\begin{subfigure}{0.47\textwidth}
		\includegraphics[width=\textwidth]{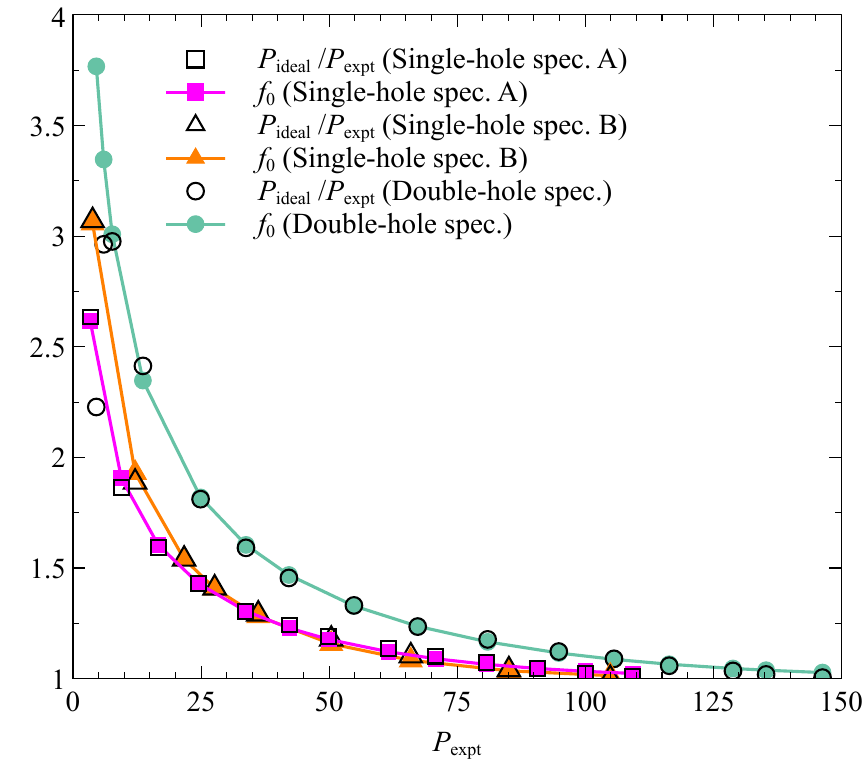}
		\caption{}
	\end{subfigure} \hspace{1mm}
	\begin{subfigure}{0.47\textwidth}
		\includegraphics[width=\textwidth]{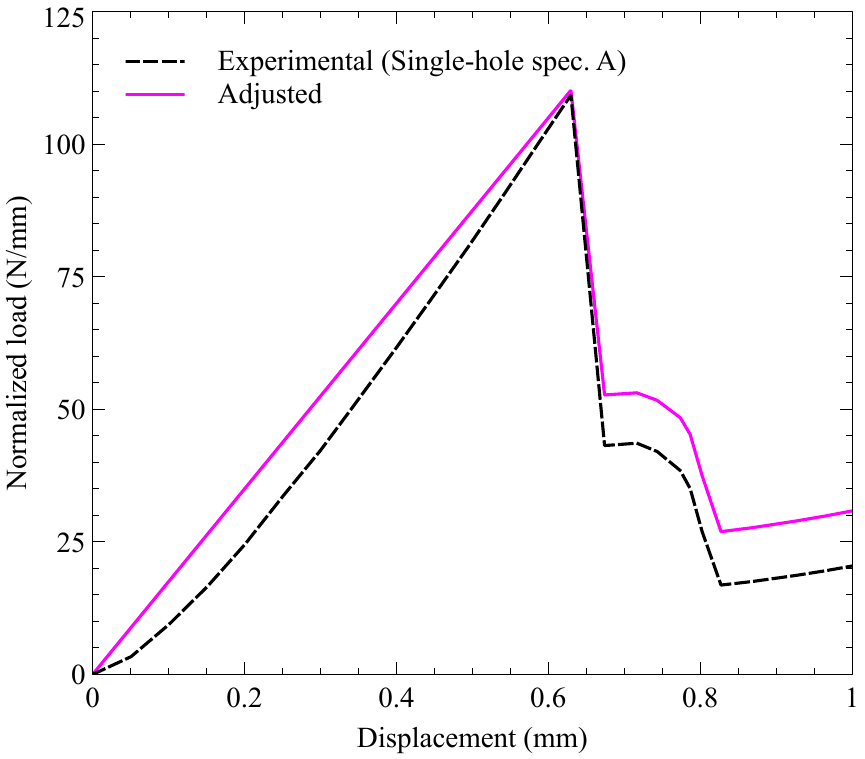}
		\caption{}
	\end{subfigure} \\
	\begin{subfigure}{0.47\textwidth}
		\includegraphics[width=\textwidth]{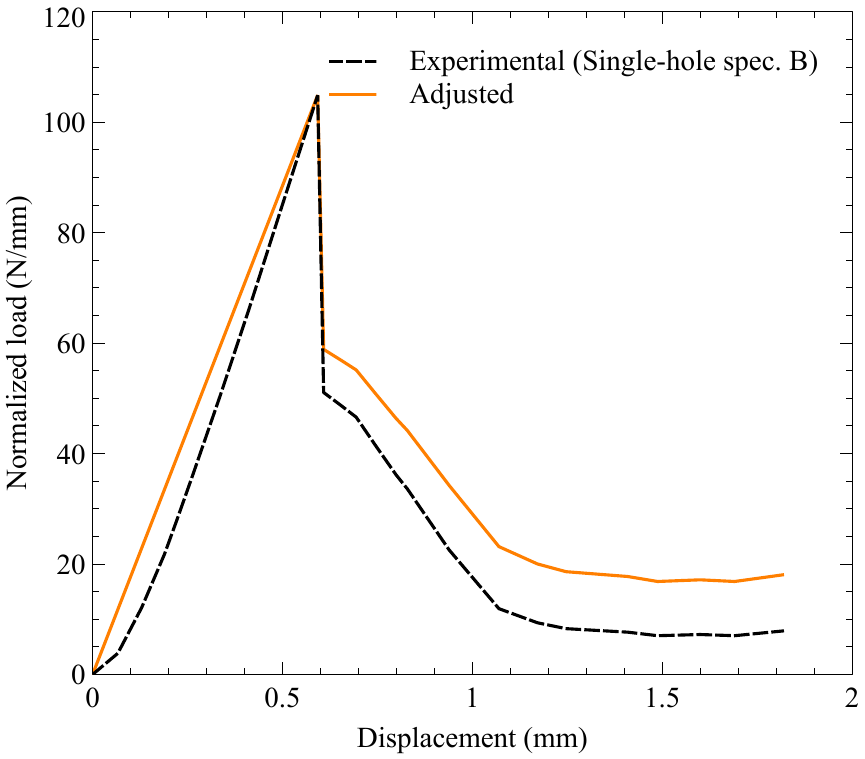}
		\caption{}
	\end{subfigure} \hspace{1mm}
	\begin{subfigure}{0.47\textwidth}
		\includegraphics[width=\textwidth]{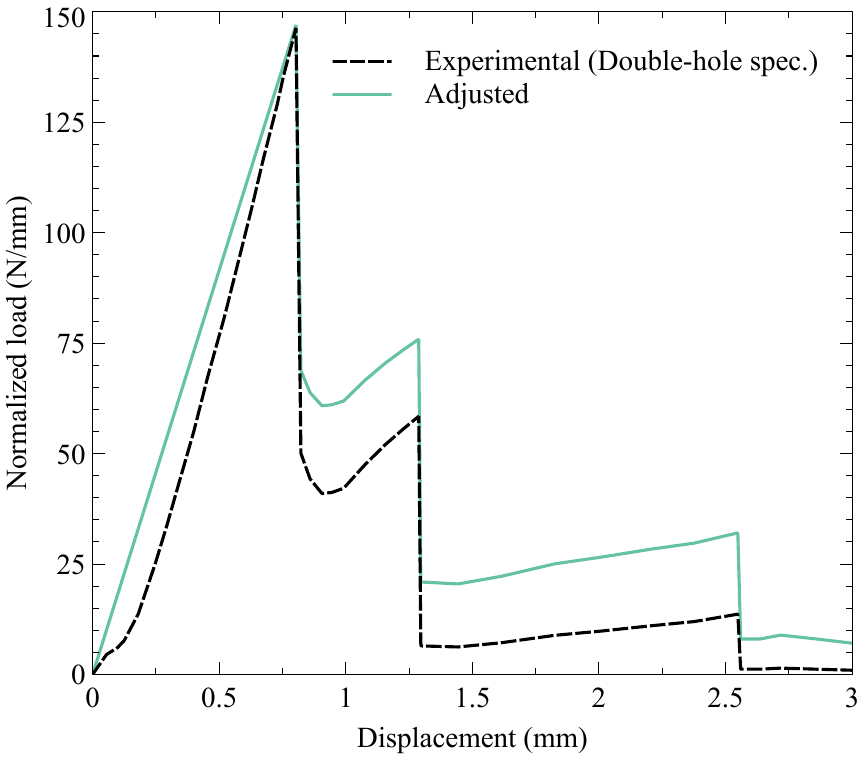}
		\caption{}
	\end{subfigure}
	\caption{Adjustment of experimental results for modified CT specimens. The fitted function $f_0$ for the different specimens is shown in (a), while the adjusted load-displacement curves are shown in (b)--(d), together with the original data from \cite{Cavuoto2022}.\label{fig:experimentalAdjustment}}
\end{figure}

We apply the procedure described above to adjust the experimental curves associated with each of the specimens shown in Figure \ref{fig:specimenGeometries}. A summary of the calculated parameters is given in Table \ref{tab:adjustmentParameters}, while the individual fits for the transfer function $f_0 \left( x \right)$ and the resulting adjusted curves are shown in Figure \ref{fig:experimentalAdjustment}.
\begin{table}
	\caption{Calculated parameters for adjustment of experimental load-displacement curves \label{tab:adjustmentParameters}}
	\centering
	\begin{tabular}{ccccc}
		\toprule
		CT Specimen & $P_\text{peak}$ & $\Delta_\text{peak}$ & $\alpha$ & $\beta$ \\
		& (N/mm) & (mm) & & \\
		\midrule
		Single-Hole A & 110.20 & 0.630 & 0.0117 & 0.3085 \\
		Single-Hole B & 104.90 & 0.593 & 0.0144 & 0.4025 \\
		Double-Hole & 147.01 & 0.803 & 0.0091 & 0.4324 \\
		\bottomrule 
	\end{tabular}
\end{table}

\subsection{Compact tension test on single-hole specimen A \label{sec:numEx_SpecA}}
\begin{figure}
	\centering
	\begin{subfigure}{0.45\textwidth}
		\includegraphics[width=\textwidth]{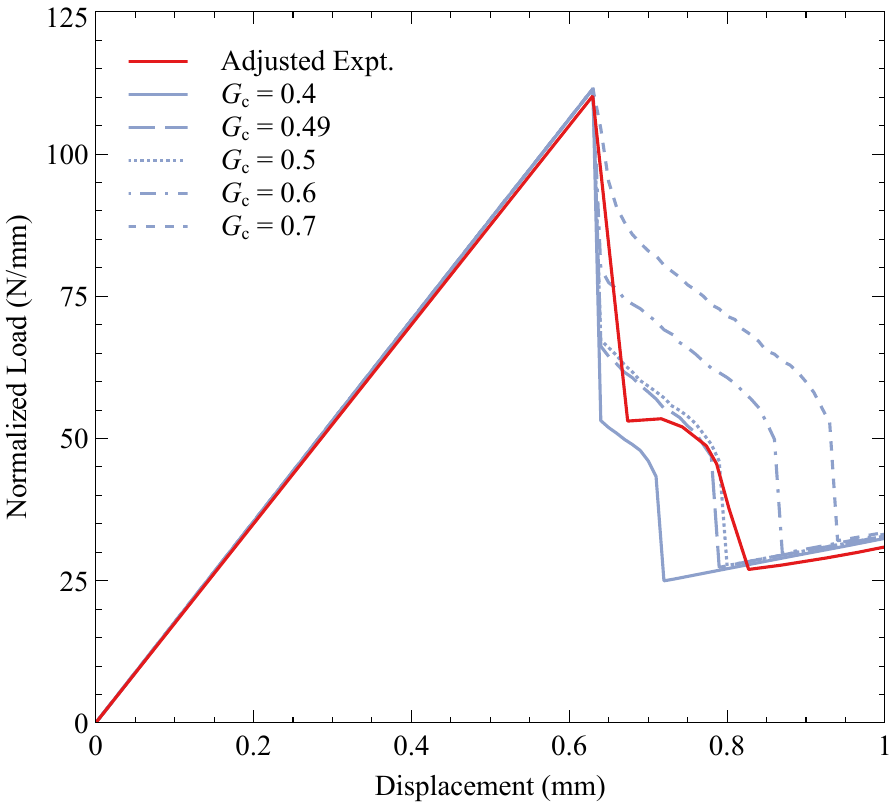}
		\caption{}
	\end{subfigure} \hspace{5mm}
	\begin{subfigure}{0.45\textwidth}
		\includegraphics[width=\textwidth]{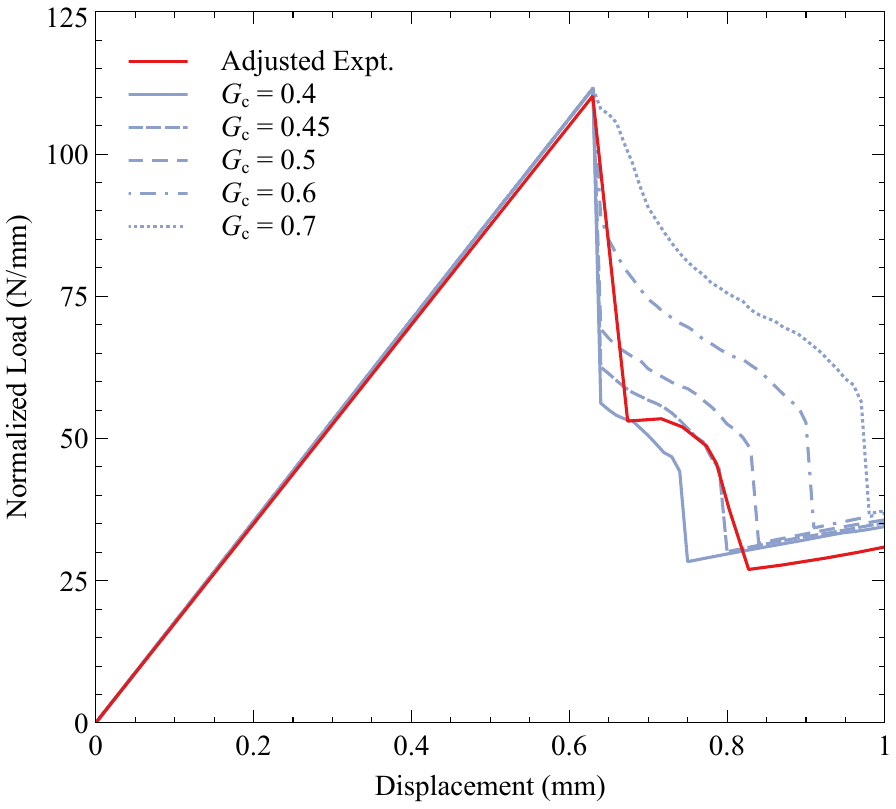}
		\caption{}
	\end{subfigure}	
	\caption{Simulated load-displacement curves for the modified CT specimen in Section \ref{sec:numEx_SpecA} using mesh refinement ratios of (a) $\ell/h^e  = 5$, and (b) $\ell/h^e = 2.5$. Note that the listed values for $G_c$ (MPa-mm) are the actual numbers provided as input to the simulations; their corresponding true (effective) values can be found in Table \ref{tab:CTSpec01A}. \label{fig:ldCurves_CTSpec01A}}
\end{figure}

We begin by modeling crack growth in the modified compact tension (CT) specimen shown in Figure \ref{fig:geom_spec_1hole_A}. In a standard CT test, the crack typically propagates in a straight, horizontal path due to the symmetry of the specimen geometry and applied loading. However, the introduction of the hole causes the crack trajectory to curve towards it. As mentioned earlier, we choose the appropriate level of mesh refinement at the loading points to accurately capture the initial slope of the adjusted experimental load-displacement curve. Following this, we proceed with our calculations as outlined below:
\begin{enumerate}[(a)]
	\item As a first step, we perform simulations using the standard quasi-static (fully energy-dissipating) phase-field model, varying $G_c$ while keeping $\ell$ constant, and taking care to match the peak load in the adjusted experimental L-D curve. We take the correct value of $G_c$ as the one that reproduces the \emph{softening region} observed in the experimental results.
	\item Next, we compute the effective critical energy release rate using the formula in \eqref{eq:effectiveToughness}. By examining photographs provided in \cite{Cavuoto2022} of the specimen at various loading stages, we can deduce the length $\Delta\Gamma_\text{init}$ of the initial crack jump immediately following the peak load. The corresponding change in surface energy is then given by
	\begin{linenomath}
	\begin{equation}
		\Delta \Psi_s = \GcEff \, \Delta\Gamma_\text{init}.
	\end{equation}
	\end{linenomath}
	\item Finally, we switch to the proposed pseudo-dynamic framework and conduct several simulations, adjusting the loss coefficient $\zeta$ until the change in surface energy associated with the initial crack jump matches the value calculated in the previous step.
\end{enumerate}
As mentioned previously, we set $E = 3000$ MPa and $\nu = 0.36$ for the simulated PMMA specimens. By tuning the parameters of the degradation function given in \eqref{eq:exponentialDegradation}, we can vary the critical load at crack nucleation for different ratios of the characteristic element size to the phase-field length scale. For this problem, we set $\ell = 0.5$ mm and conduct two sets of simulations: one with $h^e = 0.1$ mm and another with $h^e = 0.2$ mm. To minimize computational cost, we refine the mesh a priori around the expected crack trajectory and choose appropriate values for the degradation function parameters to match the correct peak load for each assumed $G_c$. The details are provided in Table \ref{tab:CTSpec01A}, while the resulting simulated L-D curves are compared with the adjusted experimental data in Figure \ref{fig:ldCurves_CTSpec01A}.
\begin{table}
	\centering
	\caption{Summary of data for simulation runs pertaining to the modified CT specimen in Section \ref{sec:numEx_SpecA}. \label{tab:CTSpec01A}}
	\begin{tabular}{ccccccc}
	\toprule 
	Simul. & $\ell$ & $h^e$ & \multicolumn{2}{c}{Deg.\ fcn.\ param.} & $G_c$ & $\GcEff$ \\
	No. & (mm) & (mm) & $n$ & $w$ & (MPa-mm) & (MPa-mm) \\
	\midrule
	1 & 0.5 & 0.1 & 3.6 & 0.1 & 0.40 & 0.44 \\
	2 & 0.5 & 0.1 & 4.1 & 0.1 & 0.49 & 0.539 \\
	3 & 0.5 & 0.1 & 4.2 & 0.1 & 0.50 & 0.55 \\
	4 & 0.5 & 0.1 & 4.7 & 0.1 & 0.60 & 0.66 \\
	5 & 0.5 & 0.1 & 5.2 & 0.1 & 0.70 & 0.77 \\
	\midrule
	6 & 0.5 & 0.2 & 3.7 & 0.1 & 0.40 & 0.48 \\
	7 & 0.5 & 0.2 & 4.0 & 0.1 & 0.45 & 0.54 \\
	8 & 0.5 & 0.2 & 4.3 & 0.1 & 0.50 & 0.60 \\
	9 & 0.5 & 0.2 & 4.8 & 0.1 & 0.60 & 0.72 \\
	10 & 0.5 & 0.2 & 5.3 & 0.1 & 0.70 & 0.84 \\
	\bottomrule
	\end{tabular}
\end{table}

Simulations with a ratio of $\ell/h^e = 2.5$ show stiffer behavior in the final elastic segment of the load-displacement curve compared to those with $\ell/h^e = 5$. This could be explained by the fact that for smaller $h^e$, elements at the crack tip region generally end up with higher values of $\phi$, leading to more compliant overall behavior. Alternatively, it may stem from limitations inherent in the spectral decomposition model of \citet{Miehe2010_ijnme}, which is not always able to enforce zero residual stresses at open cracks \citep{Vicentini2024}. A well-known example where this can be observed is in the shearing of existing crack surfaces \citep{Strobl2016}. In our case, the phenomenon seems to be similar, but it is likely mitigated at larger $\ell/h^e$ ratios where more elements experience near-complete material degradation.

We can observe that the adjusted experimental behavior is best matched by the simulation where $G_c = 0.49$ MPa-mm when $\ell/h^e = 5$, and with $G_c = 0.45$ MPa-mm when $\ell/h^e = 2.5$. These two cases correspond to the same effective critical energy release ($G_c^\text{eff} = 0.54$ MPa-mm) when we apply the correction given in \eqref{eq:effectiveToughness}, rather than 0.7 MPa-mm as was originally assumed in \cite{Cavuoto2022}. Due to the lack of enforced global energy balance, the standard phase-field approach is unable to properly model the drops in applied force.

\begin{figure}
	\centering
	\begin{subfigure}{0.45\textwidth}
		\includegraphics[width=\textwidth]{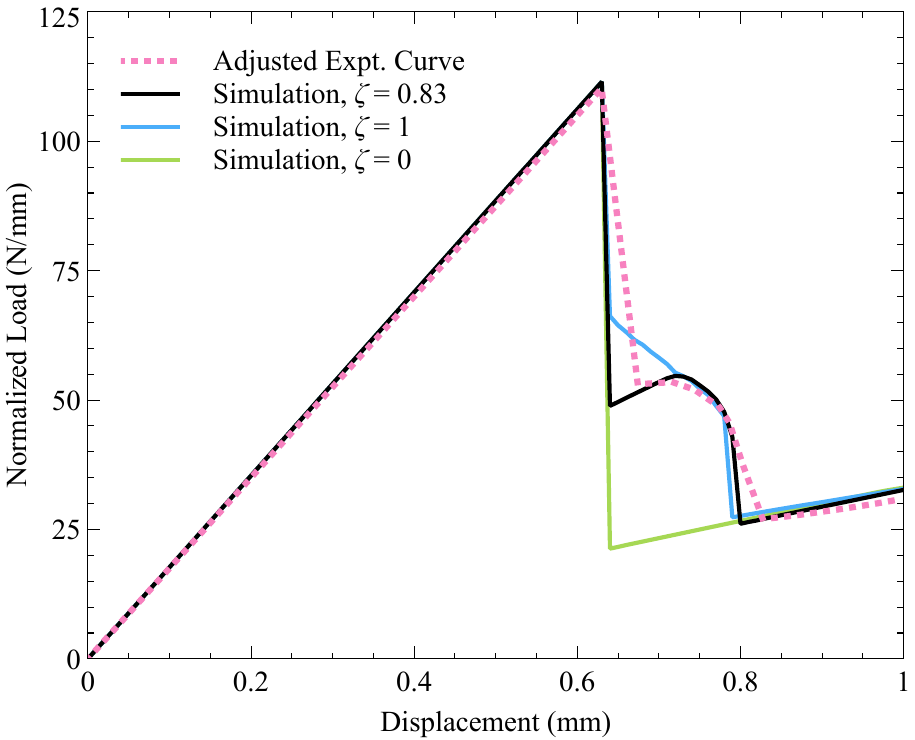}
		\caption{\label{fig:CTSpec01_A_LDCurve}}
	\end{subfigure} \hspace{5mm}
	\begin{subfigure}{0.46\textwidth}
		\includegraphics[width=\textwidth]{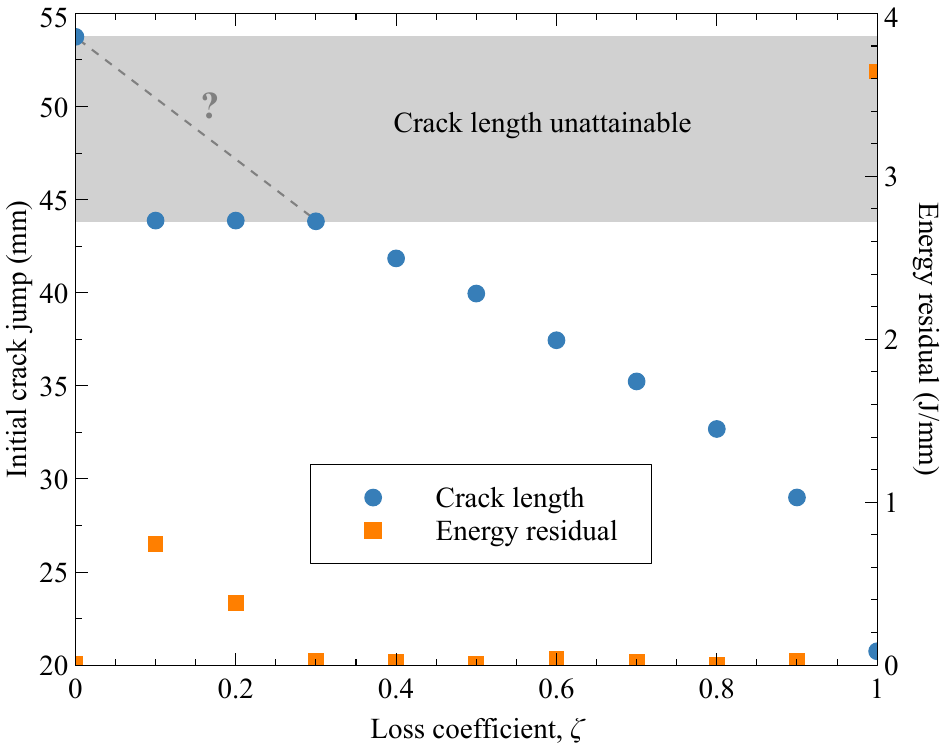}
		\caption{\label{fig:CTSpec01_A_ForbiddenRegion}}
	\end{subfigure} \\[2mm]
	\begin{subfigure}{0.3\textwidth}
		\includegraphics[width=\textwidth]{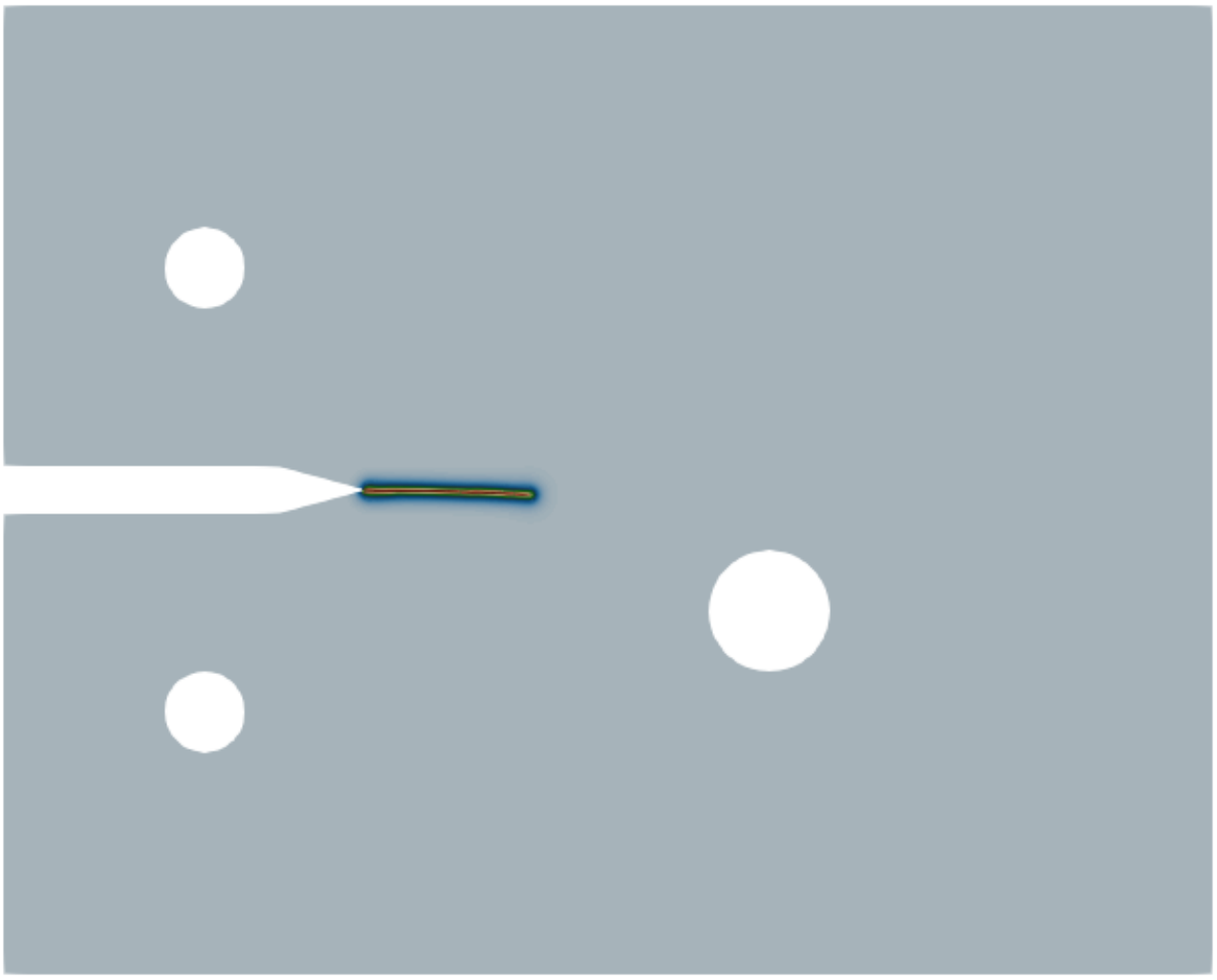}
		\caption{\label{fig:SpecA_InitJump_zeta_1_0}}
	\end{subfigure} \hspace{2mm}
	\begin{subfigure}{0.3\textwidth}
		\includegraphics[width=\textwidth]{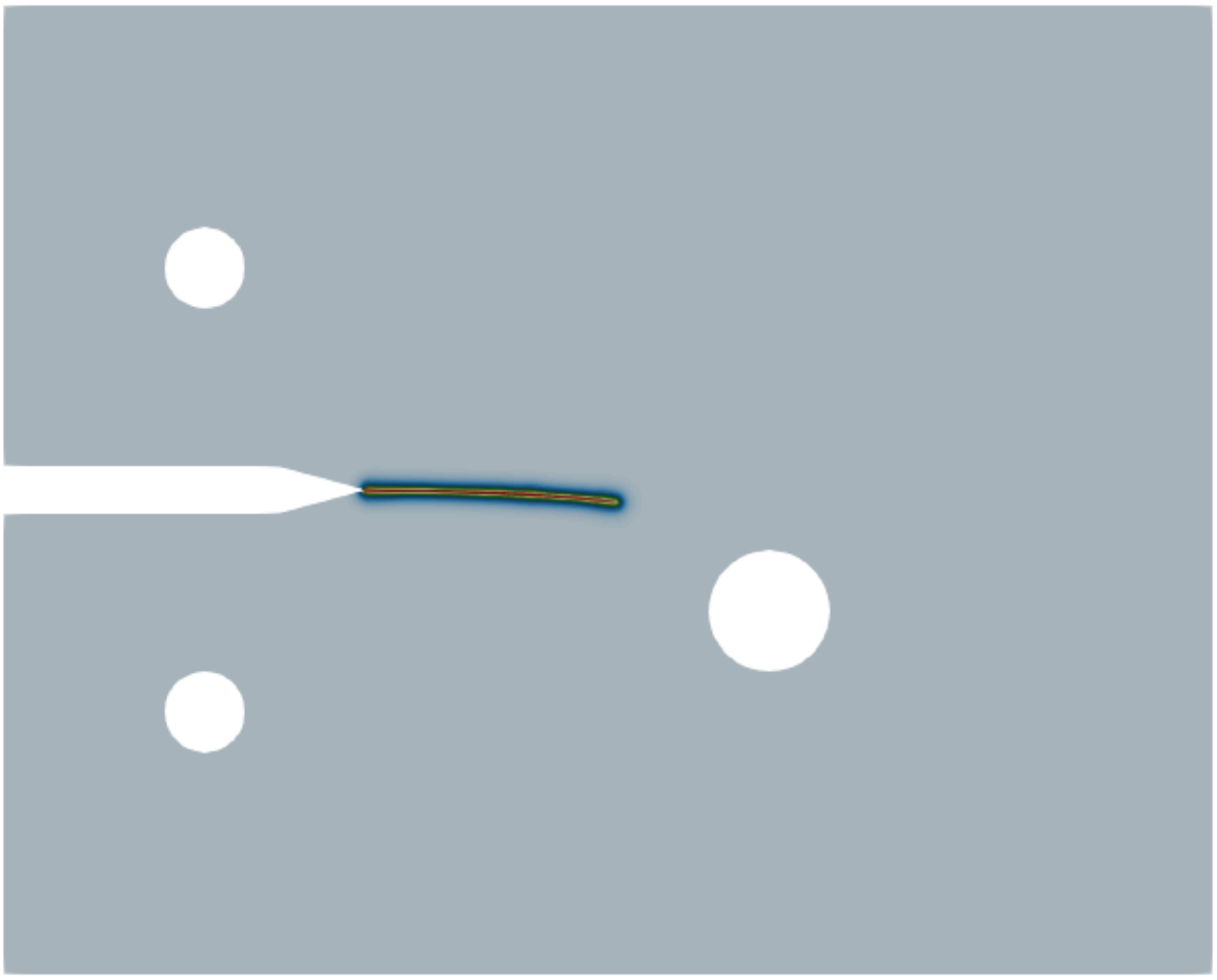}
		\caption{\label{fig:SpecA_InitJump_zeta_0_83}}
	\end{subfigure} \hspace{2mm}
	\begin{subfigure}{0.3\textwidth}
		\includegraphics[width=\textwidth]{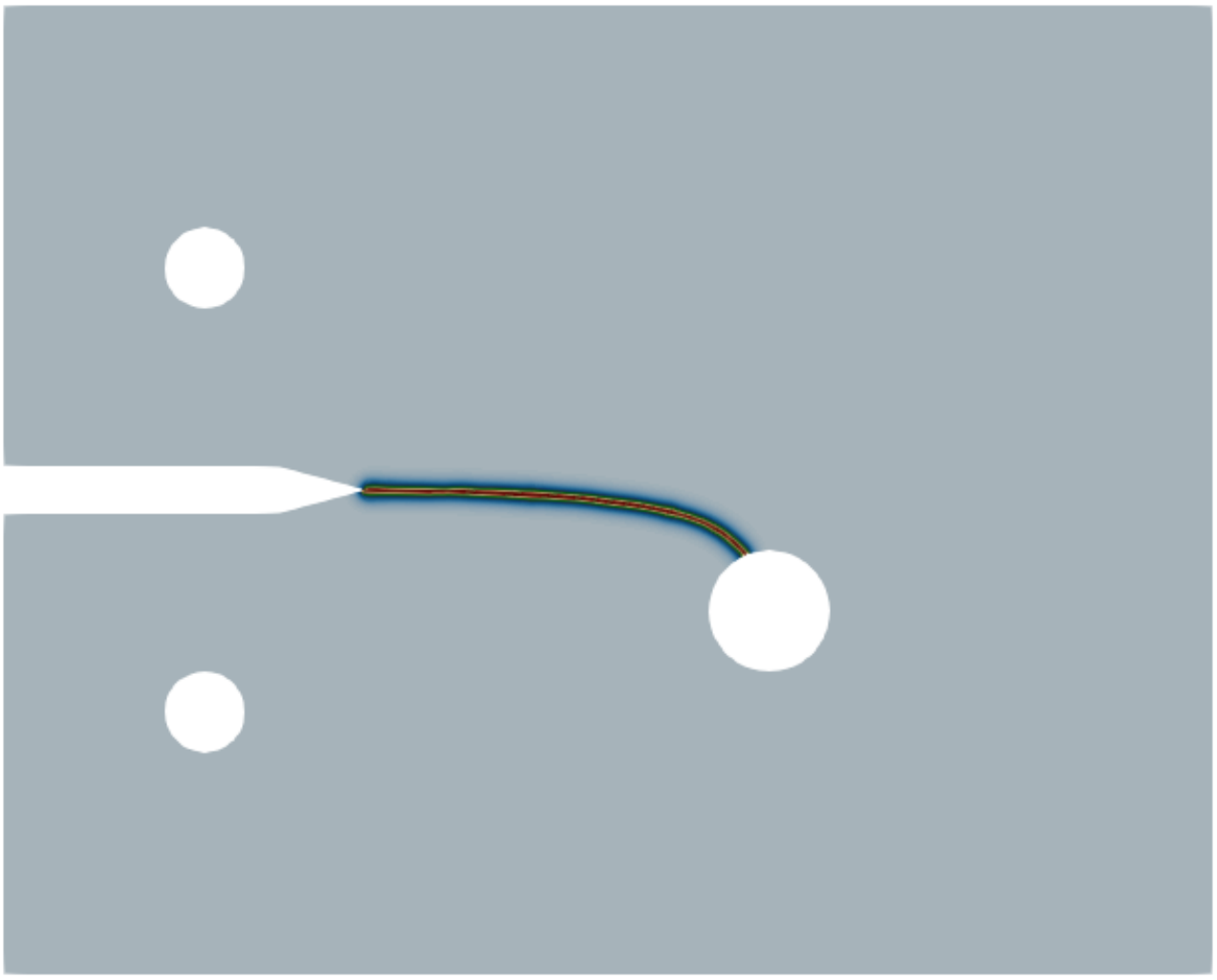}
		\caption{\label{fig:SpecA_InitJump_zeta_0_0}}
	\end{subfigure}
	\caption{Results for the modified CT test on single-hole specimen A ($\ell/h^e = 5$), showing (a) load-displacement curves for different loss coefficients together with the adjusted experimental data, and (b) the apparent dependence of the initial crack jump on $\zeta$, together with their associated energy residuals. The initial crack jump trajectory corresponding to various loss coefficients is also shown, and is equal to (c) 20.7 mm when $\zeta = 1$, (d) 32.0 mm when $\zeta = 0.83$, and (e) 53.6 mm when $\zeta = 0$.}
\end{figure}

\begin{figure}
	\centering
	\begin{subfigure}{0.31\textwidth}
		\includegraphics[width=\textwidth]{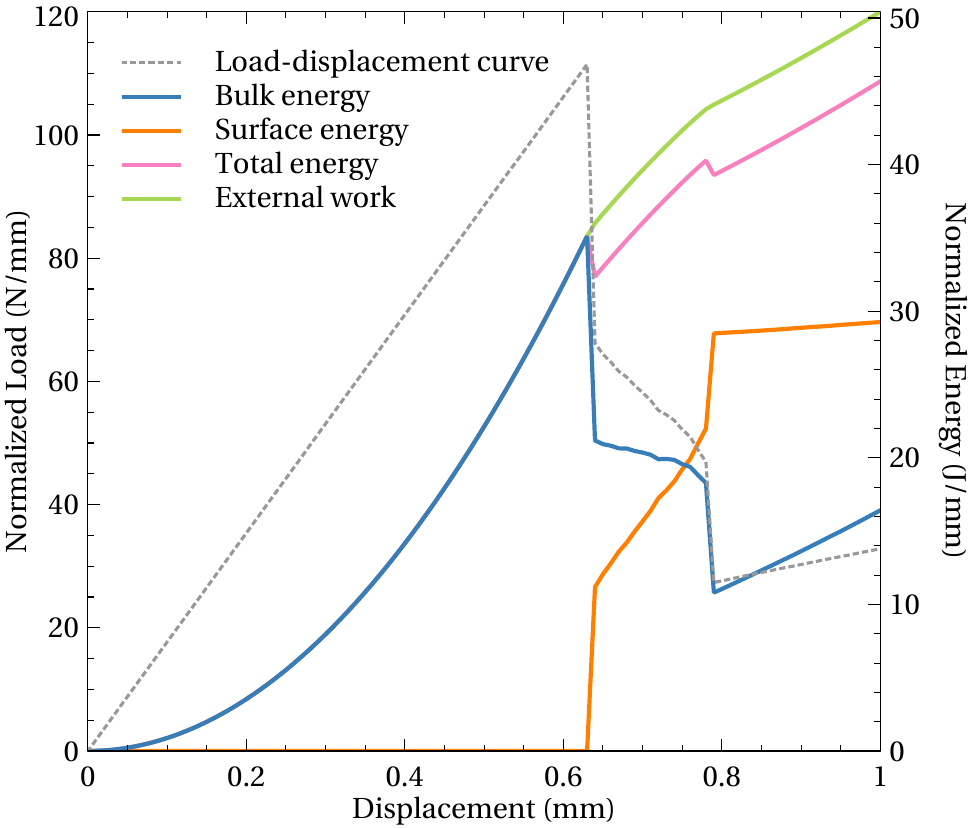}
		\caption{\label{fig:SpecA_EgyEvol_zeta_1_0}}
	\end{subfigure} \hspace{2mm}
	\begin{subfigure}{0.31\textwidth}
		\includegraphics[width=\textwidth]{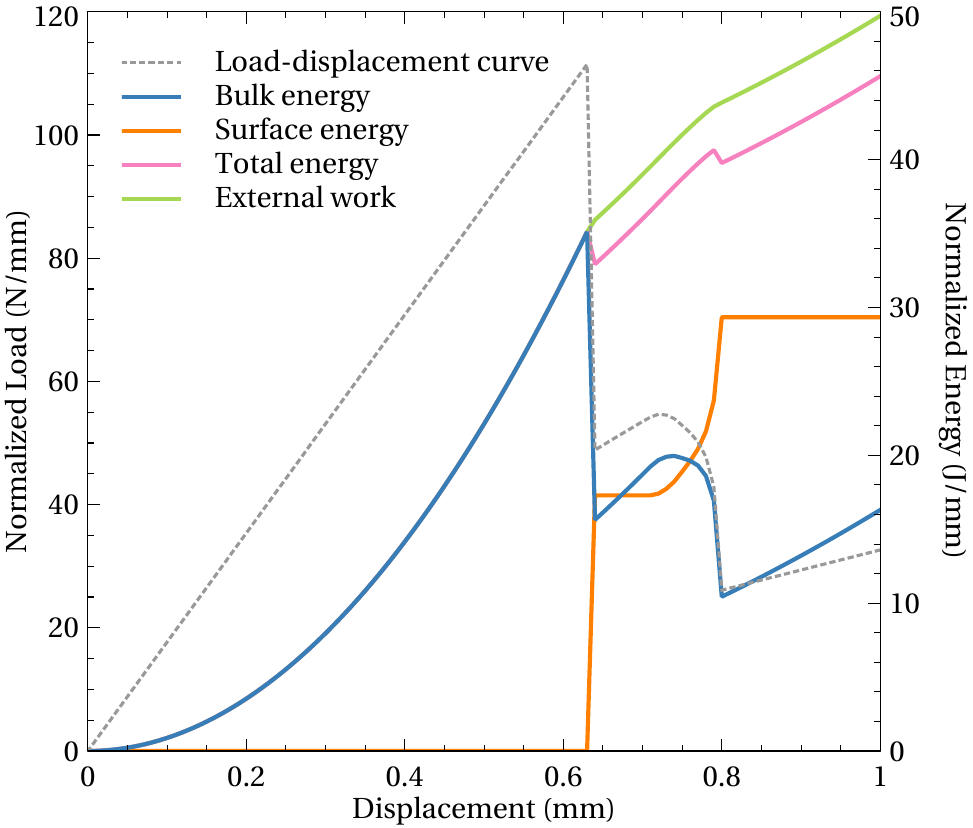}
		\caption{\label{fig:SpecA_EgyEvol_zeta_0_83}}
	\end{subfigure} \hspace{2mm}
	\begin{subfigure}{0.31\textwidth}
		\includegraphics[width=\textwidth]{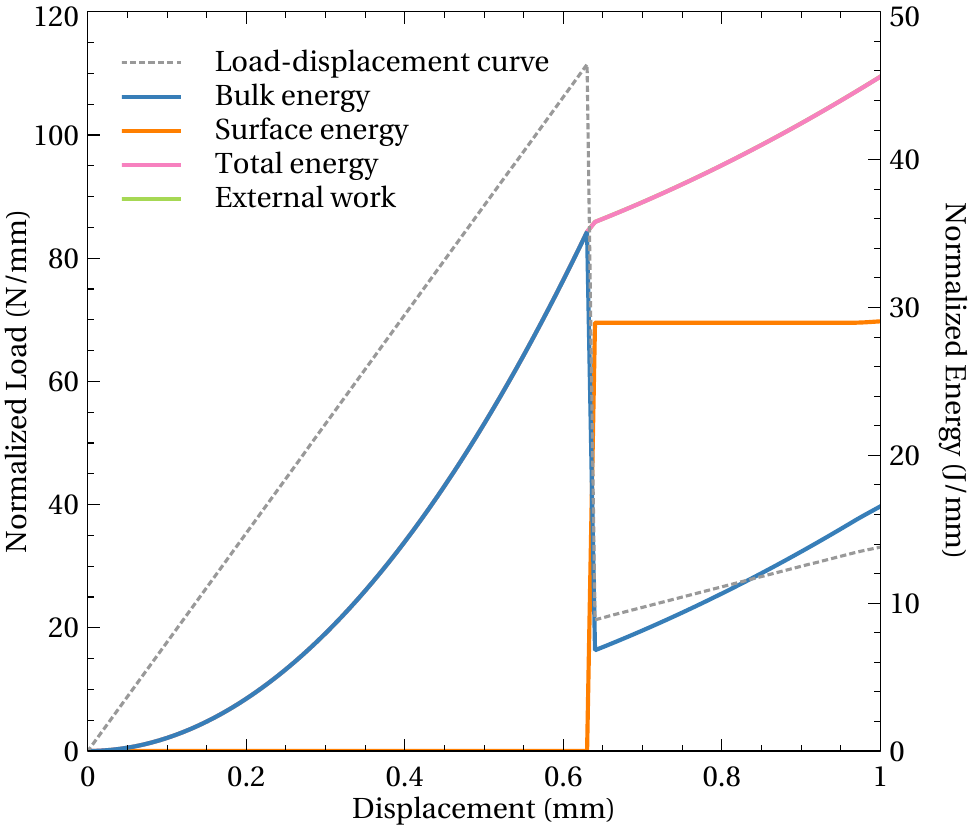}
		\caption{\label{fig:SpecA_EgyEvol_zeta_0_0}}
	\end{subfigure}
	\caption{Evolution of energy quantities with respect to displacement in the simulated compacted tension test on single-hole specimen A, corresponding to a loss coefficient of (a) $\zeta = 1$, (b) $\zeta = 0.83$, and (c) $\zeta = 0$. \label{fig:SpecA_EgyEvol}}
\end{figure}

Photographs of different stages in the fracture evolution are also reported in \cite{Cavuoto2022} (see Figure 3 of said work), from which we can determine the initial crack advance to be around 32 mm. The associated change in surface energy is then given by $\Delta \Psi_s =  0.54 \cdot 32 \approx 17.28$ N-mm per millimeter thickness of the specimen. Switching to the proposed pseudo-dynamic framework, we iterate over different values of the loss coefficient $\zeta$ until the jump in surface energy at the initial brutal crack advance matches the value calculated previously. For the current problem, we found that the condition is reasonably satisfied for a loss coefficient of $\zeta = 0.83$. The corresponding results are reported in Figure \ref{fig:CTSpec01_A_LDCurve}, where we can observe that the simulated L-D curve matches the adjusted experimental data quite well.
For comparison, we also plot the fully dissipative ($\zeta = 1$) and fully energy-conserving ($\zeta = 0$) solutions. As the loss coefficient is designed to account for \emph{all} of the dissipated energy, it generally cannot be considered a material parameter: in addition to internal friction within the specimen, some energy dissipation may also occur at specimen boundaries, and furthermore within the testing machine itself. As expected, the simulated initial crack jump is heavily influenced by the choice of $\zeta$, as can be observed in in Figures  \ref{fig:SpecA_InitJump_zeta_1_0}-\ref{fig:SpecA_InitJump_zeta_0_0}. The effect of the loss coefficient on the evolution of the different energy quantities is shown in Figure \ref{fig:SpecA_EgyEvol}.

It can be seen that for the current example, we are able to obtain fully energy conservation in the simulation for which we have specified $\zeta = 0$. However this is not always guaranteed to be the case. In fact, an intriguing phenomenon emerges when examining the specific relationship between the initial jump magnitude and the loss coefficient. Figure \ref{fig:CTSpec01_A_ForbiddenRegion} illustrates the simulated crack jumps for various values of $\zeta$ ranging from 0 to 1, with increments of 0.1. Notably, for 
$\zeta$ between 0.1 and 0.3, the predicted jump magnitude remains identical. However, as $\zeta$ decreases, the corresponding energy residuals increase, and in particular remain the same order of magnitude as the residual associated with the classical phase-field model ($\zeta = 1$). This indicates a failure of the algorithm to enforce energy balance. It is partially due to the design of our procedure for determining the overload factor, which only accepts a final solution of $\bar{\eta}$ for which the corresponding residual $r^\eta$ is non-negative. In particular when a discontinuity in the behavior of $r^\eta \left( \bar{\eta} \right)$ is detected, the algorithm terminates and returns the overload factor corresponding to the smallest positive residual, regardless of whether the latter has converged or not with respect to the initial residual according to the specified convergence criteria. The discontinuity itself can be explained by the fact that cracks may intersect boundaries, in this case the added hole in the specimen. \cite{Zak1963} showed analytically that when a crack propagates from a stiffer toward a more compliant material, the energy release rate becomes infinite when the crack tip is impinging on the interface between these two materials. Furthermore, the increase in energy release rate—compared to a single-material case—depends on the distance between the crack tip and the interface \citep{Sargado2024}. In our proposed framework, dynamic forces are essentially replaced by equivalent static loads. Within a quasi-static context, crack arrest occurs at a configuration corresponding to a local minimum of the total potential energy. However, the increased energy release rate near a material interface may lead to a situation where no local minima exist, so that the crack grows until it penetrates the interface. This leads to the formation of ``forbidden'' regions around boundaries and interfaces within which crack arrest cannot be predicted by the pseudo-dynamic framework, but may otherwise be accessible in a fully dynamic model.

Numerically, the limitation arises from the nested approach employed to solve the discrete equations in \eqref{eq:residuals}. Specifically, we use an alternate minimization algorithm (AM) within an inner loop to solve \eqref{eq:residual_u} and \eqref{eq:residual_phi}, and then apply the procedure outlined in Section \ref{sec:etaSolutionProcedure} to solve \eqref{eq:residual_energyBalance} for $\bar{\eta}$ in an outer loop. In this setup, the inner AM loop assumes a fixed value for $\bar{\eta}$. While it is possible to allow the overload factor to vary within the inner loop, doing so would impact the convergence properties of the AM algorithm and would require a redesign of the solution scheme. Moreover, it would necessitate extensive numerical studies to assess its feasibility and robustness. An alternative approach could be to abandon the assumption of a uniform $\eta$ over the entire domain, and instead allow for a spatially and temporally varying overload factor, which primarily affects only the crack tip regions. This would imply a dependence of $\eta$ on $\phi$ and may potentially suppress the occurrence of forbidden regions as mentioned earlier. Although this idea has not been implemented in the current study, it holds promise and is intended for exploration in a future work.

\subsection{Compact tension test on single-hole specimen B (alternative hole placement) \label{sec:numEx_SpecB}}
The second modified CT test investigated in \cite{Cavuoto2022} involved a specimen with the same overall dimensions as in the previous example, but with the hole placed further from the initial notch, as shown in Figure \ref{fig:geom_spec_1hole_B}. As all the CT specimens are manufactured with the same material, we can directly use the value of $G_c^\text{eff}$ obtained in the previous example. Due to the added hole being placed further away from the notch in this specimen, the crack tip has a greater distance to travel before intersecting the hole boundary, resulting in a more extended softening region in the load-displacement curve.

\begin{figure}
	\centering
	\begin{subfigure}{0.45\textwidth}
		\includegraphics[width=\textwidth]{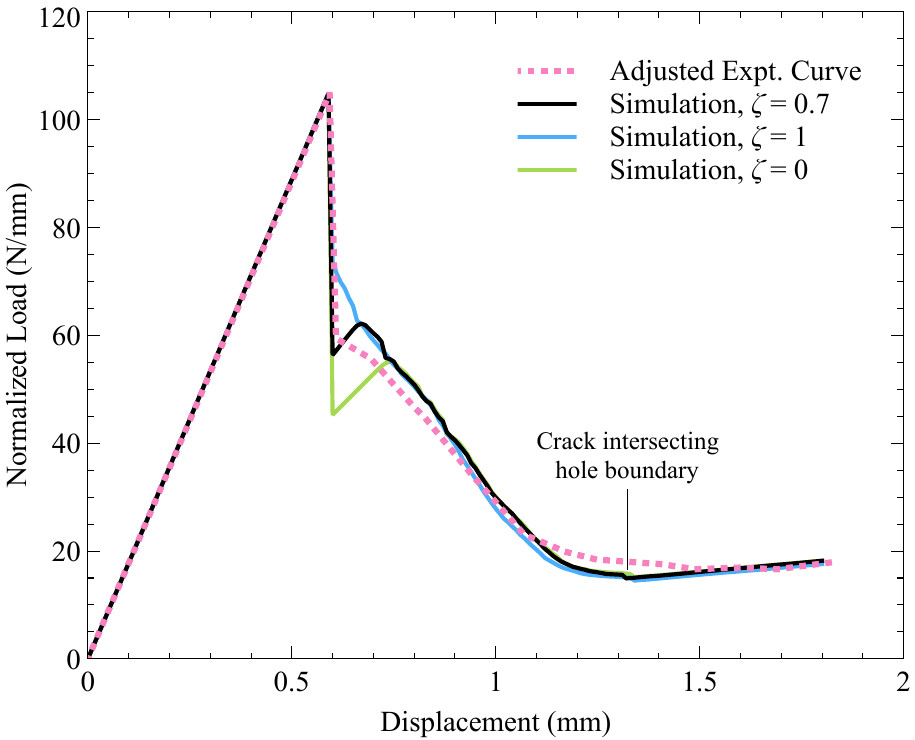}
		\caption{\label{fig:CT_Spec_1Hole_B_Summary}}
	\end{subfigure} \hspace{2mm}
	\begin{subfigure}{0.40\textwidth}
		\includegraphics[width=\textwidth]{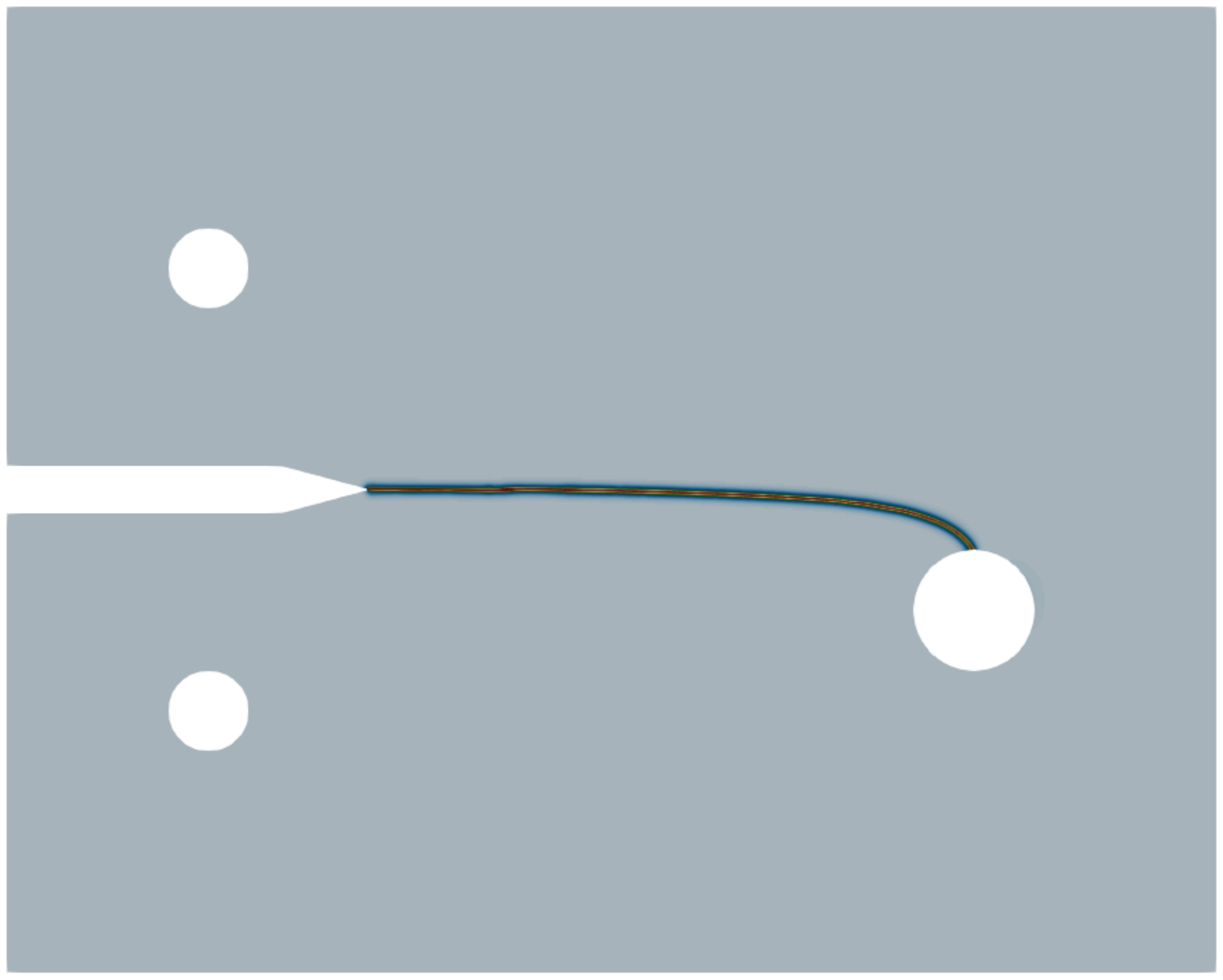} \vspace{1mm}
		\caption{\label{fig:SpecB_FinalCrackPath}}
	\end{subfigure} \\[2mm]
	\begin{subfigure}{0.3\textwidth}
		\includegraphics[width=\textwidth]{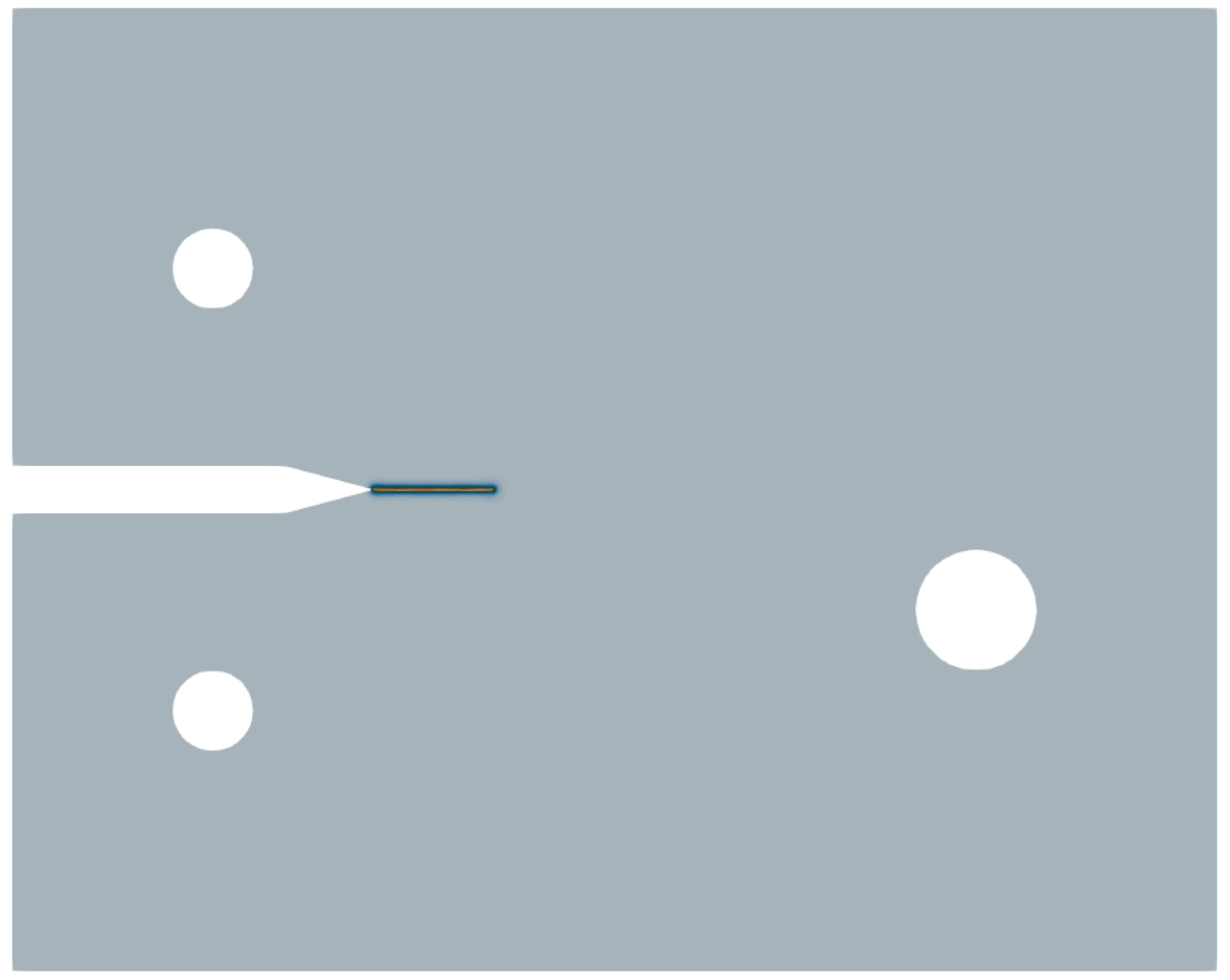}
		\caption{\label{fig:SpecB_InitJump_zeta_1_0}}
	\end{subfigure} \hspace{2mm}
	\begin{subfigure}{0.3\textwidth}
		\includegraphics[width=\textwidth]{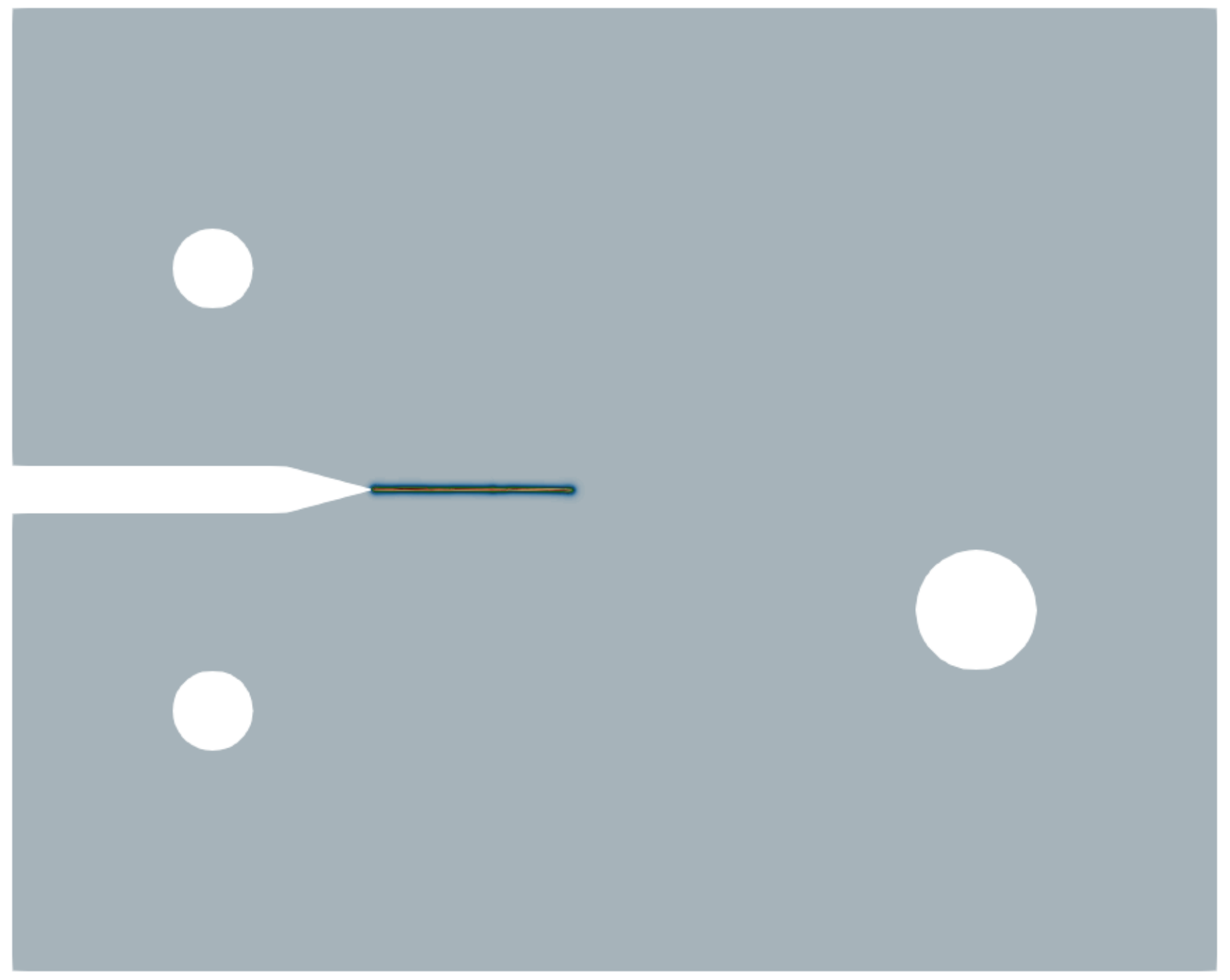}
		\caption{\label{fig:SpecB_InitJump_zeta_0_7}}
	\end{subfigure} \hspace{2mm}
	\begin{subfigure}{0.3\textwidth}
		\includegraphics[width=\textwidth]{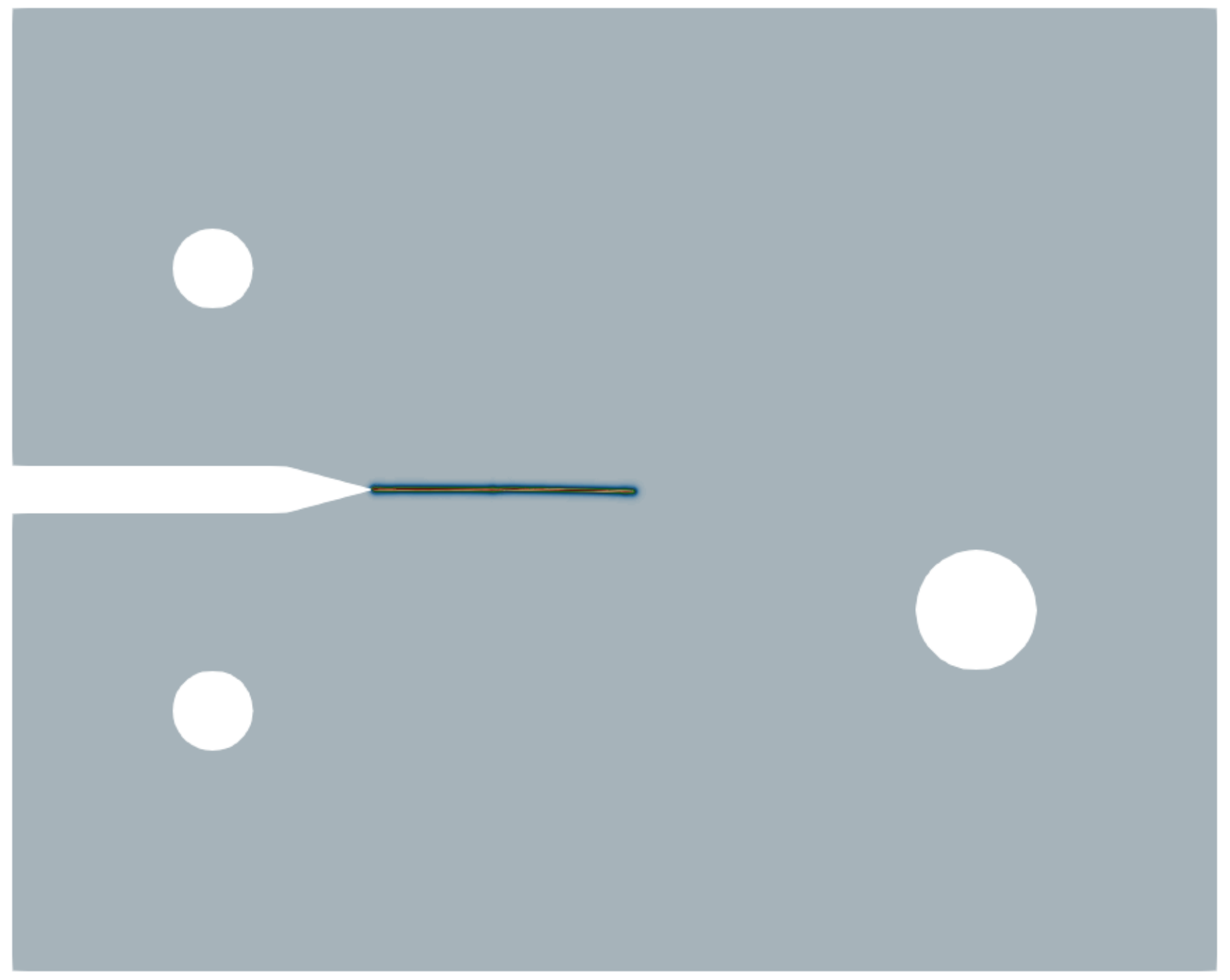}
		\caption{\label{fig:SpecB_InitJump_zeta_0_0}}
	\end{subfigure}
	\caption{Results for the modified CT test on single-hole specimen B ($\ell/h^e = 0.5$), showing (a) load-displacement curves for different loss coefficients together with the adjusted experimental data, and (b) the final crack path at $u = 1.8$ mm. The initial crack jump trajectory following the peak load is equal to (c) 15.0 mm when $\zeta = 1$, (d) 25.3 mm when $\zeta = 0.7$, and (e) 33.5 mm when $\zeta = 0$.}
\end{figure}

From the published photos in \cite{Cavuoto2022}, we can estimate the initial crack jump occurring after the peak load to be 25 mm. By setting $\zeta = 0.7$, we are able to obtain a simulated initial crack of length 25.3 mm. The corresponding load-displacement curve is shown in Figure \ref{fig:CT_Spec_1Hole_B_Summary}; for comparison, we have also plotted the curves obtained with $\zeta = 0$ and $\zeta = 1$.
It can be observed that utilizing the value $G_c^\text{eff} = 0.54$ MPa-mm obtained from the previous example results in a satisfactory reproduction of the softening curve for the current problem. 

\begin{figure}
	\centering
	\begin{subfigure}{0.31\textwidth}
		\includegraphics[width=\textwidth]{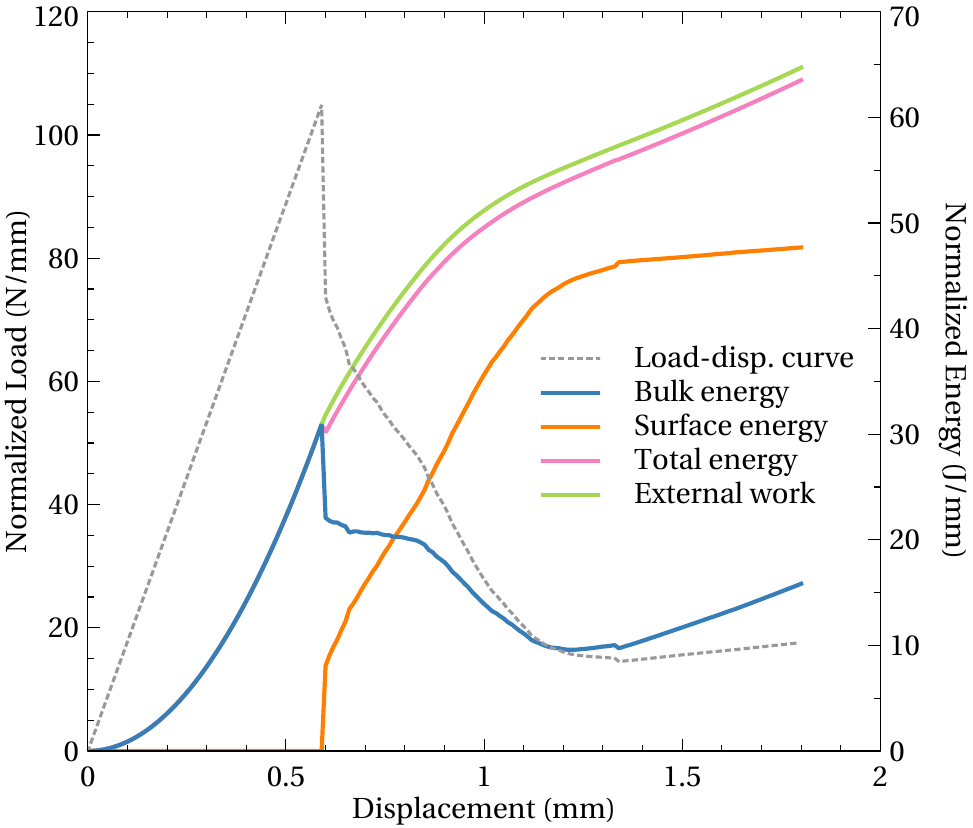}
		\caption{\label{fig:SpecB_EgyEvol_zeta_1_0}}
	\end{subfigure} \hspace{2mm}
	\begin{subfigure}{0.31\textwidth}
		\includegraphics[width=\textwidth]{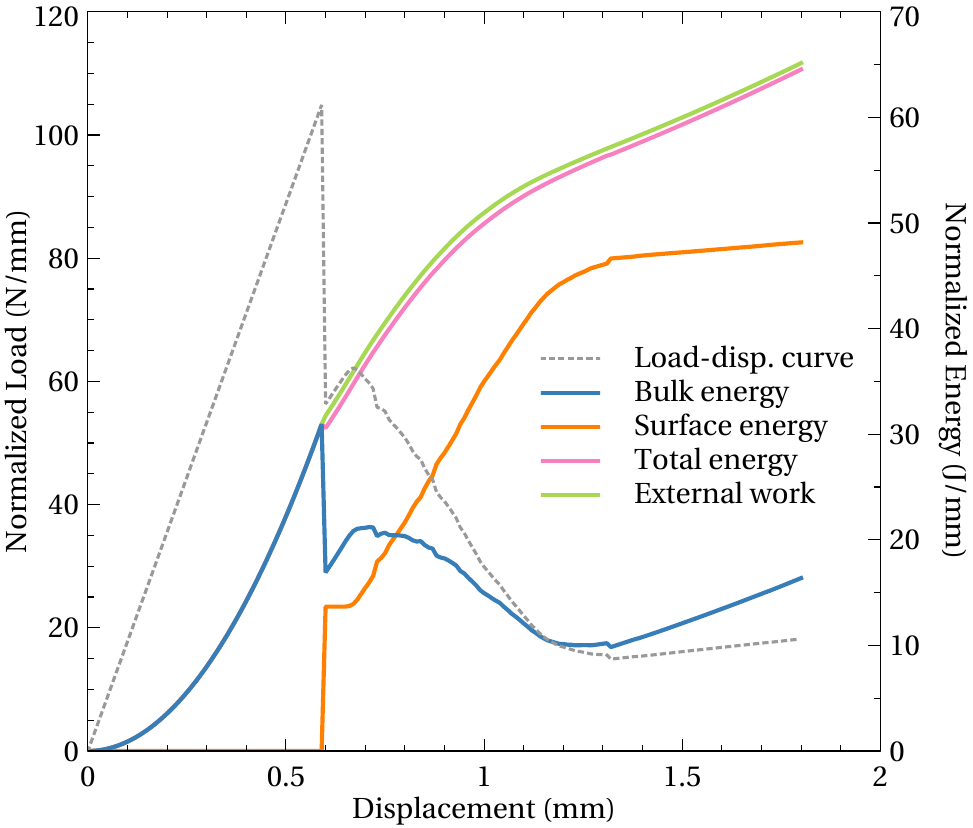}
		\caption{\label{fig:SpecB_EgyEvol_zeta_0_7}}
	\end{subfigure} \hspace{2mm}
	\begin{subfigure}{0.31\textwidth}
		\includegraphics[width=\textwidth]{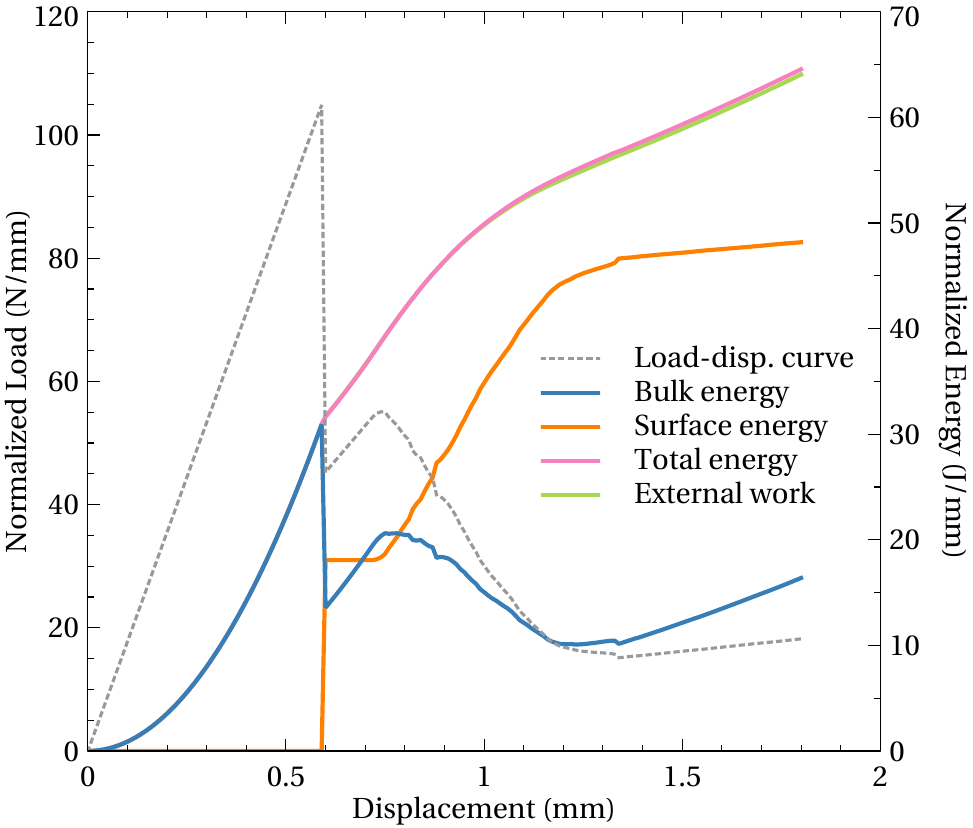}
		\caption{\label{fig:SpecB_EgyEvol_zeta_0_0}}
	\end{subfigure}
	\caption{Evolution of energy quantities with respect to displacement in the simulated compacted tension test on single-hole specimen A, corresponding to a loss coefficient of (a) $\zeta = 1$, (b) $\zeta = 0.83$, and (c) $\zeta = 0$. \label{fig:SpecB_EgyEvol}}
\end{figure}

As with the previous example, we plot the evolution of the different energy components for each assumed value of the loss coefficient. We can observe from Figure \ref{fig:SpecB_EgyEvol} that unlike in the previous example, here the crack breakthrough to the boundary is not an unstable event, as no additional energy imbalance takes place when the crack finally intersects the hole, even when simulations are carried out using the fully-dissipative model (i.e. $\zeta = 1$). 
This finding is in line with the adjusted experimental results, which show no sign of a finite crack jump at the said instance. Nevertheless, we notice a slight drop in the simulated load-displacement curve at crack breakthrough, as can be seen in Figure \ref{fig:CT_Spec_1Hole_B_Summary}, even though the total energy exhibits smooth behavior. We have found that this drop is in fact a numerical artifact brought about by the diffuse representation of the crack, i.e. for larger values of $\ell$, the drop happens at a lower magnitude of the imposed displacement and is also larger. This highlights the need for $\ell$ to be sufficiently small so as to avoid the emergence of such unrealistic behavior in the simulated specimen response.

\subsection{Compact tension test on double-hole specimen \label{sec:numEx_SpecC}}
In the final example, we simulate crack evolution in the double-hole specimen shown in Figure \ref{fig:geom_spec_2hole}. In order to fix the location of crack initiation and initial direction of the resulting crack paths, the two added holes are notched. The specific dimensions of said notches are not given in the original references, however we found that these significantly influence the instance of crack nucleation. That is, smaller notch areas and wider notch opening angles have the effect of delaying the occurrence of fracture initiation compared to notches with larger areas and narrower opening angles.  Here we have decided to set the said notch opening angles to 50$^\circ$ for both holes, and the ratio of notch depth relative to the radius of the hole that is intersected by them as close to the ratios that can be gleaned from the published photos in \cite{Cavuoto2022}.

As in the previous two examples, we make use of an effective critical energy release rate of 0.54 MPa-mm. However, we initially encountered convergence problems with the AM algorithm when applying the pseudo-dynamic model to the discrete problem in which $\ell/h^e = 5$. We found that by setting $\ell = 0.4$ mm and $h^e = 0.18$ mm, we are able to carry out the necessary simulations without running into the convergence issues we had previously. Applying the necessary correction to account for the mesh refinement ratio in the effective energy release rate, we obtain $G_c = 0.54 / \left[ 1 + 0.5 \left( 0.18 / 0.4 \right) \right] = 0.44$ MPa-mm. Due to the lack of a definitive softening region in the adjusted experimental load-displacement curve, it is not possible to verify our choice of $G_c$ for this example. Nevertheless, as the three specimens analyzed in the current study are all made out of the same PMMA, it seems reasonable to assume that the adopted value for the material toughness should be appropriate for the current problem based on the goodness of fit of numerical results to the adjusted experimental data in the prior two examples.

\begin{figure}
	\centering
	\begin{subfigure}{0.4\textwidth}
		\includegraphics[width=\textwidth]{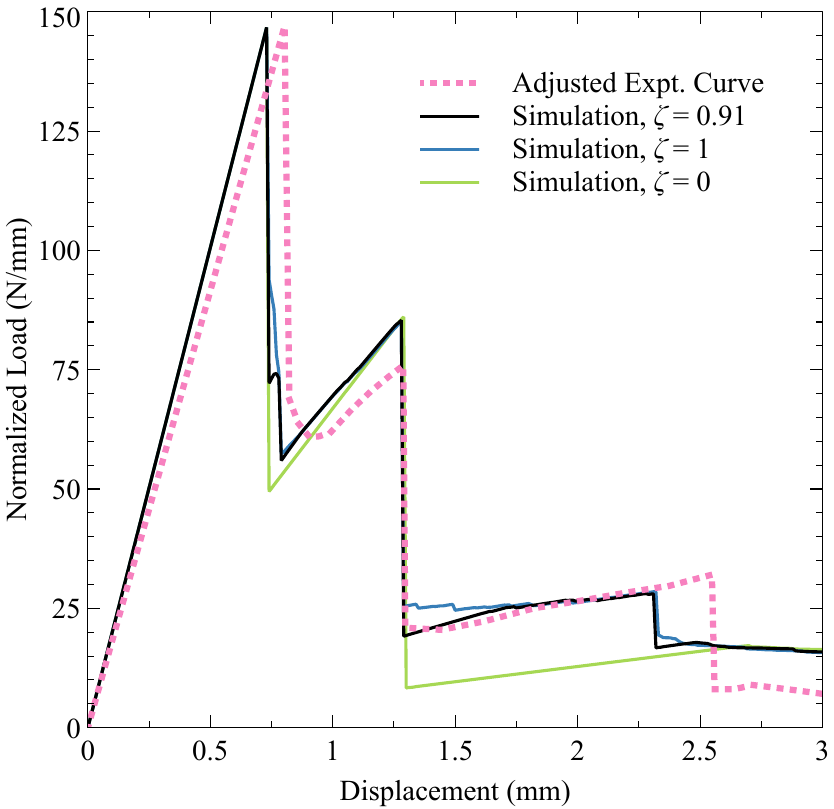}
		\caption{\label{fig:SpecC_LDCurvesSummary}}
	\end{subfigure} \hspace{2mm}
	\begin{subfigure}{0.455\textwidth}
		\includegraphics[width=\textwidth]{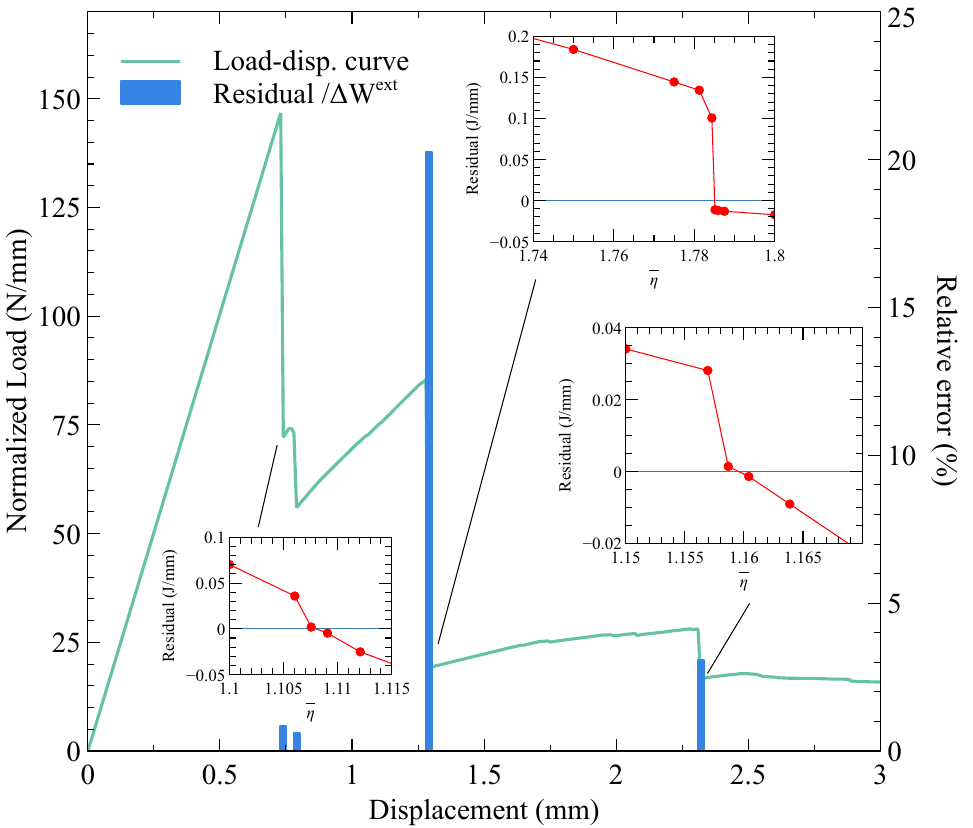}
		\caption{\label{fig:SpecC_Residuals}}
	\end{subfigure} \\
	\begin{subfigure}{0.31\textwidth}
		\includegraphics[width=\textwidth]{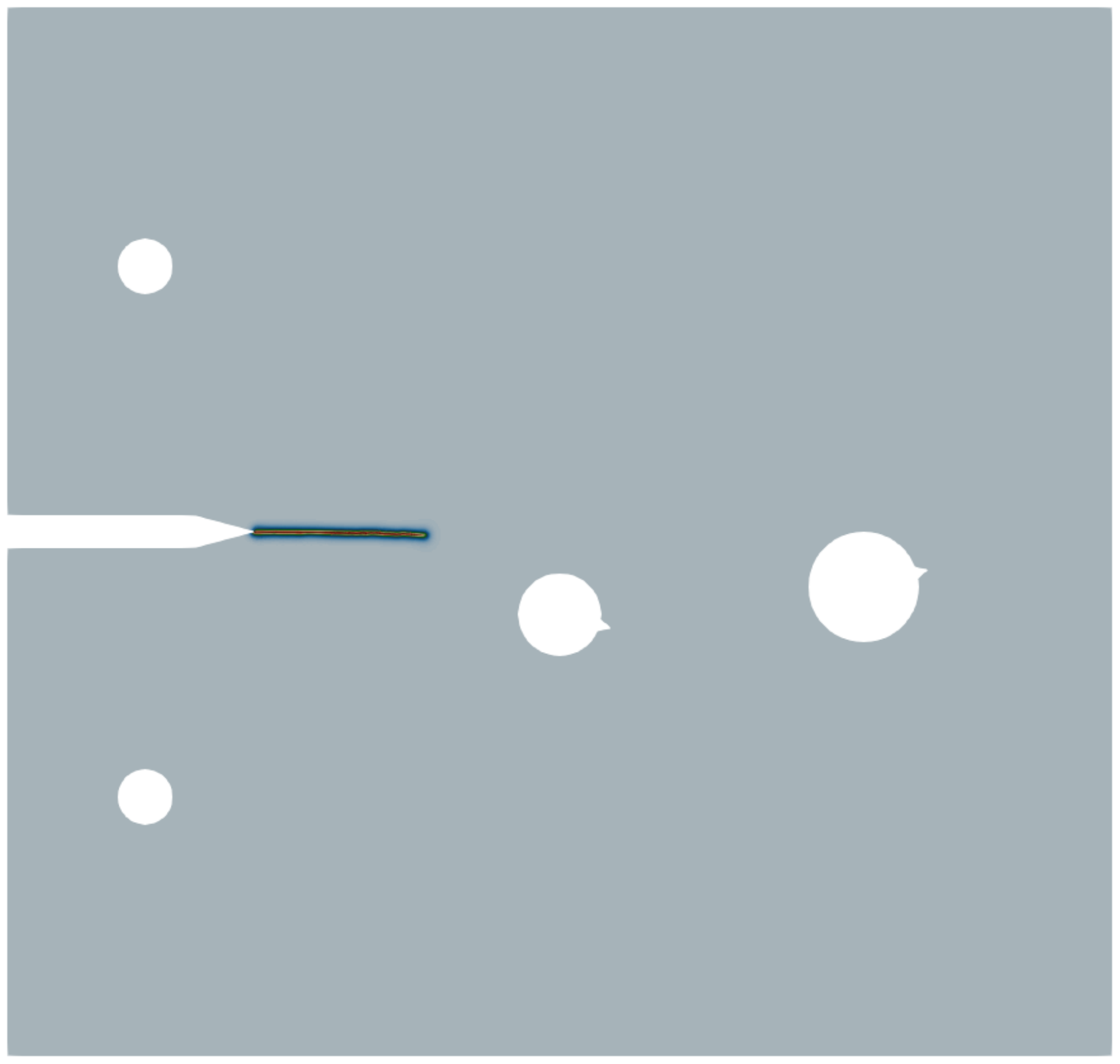}
		\caption{\label{fig:SpecC_InitJump_1_0}}
	\end{subfigure} \hspace{2mm}
	\begin{subfigure}{0.31\textwidth}
		\includegraphics[width=\textwidth]{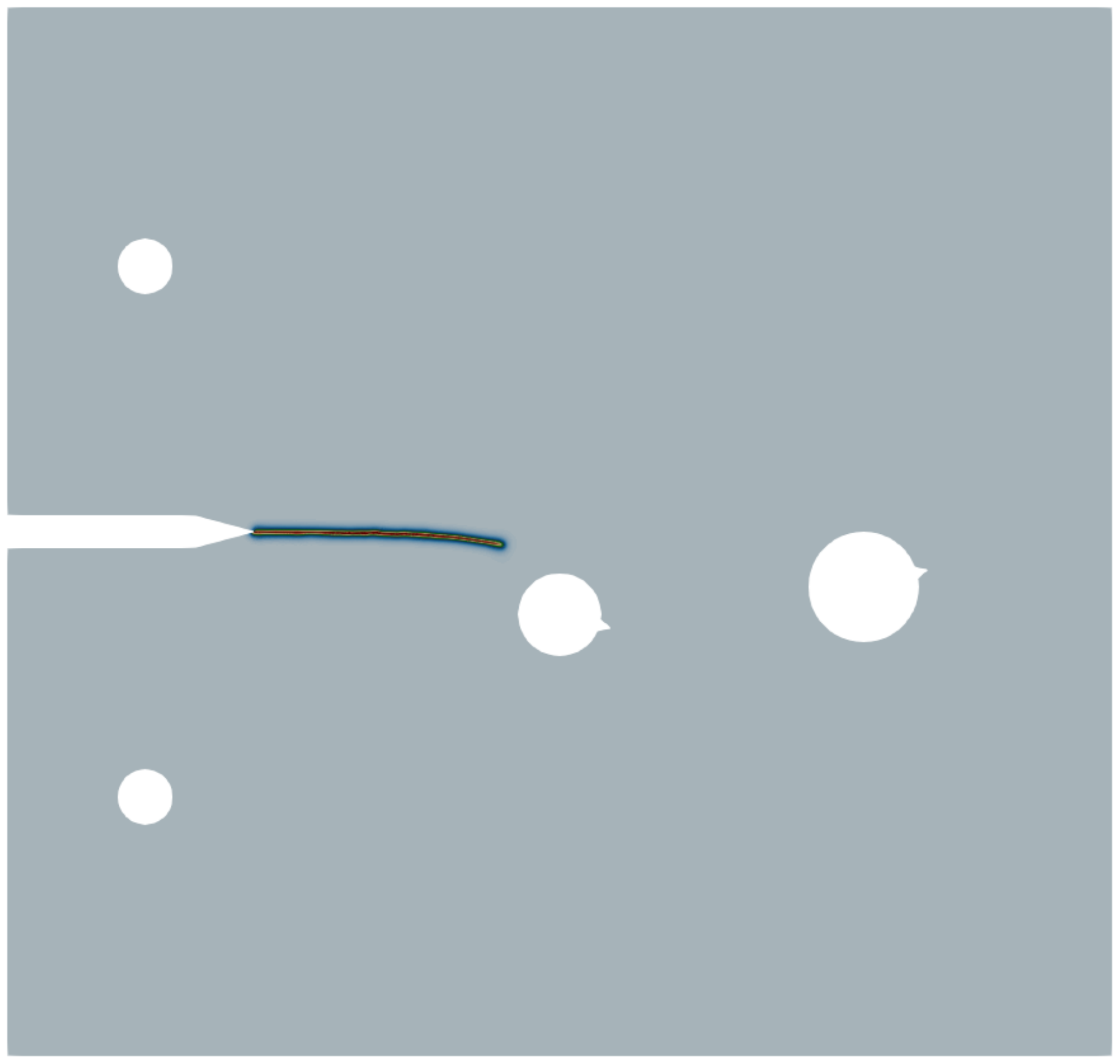}
		\caption{\label{fig:SpecC_InitJump_0_91}}
	\end{subfigure} \hspace{2mm}
	\begin{subfigure}{0.31\textwidth}
		\includegraphics[width=\textwidth]{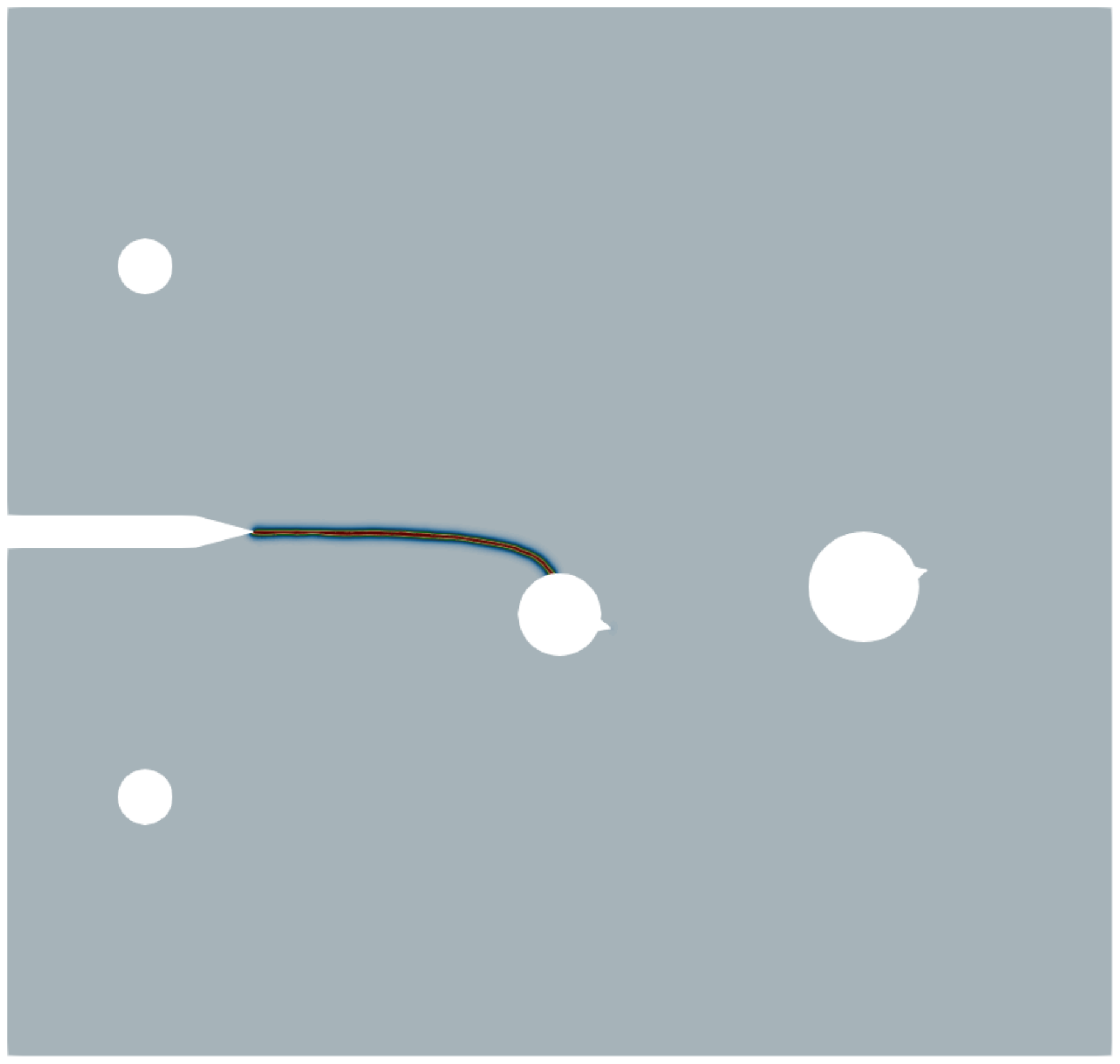}
		\caption{\label{fig:SpecC_InitJump_0_0}}
	\end{subfigure}
	\caption{Results for the modified CT test on the double-hole specimen ($\ell/h^e = 2.22$): (a) load-displacement curves for different values of $\zeta$ compared with the adjusted experimental data. Relative errors of the energy residuals at unstable fracture events are shown in (b), where the inset graphs plot the dependence of the energy residual on $\bar{\eta}$. The initial crack occurring immediately after the peak load is also shown, and is equal to (c) 30.6 mm when $\zeta = 1$, (d) 45.8 mm when $\zeta = 0.91$, (e) 64.0 mm when $\zeta = 0$.}
\end{figure}

\begin{figure}
	\centering
	\begin{subfigure}{0.31\textwidth}
		\includegraphics[width=\textwidth]{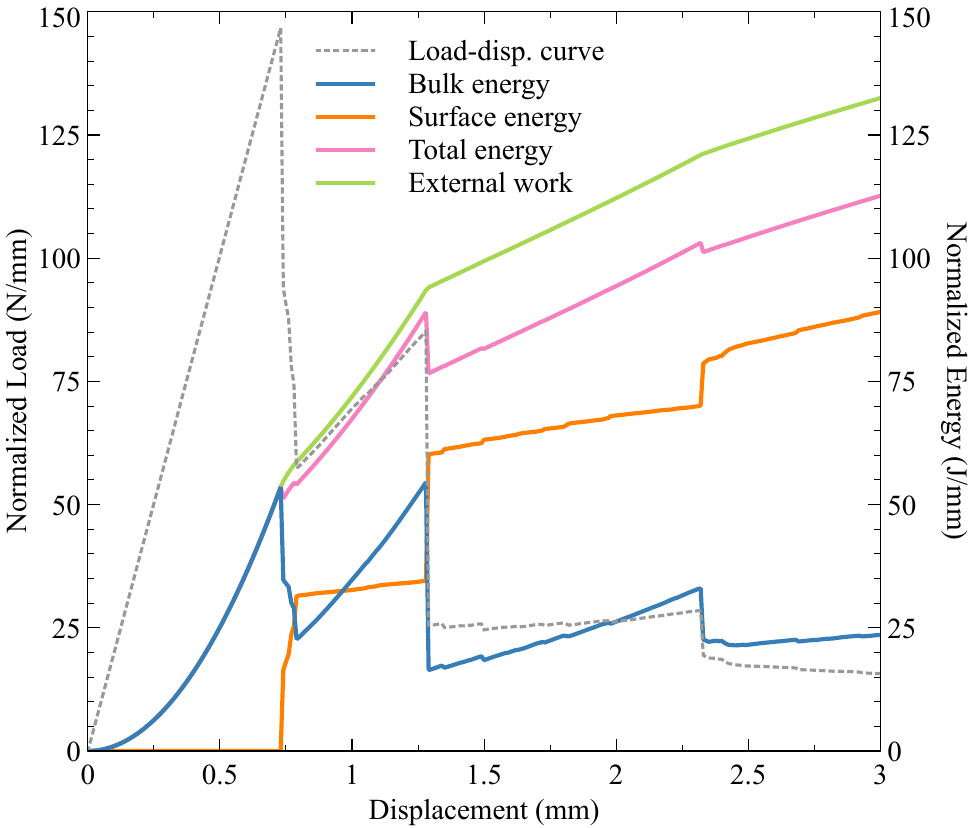}
		\caption{\label{fig:SpecC_EgyEvol_zeta_1_0}}
	\end{subfigure} \hspace{2mm}
	\begin{subfigure}{0.31\textwidth}
		\includegraphics[width=\textwidth]{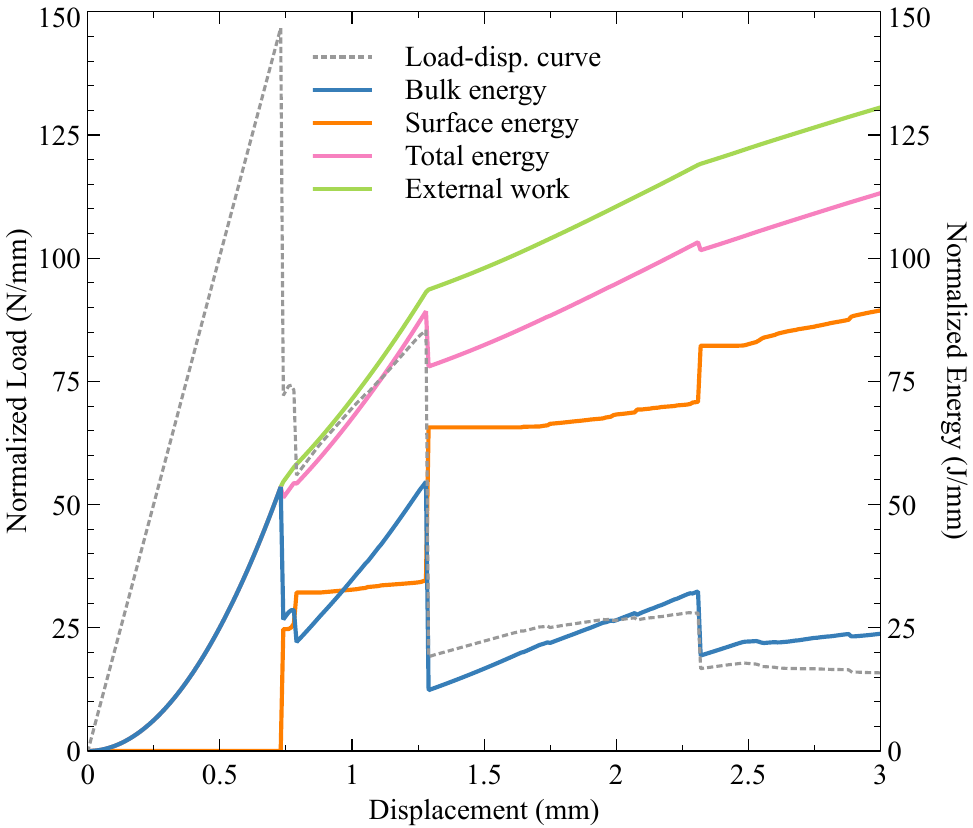}
		\caption{\label{fig:SpecC_EgyEvol_zeta_0_7}}
	\end{subfigure} \hspace{2mm}
	\begin{subfigure}{0.31\textwidth}
		\includegraphics[width=\textwidth]{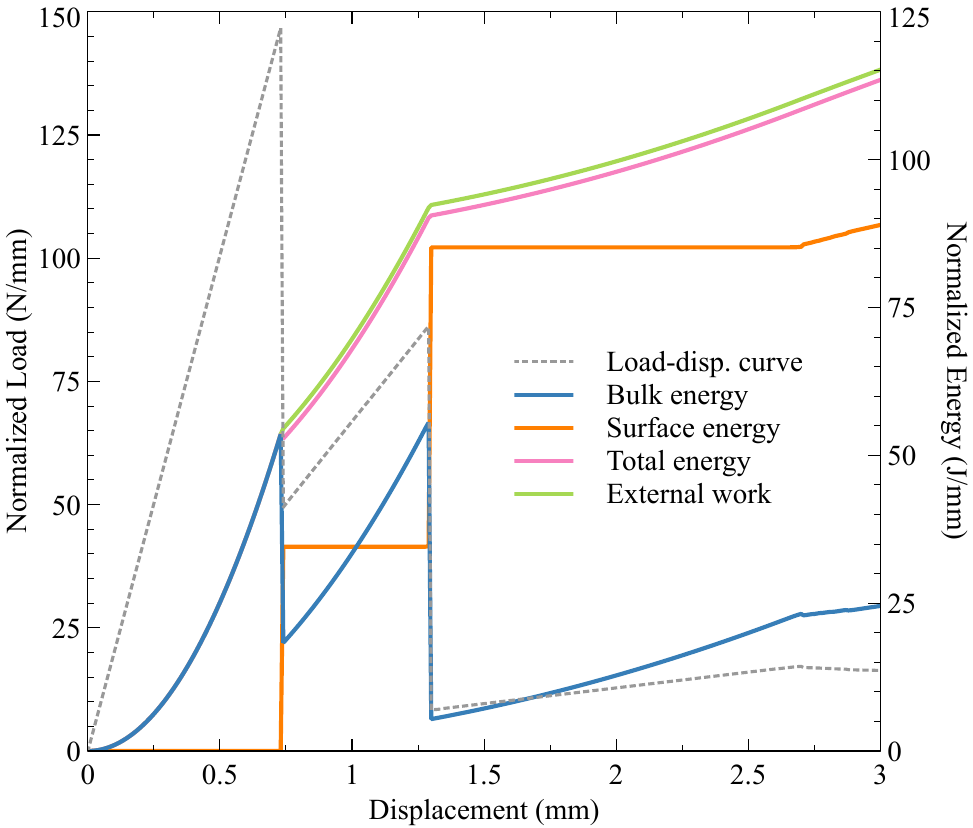}
		\caption{\label{fig:SpecC_EgyEvol_zeta_0_0}}
	\end{subfigure}
	\caption{Evolution of energy quantities with respect to displacement in the simulated compacted tension test on the double hole specimen, corresponding to a loss coefficient of (a) $\zeta = 1$, (b) $\zeta = 0.91$, and (c) $\zeta = 0$. \label{fig:SpecC_EgyEvol}}
\end{figure}

A key distinction between the current example and the previous two is that here we encounter the presence of more than one unstable fracture event. Using measurements made on the published images from \cite{Cavuoto2022}, we determine the length of the initial crack jump occurring immediately after the peak load to be 48 mm. On the other hand, the final unstable fracture that nucleates from the rightmost hole in the specimen results in a crack segment 23 mm long.  Assuming the loss coefficient remains constant throughout the simulation, we adjust the value of $\zeta$ by considering the error in both the initial and final crack jump lengths. Figure \ref{fig:SpecC_LDCurvesSummary} displays the simulated load-displacement curve for a loss coefficient $\zeta = 0.91$, alongside curves for $\zeta = 1$ and $\zeta = 0$.
As with the previous examples, we plot the evolution of energy components for different assumed values of the loss coefficient (see Figure \ref{fig:SpecC_EgyEvol}).
The numerical results show a slightly stiffer material behavior compared to the experimental data in the initial linear elastic portion of the load-displacement curve. However, this is the closest we can make by adjusting element sizes at the loading regions and still avoid element interpenetration. For the case where $\zeta = 0.91$, the resulting initial and final crack jump lengths are found to be 46 mm and 21 mm respectively, thus  2 mm short of experimental observations. Further improvements may be possible, but would likely require a reduction of the phase-field length scale $\ell$ due to the proximity of crack tip arrest to a hole boundary, i.e. with the currently utilized value of $\ell$, any further decrease in $\zeta$ leads to an initial crack that intersects the hole boundary without any intermediate arrest of the crack tip, suggesting the existence of a forbidden region of crack arrest similar to that described in Section \ref{sec:numEx_SpecA}.

While the simulations predict a premature occurrence of the final unstable fracture, the numerical results are nevertheless able to reproduce the temporary linear elastic phase in the overall material response just after the last crack jump, before the crack resumes stable propagation. These last two phases are not recognizable from the original (unadjusted) experimental load-displacement curve published in \cite{Cavuoto2022}, however the existence of an in-between linear elastic phase is supported by the reported crack tip velocities for different portions of the L-D curve reported in the same reference. They are also more or less evident in videos of the experiments that are included in supplementary material downloadable from the journal website.

Figure \ref{fig:SpecC_Residuals} plots the relative error of the energy residual with respect to the external work increment at load steps where unstable fracture occurs in the simulation with $\zeta = 0.91$. Four such events are identified, occurring at imposed displacements of 0.74 mm, 0.79 mm, 1.29 mm, and 2.32 mm. Examining the behavior the residual dependency on the overload factor at each event reveals that in cases where the residuals are small, $r^\eta \left( \bar{\eta} \right)$ appears continuous across $r^\eta = 0$. However, a clear discontinuity in $r^\eta \left( \bar{\eta} \right)$ is observed when the crack initiates and grows from the smaller to the larger hole in a single load step. This is similar to cases where a forbidden regions of crack arrest is encountered, in which the algorithm accepts solutions corresponding to the smallest positive residual even if it is not negligible in relation to the external work increment. However while the former may be a numerical artifact, the latter reflects a physical phenomenon where the available energy is simply insufficient to initiate a crack from the second hole.

\begin{figure}
	\centering
	\begin{minipage}[b]{0.45\textwidth}
		\includegraphics[width=\textwidth]{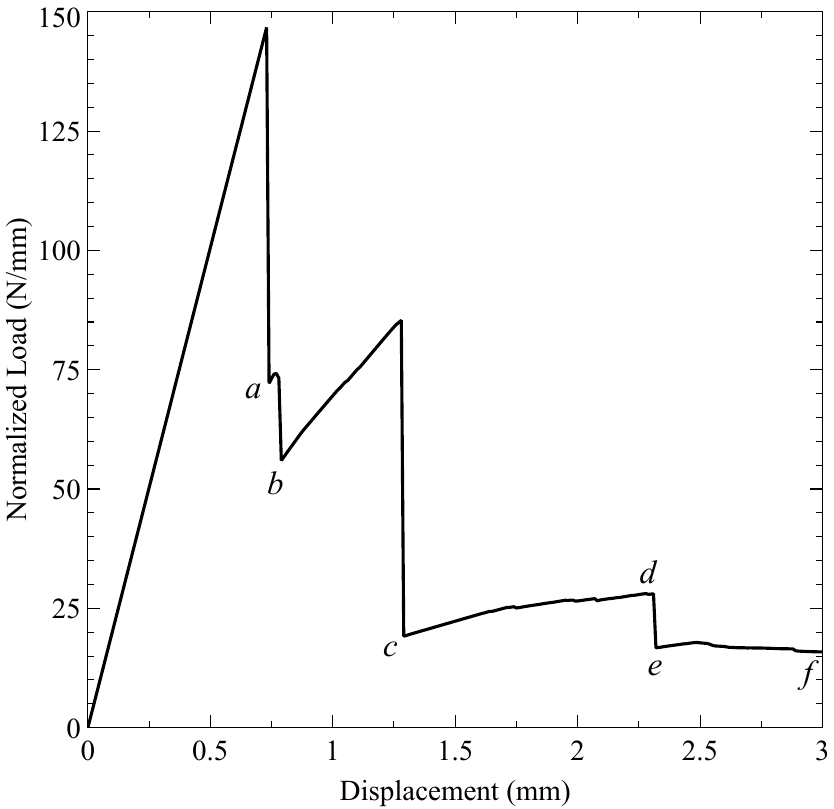} \\[10mm]
		\begin{subfigure}{\textwidth}
			\includegraphics[width=\textwidth]{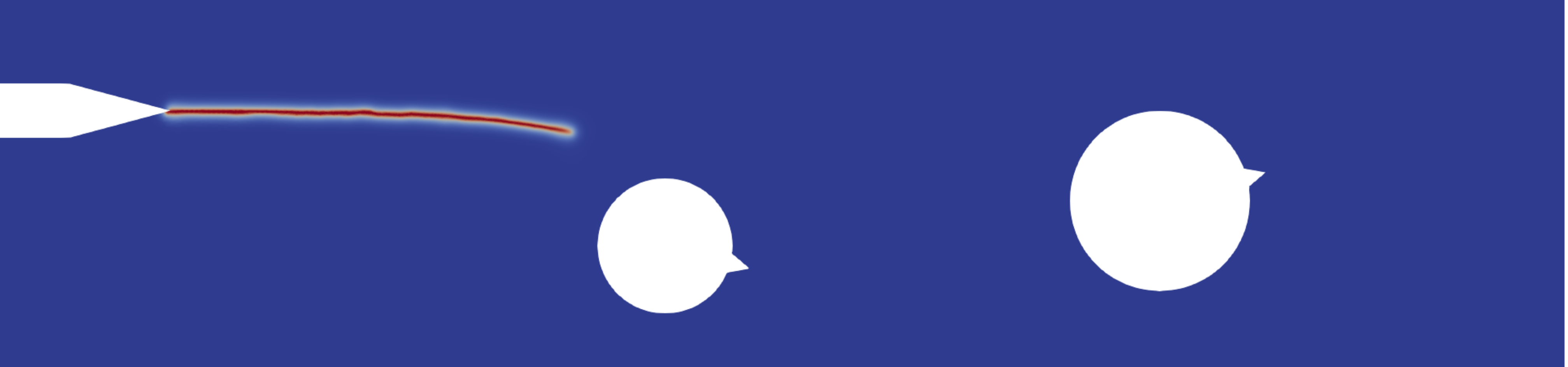}
			\caption{}
		\end{subfigure}
	\end{minipage} \hspace{5mm}
	\begin{minipage}[b]{0.45\textwidth}
		\begin{subfigure}{\textwidth}
			\includegraphics[width=\textwidth]{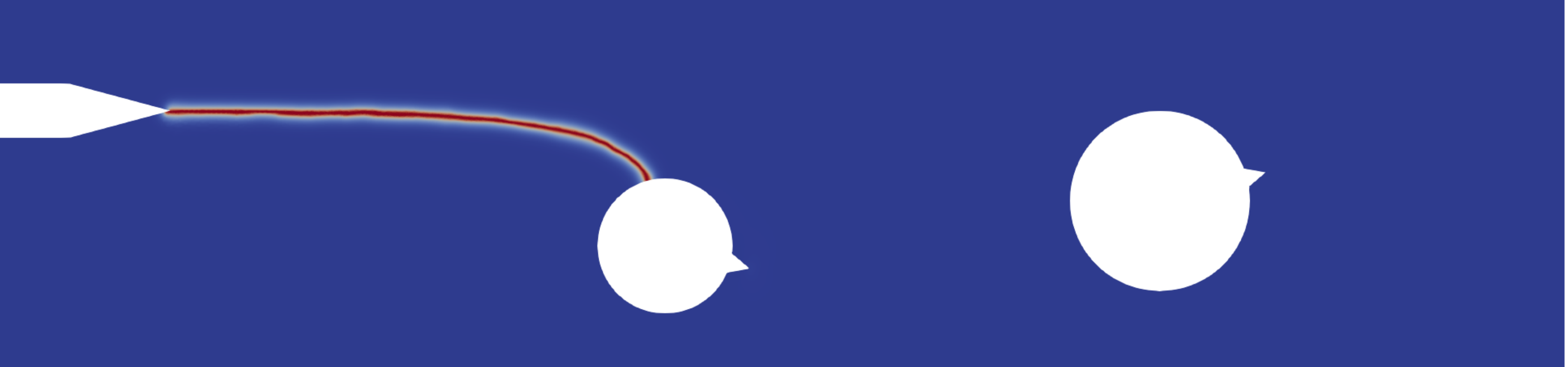}
			\caption{\label{fig:firstCrackSegment}}
		\end{subfigure}
		\begin{subfigure}{\textwidth}
			\includegraphics[width=\textwidth]{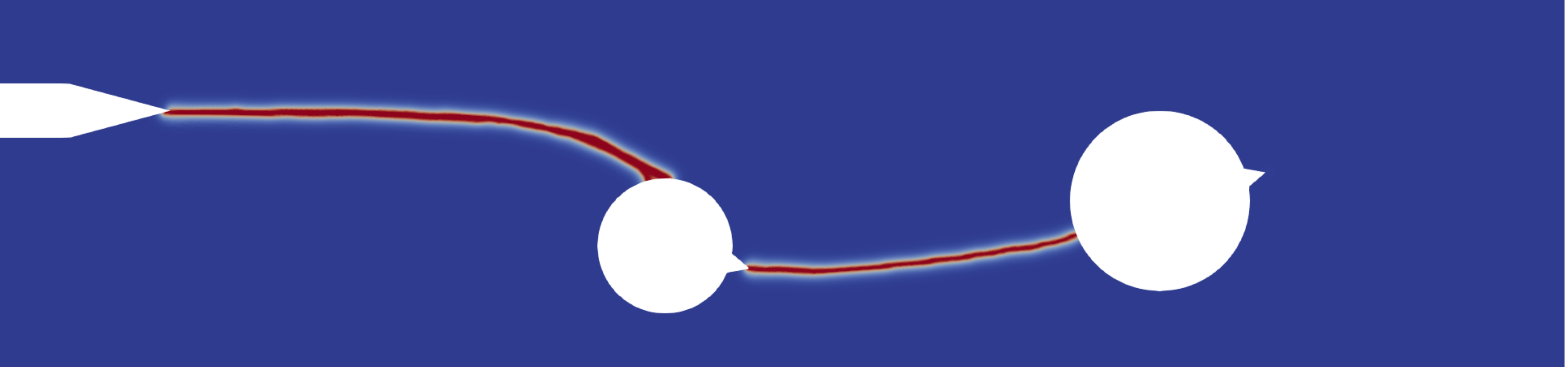}
			\caption{\label{fig:recrack}}
		\end{subfigure}
		\begin{subfigure}{\textwidth}
			\includegraphics[width=\textwidth]{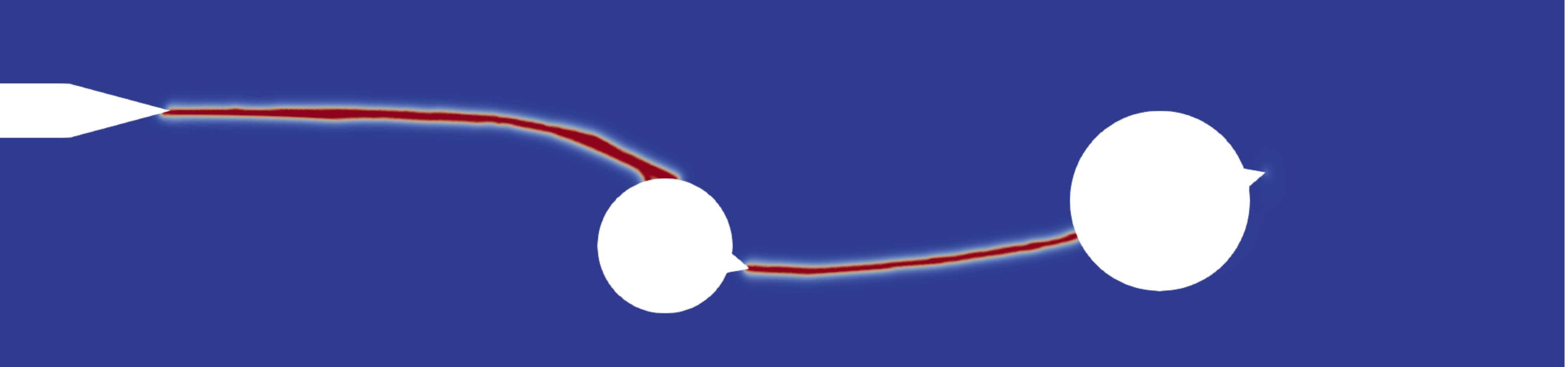}
			\caption{\label{fig:thickening}}
		\end{subfigure}
		\begin{subfigure}{\textwidth}
			\includegraphics[width=\textwidth]{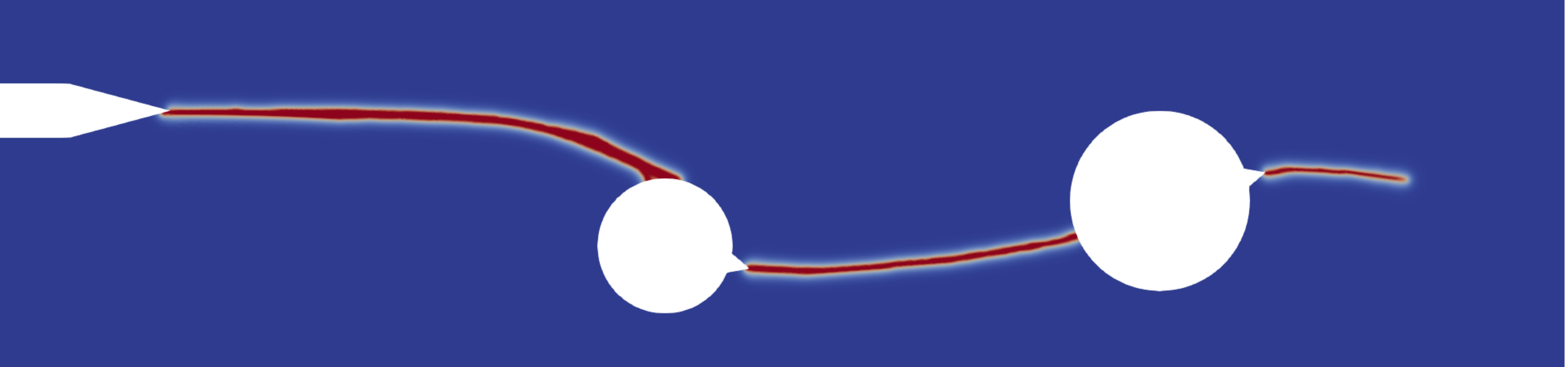}
			\caption{}
		\end{subfigure}
		\begin{subfigure}{\textwidth}
			\includegraphics[width=\textwidth]{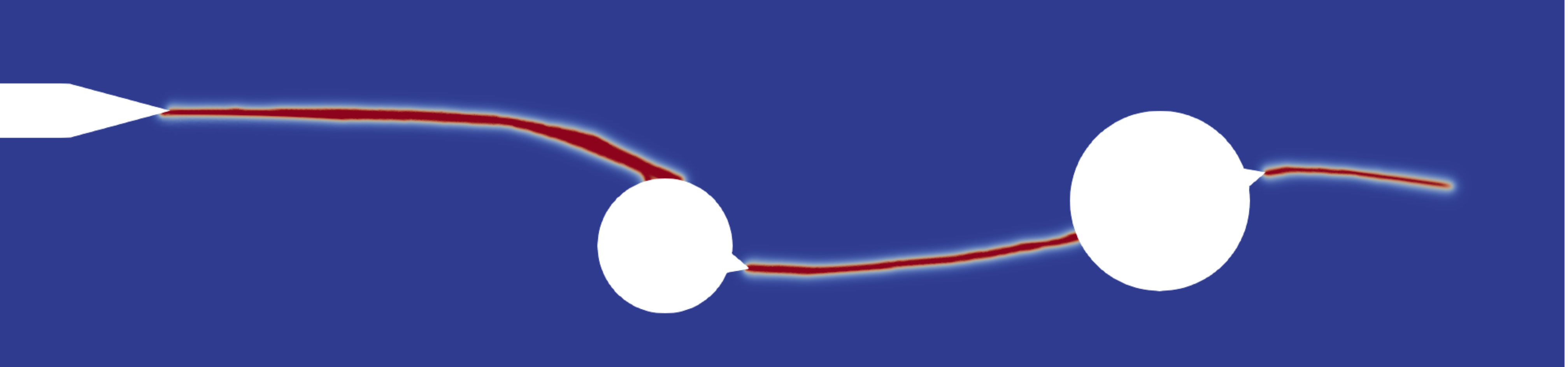}
			\caption{}
		\end{subfigure}
	\end{minipage}
	\caption{Crack evolution in the double-hole CT specimen. Close-up views of the crack phase-field are taken at imposed displacement values of (a) $u = 0.74$ mm, (b) $u = 0.79$ mm, (c) $u = 1.29$ mm, (d) $u = 2.31$ mm, (e) $u = 2.32$ mm, and (f) $u = 3.$ mm. \label{fig:SpecC_crackEvol}}
\end{figure}

Figure \ref{fig:SpecC_crackEvol} tracks the phase-field evolution at critical instances throughout the CT test simulation.
An interesting numerical artifact is observed in the analysis. At a displacement of 0.79 mm at the upper loading point, the crack propagates into the first (smaller) hole. This is followed by an elastic phase, after which the crack initiates from the smaller hole and penetrates into the larger hole within a single load step ($u = 1.29$ mm). However, a closer inspection of Figure \ref{fig:recrack} reveals a subtle issue: the pre-existing crack segment (from the notch to the smaller hole) appears to have been ``refractured'', such that a second crack can be discerned overlaying the first, most clearly near the hole boundary where the crack paths diverge slightly, so that now there are two cracks entering the hole.

We attribute such spurious behavior to our use of the spectral decomposition model of \cite{Miehe2010_ijnme}, which splits the bulk energy according to the spectral components of the strain tensor. This approach does not incorporate information from the phase-field to determine crack orientation, and instead relies solely on the relative magnitudes of the strain components. While the model is effective for describing predominantly tensile fractures where crack propagation follows the principal strain directions, it encounters problems in the present case. This is because the formation of a second crack segment between the two holes causes a significant shift in the center of rotation of the top half of the specimen, altering the orientation of the major principal strain. The misalignment between the normal to the existing crack path and the direction of the major principal strain generates residual stresses, which can then lead to further evolution of the phase-field when the associated driving force becomes sufficiently large. This phenomenon also accounts for the apparent reduction in slope of the numerical load-displacement curve between $u = 1.29$ mm and $u = 2.31$ mm. Under normal conditions, this range should exhibit a constant slope, as the bulk material is linear elastic prior to fracture, and no crack growth occurs during this phase. However, a detailed examination of Figure \ref{fig:thickening} reveals a slight thickening of the initial crack segment (from the notch to the first hole) compared to Figure \ref{fig:recrack}, which has the effect of reducing the spurious residual stresses. Needless to say, such phenomena compromise the energy balance calculations by introducing spurious growth of the surface energy. Moreover, we suspect that the same issues contributed to the convergence problems we initially encountered when carrying out simulations on the specimen discretized with mesh refinement ratio $\ell / h^e = 5$, since convergence difficulties only arose when spurious refracturing or thickening of cracks occurred.

%% file: concluding_remarks.tex
\section{Conclusion and outlook} \label{sec:concluding_remarks}
A pseudo-dynamic phase-field model for brittle fracture is presented which enables the accurate simulation of finite crack growth associated with unstable/brutal fracture. To achieve this, the driving force in the phase-field equation is modified by incorporating an additional unknown in the form of an overload factor, whose role is to heuristically account for the effects of dynamic forces on crack evolution. The discrete system is closed by adding an equation that enforces global energy balance according to a chosen loss coefficient, which governs the amount of energy dissipated during time steps containing unstable fracture events. This allows the model to simulate crack growth across a spectrum, from complete energy conservation at one extreme to maximal dissipation at the other.  Numerical examples show that the pseudo-dynamic model is able to model the length of unstable crack growth with reasonable accuracy and also reproduce key features from the experimental load-displacement behavior. Several important points became clear to us in the course of conducting the study. One is that the material fracture toughness has a very significant effect on the load-deformation response, and as such its value cannot simply be assumed. Secondly, it is important to capture accurately both the peak load and the onset of crack evolution. In the classical quasi-static phase-field model, this is perhaps not as important since the solution converges to a stationary point of the regularized potential energy. However in the pseudo-dynamic model, further crack evolution depends on the magnitude of net energy imbalance after the stationary solutions have been attained. Thus, errors in prediction of the peak loads and their corresponding displacements will spill over into the global energy balance calculations. The examples also underscore the need to account for the effect of mesh refinement with respect to the phase-field length scale on the simulated toughness of the material, in order to get meaningful and consistent results. Finally, the last example exposes the inadequacy of the popular spectral decomposition scheme of \cite{Miehe2010_ijnme} for modeling stress release across open crack surfaces, especially in cases where the damaged elements undergo motion that leads to complex strain paths. Although alternative models have been proposed, research continues in the development of constitutive models that accurately approximate unilateral contact in phase-field fracture modeling. On the other hand, the specific form of the overload factor can be further improved by relaxing the assumption of uniformity over the entire domain. A promising approach involves making the overload factor dependent on the phase-field, thus $\eta \left( \mathbf{x} \right) = 1 + C f \left( \phi \left( \mathbf{x}\right) \right)$ where $C$ is a constant and $f : \left[ 0,1 \right] \mapsto \left[ 0,1 \right]$ is a monotonically increasing function. This modification may be sufficient to address the issue regarding forbidden regions of crack arrest that have been noted in the numerical examples, and will be explored in future work.

\section*{Acknowledgments}
JMS and JM acknowledge support from the ESS lighthouse SOLID, funded by the Danish Agency for Science and Higher Education (Grant 8144‐00002B), and the Novo Nordisk Foundation project PRECISE (NNF23OC0081251). JMS would also like to thank the authors of \citet{Cavuoto2022} for responding to questions and clarifications in connection with the experimental and numerical results published in their paper.

%% file: spectral_decomposition.tex
\section{Derivation of constitutive tensor for spectral decomposition model} \label{sec:modulusTensor_derivation}
\subsection{Full 3-D case}
In the spectral decomposition model of \citet{Miehe2010_ijnme}, the strain tensor $\straintensor$ is defined in terms of its principal components $\strain_a$, $a \in \left\{ 1,2,3 \right\}$. In order to derive the correct form of the modulus tensor with respect to the Cartesian components of the strain, we require the derivatives of eigenvalues and eigenvectors with respect to the tensor components. These are given in \cite{Miehe2001} for the case of distinct eigenvalues as
\begin{linenomath}
\begin{align}
	\frac{\partial \strain_a}{\partial\straintensor} &= \mathbf{n}_a \otimes \mathbf{n}_a \qquad \frac{\partial\strain_a}{\partial\strain_{ij}} = n^a_i n^a_j \\
	\frac{\partial \mathbf{n}_a}{\partial\straintensor} &= \sum_{\substack{b=1 \\ b\neq a}}^3 \frac{1}{2 \left( \strain_a - \strain_b \right)} \mathbf{n}_b \otimes \left( \mathbf{n}_a \otimes \mathbf{n}_b + \mathbf{n}_b \otimes \mathbf{n}_a \right) \qquad \frac{\partial n^a_i}{\partial\strain_{kl}} = D_{ikl} = \sum_{\substack{b=1 \\ b\neq a}}^3 \frac{n^b_i n^a_k n^b_l + n^b_i n^b_k n^a_l}{2 \left( \strain_a - \strain_b \right)},
\end{align}
\end{linenomath}
where $\mathbf{n}_a$ is the unit eigenvector associated with the principal component $\strain_a$. For convenience we adopt the notation $\mathbf{M}_a = \mathbf{n}_a \otimes \mathbf{n}_a$, with components $M^a_{ij} = n^a_i n^a_j$. Its derivative is then given in component form by
\begin{linenomath}
\begin{align}
	\frac{\partial M^a_{ij}}{\partial \strain_{kl}} &= \frac{\partial}{\partial\strain_{kl}} \left( n^a_i n^a_j \right) = \frac{\partial n^a_i}{\partial\strain_{kl}} n^a_j + n^a_i \frac{\partial n^a_j}{\partial\strain_{kl}} \\
	&= \sum_{\substack{b=1 \\ b\neq a}}^3 \frac{n^b_i n^a_j n^a_k n^b_l + n^b_i n^a_j n^b_k n^a_l}{2 \left( \strain_a - \strain_b \right)} + \sum_{\substack{c=1 \\ c\neq a}}^3 \frac{n^a_i n^c_j n^a_k n^c_l + n^a_i n^c_j n^c_k n^a_l}{2 \left( \strain_a - \strain_c \right)}.
\end{align}
\end{linenomath}
In symbolic form,
\begin{linenomath}
\begin{equation}
	\frac{\partial \mathbf{M}_a}{\partial \straintensor} = \sum_{\substack{b=1 \\ b\neq a}}^3 \frac{1}{2 \left( \strain_a - \strain_b \right)} \left( \mathbb{G}_{ab} + \mathbb{G}_{ba} \right)
\end{equation}
\end{linenomath}
where $\mathbb{G}_{ab} = \mathbf{n}_a \otimes \mathbf{n}_b \otimes \mathbf{n}_a \otimes \mathbf{n}_b + \mathbf{n}_a \otimes \mathbf{n}_b \otimes \mathbf{n}_b \otimes \mathbf{n}_a$.
Going back to the spectral decomposition proposed by \cite{Miehe2010_ijnme}, there the elastic strain energy is split into positive and negative parts as follows:
\begin{linenomath}
\begin{equation}
	\psi_0^\pm \left( \straintensor \right) = \frac{\lambda}{2} \left< \strain_1 + \strain_2 + \strain_3 \right>^2_\pm + \mu \left[ \left< \strain_1 \right>_\pm^2 + \left< \strain_2 \right>_\pm^2 + \left< \strain_3 \right>_\pm^2 \right],
	\label{eq:MieheSplit}
\end{equation}
\end{linenomath}
where $\left< x \right>_\pm = \left( x \pm \left| x \right| \right)/2$. It follows that the decomposed stress is given by
\begin{linenomath}
\begin{equation}
	\stresstensor_0^\pm \left( \straintensor \right) = \frac{\partial\psi_0^\pm}{\partial\straintensor} = \lambda \left< \strain_1 + \strain_2 + \strain_3 \right>_\pm \sum_{a=1}^3 \mathbf{M}_a + 2\mu \sum_{a=1}^3 \left< \strain_a \right>_\pm \mathbf{M}_a = \sum_{a=1}^3 \left( \lambda \left< \strain_\text{vol} \right>_\pm + 2\mu \left< \strain_a \right>_\pm \right) \mathbf{M}_a = \sum_{a=1}^3 \left< \sigma_a \right>_\pm \mathbf{M}_a
\end{equation}
\end{linenomath}
where $\strain_\text{vol} = \strain_1 + \strain_2 + \strain_3$. The decomposed modulus tensor is then
\begin{linenomath}
\begin{equation}
	\mathbb{C}_\pm \left( \straintensor \right) = \sum_{a=1}^3 \left\{ \left[ \lambda \left< \strain_\text{vol} \right>_\pm^0 \sum_{b=1}^3 \mathbf{M}_b + 2\mu \left< \strain_a \right>_\pm^0 \mathbf{M}_a \right] \mathbf{M}_a + \left( \lambda \left< \strain_\text{vol} \right>_\pm + 2\mu \left< \strain_a \right>_\pm \right) \sum_{\substack{b=1 \\ b\neq a}}^3 \frac{1}{2 \left( \strain_a - \strain_b \right)}  \left( \mathbb{G}_{ij} + \mathbb{G}_{ji} \right) \right\}
\end{equation}
\end{linenomath}
in which $\left< x \right>^0_\pm = \left[ 1 \pm \sgn \left( x \right) \right]/2$. The above expression can be simplified to
\begin{linenomath}
\begin{equation}
 	\mathbb{C}_\pm \left( \straintensor \right) = \sum_{a=1}^3 \sum_{b=1}^3 \left( \lambda\sgn \left( \strain_\text{vol} \right) + 2\mu \delta_{ab} \sgn \left( \strain_b \right) \right) \mathbf{M}_a \mathbf{M}_b + \sum_{a=1}^3 \sum_{\substack{b=1 \\ b\neq a}}^3 \frac{\left< \sigma_a \right>_\pm}{2 \left( \strain_a - \strain_b \right)} \left( \mathbb{G}_{ab} + \mathbb{G}_{ba} \right).
\end{equation}
\end{linenomath}
It is possible to further simplify the second term in the above expression by taking advantage of the fact that both $a$ and $b$ are dummy indices. Letting $\overline{\mathbb{C}}_\pm \left( \straintensor \right) = \sum_{a=1}^3 \sum_{\substack{b=1 \\ b\neq a}}^3 \frac{\left< \sigma_a \right>_\pm}{2 \left( \strain_a - \strain_b \right)} \left( \mathbb{G}_{ab} + \mathbb{G}_{ba} \right)$, its components are given by
\begin{linenomath}
\begin{align}
	\overline{C}_{ijkl} &= \sum_{a=1}^3 \sum_{\substack{b=1 \\ b\neq a}}^3 \frac{\left< \sigma_a \right>_\pm}{2 \left( \strain_a - \strain_b \right)} \left( n^a_i n^b_j n^a_k n^b_l + n^a_i n^b_j n^b_k n^a_l \right) + \sum_{a=1}^3 \sum_{\substack{b=1 \\ b\neq a}}^3 \frac{\left< \sigma_a \right>_\pm}{2 \left( \strain_a - \strain_b \right)} \left( n^b_i n^a_j n^b_k n^a_l + n^b_i n^a_j n^a_k n^b_l \right) \nonumber \\
	&= \sum_{a=1}^3 \sum_{\substack{b=1 \\ b\neq a}}^3 \frac{\left< \sigma_a \right>_\pm}{2 \left( \strain_a - \strain_b \right)} \left( n^a_i n^b_j n^a_k n^b_l + n^a_i n^b_j n^b_k n^a_l \right) + \sum_{b=1}^3 \sum_{\substack{a=1 \\ a\neq b}}^3 \frac{\left< \sigma_b \right>_\pm}{2 \left( \strain_b - \strain_a \right)} \left( n^a_i n^b_j n^a_k n^b_l + n^a_i n^b_j n^b_k n^a_l \right) \nonumber \\
	&= \sum_{a=1}^3 \sum_{\substack{b=1 \\ b\neq a}}^3 \frac{\left< \sigma_a \right>_\pm}{2 \left( \strain_a - \strain_b \right)} \left( n^a_i n^b_j n^a_k n^b_l + n^a_i n^b_j n^b_k n^a_l \right) - \sum_{b=1}^3 \sum_{\substack{a=1 \\ a\neq b}}^3 \frac{\left< \sigma_b \right>_\pm}{2 \left( \strain_a - \strain_b \right)} \left( n^a_i n^b_j n^a_k n^b_l + n^a_i n^b_j n^b_k n^a_l \right) \nonumber \\
	&= \sum_{a=1}^3 \sum_{\substack{b=1 \\ b\neq a}}^3 \frac{\left< \sigma_a \right>_\pm - \left< \sigma_b \right>_\pm}{2 \left( \strain_a - \strain_b \right)} \left( n^a_i n^b_j n^a_k n^b_l + n^a_i n^b_j n^b_k n^a_l \right)
\end{align}
\end{linenomath}
Finally, $\left< \sigma_a \right>_\pm = \lambda \left< \strain_\text{vol} \right>_\pm + 2\mu \left< \strain_a \right>_\pm$ so that $\left< \sigma_a \right>_\pm - \left< \sigma_b \right>_\pm = 2\mu \left( \left< \strain_a \right>_\pm - \left< \strain_b \right>_\pm \right)$. Then
\begin{linenomath}
\begin{equation}
	\mathbb{C}_\pm \left( \straintensor \right) = \sum_{a=1}^3 \sum_{b=1}^3 \left[ \lambda\sgn \left( \strain_\text{vol} \right) + 2\mu \delta_{ab} \sgn \left( \strain_b \right) \right] \mathbf{M}_a \mathbf{M}_b + \sum_{a=1}^3 \sum_{\substack{b=1 \\ b\neq a}}^3 \mu \frac{\left< \strain_a \right>_\pm - \left< \strain_b \right>_\pm}{\strain_a - \strain_b} \mathbb{G}_{ab}.
\end{equation}
\end{linenomath}

\subsection{Specialization for plane stress}
For problems involving plane stress states, the spectral decomposition model requires careful modification in order to satisfy the constraint that $\sigma_3$ (the out-of-plane principal stress) be equal to zero as explained by \citet{Li2021}. Going back to \eqref{eq:MieheSplit} and noting that $\psi \left( \straintensor, \phi \right) = g \left( \phi \right) \psi_0^+ \left( \straintensor \right) +  \psi_0^+ \left( \straintensor \right)$, we obtain the following expression for the principal stresses in terms of the principal strains $\left\{ \strain_1, \strain_2, \strain_3 \right\}$:
\begin{linenomath}
\begin{equation}
	\sigma_i \left( \straintensor, \phi \right) = g \left( \phi \right) \left[ \lambda \left< \strain_1 + \strain_2 + \strain_3 \right>_+ + 2\mu \left< \strain_i \right>_+ \right] + \lambda \left< \strain_1 + \strain_2 + \strain_3 \right>_- + 2\mu \left< \strain_i \right>_-.
	\label{eq:principalStresses}
\end{equation}
\end{linenomath}
In the case of plane stress, $\sigma_3 = 0$. Thus,
\begin{linenomath}
\begin{equation}
	g \left( \phi \right) \left[ \lambda \left< \strain_1 + \strain_2 + \strain_3 \right>_+ + 2\mu \left< \strain_3 \right>_+ \right] + \lambda \left< \strain_1 + \strain_2 + \strain_3 \right>_- + 2\mu \left< \strain_3 \right>_- = 0
\end{equation}
\end{linenomath}
This leads to several cases based on different possibilities regarding the signs of $\strain_1 + \strain_2 + \strain_3$ and $\strain_3$.
\begin{enumerate}[a)]
	\item $\strain_1 + \strain_2 + \strain_3 \geq 0$ and $\strain_3 \geq 0$: \label{enumitem:caseA}
	\begin{linenomath}
	\[
	 	g \left( \phi \right) \left[ \lambda \left( \strain_1 + \strain_2 + \strain_3 \right) + 2\mu \strain_3 \right] = 0 \implies \strain_3 = -\frac{\lambda}{\lambda + 2\mu} \left( \strain_1 + \strain_2 \right)
	 \]
	 \end{linenomath}
 	For materials with non-negative Poisson ratios ($0 \leq \nu \leq 1/2$), both $\lambda$ and $\mu$ are guaranteed to be non-negative. Therefore the above result further implies that either $\strain_3$ has the opposite sign to $\strain_1 + \strain_2$ or both quantities are zero, and that $\left| \strain_3 \right| \leq \left| \strain_1 + \strain_2 \right|$. Thus if $\strain_1 + \strain_2 > 0$, then $\strain_1 + \strain_2 + \strain_3 \geq 0$ but $\strain_3 < 0$ which contradicts the assumptions regarding the signs of these quantities. On the other hand, the case $\lambda < 0$ arises when a material has negative Poisson ratio, which also leads to $\lambda$ and $\mu$ having opposite signs. Thus, $\lambda/ \left( \lambda + 2\mu \right)$ is positive whenever $\left| \lambda \right| > \left| 2\mu \right|$ and negative when $\left| \lambda \right| < \left| 2\mu \right|$. It can be shown that the former is true whenever $\nu > 1/3$. Thus for $\nu < 0$, $\lambda / \left( \lambda + 2\mu \right) < 0$ which implies that  $\strain_3$ has the same sign as $\strain_1 + \strain_2$, which now aligns with the initial assumptions. 	
 	\item $\strain_1 + \strain_2 + \strain_3 \geq 0$ and $\strain_3 < 0$:
 	\begin{linenomath}
	\[
		g \left( \phi \right) \lambda \left( \strain_1 + \strain_2 + \strain_3 \right) + 2\mu \strain_3 = 0 \implies \strain_3 = -\frac{g \left( \phi \right) \lambda}{g \left( \phi \right) \lambda + 2\mu} \left( \strain_1 + \strain_2 \right)
	\]
 	\end{linenomath}
 	By definition, $g \left( \phi \right) \geq 0$, and by continuing from the earlier analysis we arrive at the same relations in the case where $\lambda > 0$: $\left| \strain_3 \right| \leq \left| \strain_1 + \strain_2 \right|$, and either $\strain_3$ and $\strain_1 + \strain_2$ have opposite signs or are both zero. Thus if $\strain_1 + \strain_2 \geq 0$ then we recover the original assumptions made in the current case.
 	\item $\strain_1 + \strain_2 + \strain_3 < 0$ and $\strain_3 < 0$:
 	\begin{linenomath}
	\[
		\lambda \left( \strain_1 + \strain_2 + \strain_3 \right) + 2\mu \strain_3 = 0 \implies \strain_3 = -\frac{\lambda}{\lambda + 2\mu} \left( \strain_1 + \strain_2 \right)
	\]
 	\end{linenomath}
	The above is true when $\lambda < 0$, following the discussion in case \ref{enumitem:caseA}.
 	\item $\strain_1 + \strain_2 + \strain_3 < 0$ and $\strain_3 > 0$:
 	\begin{linenomath}
	\[
		\lambda \left( \strain_1 + \strain_2 + \strain_3 \right) + 2\mu g \left( \phi \right) \strain_3 = 0 \implies \strain_3 = -\frac{\lambda}{\lambda + 2\mu g \left( \phi \right)} \left( \strain_1 + \strain_2 \right)
	\]
 	\end{linenomath}
 	As $\lambda$, $\mu$ and $g \left( \phi \right)$ are all non-negative, it follows that $0 < \lambda / \left[ \lambda + 2\mu g \left( \phi \right) \right] < 1$. Thus $\strain_3$ is opposite in sign to $\strain_1 + \strain_2$, and also $\left| \strain_3 \right| < \left| \strain_1 + \strain_2 \right|$ so that if $\strain_1 + \strain_2 < 0$, then also $\strain_1 + \strain_2 +\strain_3 < 0$ which aligns with the initial assumptions for the case.
\end{enumerate}
As reported in \cite{Li2021}, results for the different cases above can combined into a single expression, namely
\begin{linenomath}
\begin{equation}
	\strain_3 = -\theta \left( \strain_1 + \strain_2 \right)
	\label{eq:principalStrain_3}
\end{equation}
\end{linenomath}
wherein
\begin{linenomath}
\begin{equation}
	\theta = \left\{ \begin{array}{ll}
		\dfrac{g \left( \phi \right) \lambda}{g \left( \phi \right) \lambda + 2\mu}, &\text{if } \lambda \geq 0 \text{ and } \strain_1 + \strain_2 \geq 0 \\[10pt]
		\dfrac{\lambda}{\lambda + 2\mu g \left( \phi \right)}, &\text{if } \lambda \geq 0 \text{ and } \strain_1 + \strain_2 < 0 \\[10pt]
		\dfrac{\lambda}{\lambda + 2\mu}, &\text{if } \lambda < 0 .
	\end{array} \right.
\end{equation}
\end{linenomath}
Substituting \eqref{eq:principalStrain_3} into \eqref{eq:principalStresses} yields
\begin{linenomath}
\begin{equation}
	\sigma_i = g \left( \phi \right) \left[ \lambda \left( 1 - \theta \right) \left< \strain_1 + \strain_2 \right>_+ + 2\mu \left< \strain_i \right>_+ \right] + \lambda \left( 1 - \theta \right) \left< \strain_1 + \strain_2 \right>_- + 2\mu \left< \strain_i \right>_-, \qquad i = 1,2
\end{equation}
\end{linenomath}
The positive and negative parts of the elastic strain energy, stress and elasticity tensor can thus be written as
\begin{linenomath}
\begin{align}
	\psi_0^\pm \left( \straintensor, \phi \right) &= \frac{\lambda}{2} \left( 1 - \theta \right) \left< \strain_1 + \strain_2 \right>^2_\pm + \mu \left[ \left< \strain_1 \right>_\pm^2 + \left< \strain_2 \right>_\pm^2 \right] \\
	\stresstensor_0^\pm \left( \straintensor, \phi \right) &= \lambda \left( 1 - \theta \right) \left< \strain_1 + \strain_2 \right>_\pm \sum_{a=1}^2 \mathbf{M}_a + 2\mu \sum_{a=1}^2 \left< \strain_a \right>_\pm \mathbf{M}_a \\
	\mathbb{C}^\pm_0 \left( \straintensor, \phi \right) &= \sum_{a=1}^2 \sum_{b=1}^2 \left[ \lambda \left( 1 - \theta \right) \sgn \left( \strain_1 + \strain_2 \right) + 2\mu \delta_{ab} \sgn \left( \strain_b \right) \right] \mathbf{M}_a \mathbf{M}_b + \mu \frac{\left< \strain_1 \right>_\pm - \left< \strain_2 \right>_\pm}{\strain_1 - \strain_2} \mathbb{G}_{12} + \mu \frac{\left< \strain_2 \right>_\pm - \left< \strain_1 \right>_\pm}{\strain_2 - \strain_1} \mathbb{G}_{21}.
\end{align}
\end{linenomath}